\documentclass[11pt,oneside,letterpaper]{article}
\usepackage{amssymb}
\usepackage{amsmath}
\usepackage[dvips]{graphicx}
\usepackage{setspace}
\usepackage{amsfonts}
\usepackage{fancyhdr}
\usepackage{xcolor}
\usepackage{graphicx}
\usepackage{rotating}
\usepackage{comment}
\usepackage{color}
\usepackage{cite}
\usepackage{braket}

\usepackage{tikz}

\usepackage{subfigure}
\usetikzlibrary{decorations.pathmorphing}

\definecolor{darkgreen}{rgb}{0,0.5,0}
\definecolor{darkblue}{rgb}{0,0,0.6}
\definecolor{purple}{rgb}{0.4,.2,0.7}

\newcommand{\be}{\begin{equation}}
\newcommand{\ee}{\end{equation}}

\newcommand{\XX}{{\mathcal X}}
\newcommand{\TT}{{\mathcal T}}

\newcommand{\YY}{{\mathcal Y}}
\newcommand{\PP}{{\mathcal P}}
\newcommand{\SP}{{\sf P}}
\newcommand{\HH}{{\mathcal H}}
\newcommand{\eff}{{\mathrm{eff}}}
\newcommand{\RR}{{\mathbb R}}
\newcommand{\ZZ}{{\mathbb Z}}

\usepackage[colorlinks=true,citecolor=darkgreen,linkcolor=black,urlcolor=purple]{hyperref}

\usepackage{pdfsync}

\makeatletter
\newcommand*{\defeq}{\mathrel{\rlap{%
                     \raisebox{0.3ex}{$\m@th\cdot$}}%
                     \raisebox{-0.3ex}{$\m@th\cdot$}}%
                     =} 
\makeatother

\DeclareMathOperator{\Tr}{Tr}
\def\be{\begin{eqnarray}}
\def\ee{\end{eqnarray}}

\newcommand{\la}{\langle}

\newcommand{\bea}{\begin{eqnarray}}
\newcommand{\eea}{\end{eqnarray}}
\def\ben{\begin{equation}}
\def\een{\end{equation}}
\def\half{{\textstyle{\frac{1}{2}}}}

     \let\r=v

\let\la=\label

\def\be{\begin{equation}}
\def\ee{\end{equation}}
\def\ba{\begin{array}}
\def\ea{\end{array}}

\def\ba#1\ea{\begin{align}#1\end{align}}
\def\bs#1\es{\begin{split}#1\end{split}}

\allowdisplaybreaks  

\numberwithin{equation}{section}

\def\nref#1{(\ref{#1})}
\def \la {\label}   
\def \be {\begin{equation}}
\def \ee {\end{equation}}
\def \half {{1\over 2}}	
\def \JM#1 {{\color{blue}  JM: #1 }}
\def \EW#1 {{\color{red}  EW: #1 }}
\def \YC#1 {{\color{darkgreen}  YC: #1 }}

\def \JM#1 { }
\def \EW#1 { }
\def \YC#1 { }

\def \HP{Horowitz-Polchinski } 

\begin{document}
\onehalfspacing

\begin{center}

~
\vskip5mm

{\LARGE  {   On the black hole/string transition \\
}}

\vskip10mm

Yiming Chen$^1$,\ \ Juan Maldacena$^{2}$ and \ Edward Witten$^2$

\vskip15mm

{\it $^{1}$ Jadwin Hall, Princeton University, Princeton, New Jersey, USA } \\
\vskip5mm

{\it $^{2}$ Institute for Advanced Study, Princeton, New Jersey, USA } \\
\vskip5mm

\vskip5mm

\end{center}

\vspace{4mm}

\begin{abstract}
\noindent
We discuss aspects of the possible transition between small black holes and highly excited fundamental strings. We focus on the connection between black holes and the self gravitating string solution of Horowitz and Polchinski. This solution is interesting because it has non-zero entropy at the classical level and it is natural to suspect that it might be continuously connected to the black hole.    Surprisingly, we find a different behavior for heterotic and type II cases. For the type II case we find an obstruction to the idea that the two are connected as classical solutions of string theory, while no such obstruction exists for the heterotic case. We further provide a linear sigma model analysis that suggests a continuous connection for the heterotic case. We also describe a solution generating transformation that produces a charged version of the self gravitating string. This provides a fuzzball-like construction of near extremal configurations carrying fundamental string momentum and winding charges.
We provide formulas which are exact in $\alpha'$  relating the thermodynamic properties of the charged and the uncharged solutions. 
  \end{abstract}


\pagebreak
\pagestyle{plain}

\setcounter{tocdepth}{2}
{}
\vfill
\tableofcontents

\newpage


\section{Introduction}

The Schwarzschild black hole is an important solution to Einstein's equations. Classical string theory is a certain deformation of Einstein's theory and it would be nice to be able to find the stringy version of Schwarzschild's solution, which should be described as a certain worldsheet CFT. It is simpler to think about the Euclidean problem so that time is a circle of length $\beta$,  asymptotically. For large $\beta$, $\beta\gg l_s$, we expect to have the usual black hole with some $\alpha'$ corrections \cite{Callan:1988hs,Myers:1987qx,Chen:2021qrz}. When $\beta $ approaches a special value, $\beta_H$,  called the inverse Hagedorn temperature, we expect that the thermal ensemble should not be well defined in the bulk, even far from the black hole. In low number of dimensions, Horowitz and Polchinski found an interesting solution involving a winding condensate when $\beta -\beta_H \ll l_s$ \cite{Horowitz:1997jc}. 
This solution can be viewed as a self gravitating gas of hot strings, a kind of ``string star''. We will refer to the solution as the ``\HP solution''. 

It had also been speculated that small black holes would turn into fundamental strings at this temperature \cite{Bowick:1985af,Susskind:1993ws,Horowitz:1996nw}, see also \cite{Giveon:2006pr}. Since the \HP solution already describes an oscillating string, one could wonder whether the black hole and the \HP solution are continuously connected as conformal field theories. 

One reason for asking this question is that both the \HP   and the black hole solutions have a classical entropy. However, for the \HP solution we have a clear understanding of the microstates in the Lorentzian signature theory: they are just the states of a highly oscillating (and self gravitating) string. For the black hole we do not have a similar description in the gravity theory. So one might hope that by understanding this transition one would get some insight about black hole microstates \cite{Susskind:1993ws,Horowitz:1996nw}.

We will argue that in the type II string theory case, they are {\it not} smoothly connected as worldsheet supersymmetric field theories (even if one does not assume conformal invariance).  The argument involves computing the supersymmetric index  \cite{Witten:1982df} of the two solutions and finding that it is different on the two sides. The black hole and the \HP solution  could still be continuously connected for Type II by passing through a point where strong (string) coupling effects are important.  In the heterotic case, the supersymmetric index vanishes on both sides and so presents no obstruction to a smooth interpolation,
and, as we explain in section \ref{analysis}, there are no other topological obstructions to such an interpolation.  
This difference between the type II and heterotic cases is perhaps the most unexpected result of our analysis. 

More specifically, we will discuss a linear sigma model construction which is expected to flow in the IR either to the \HP solution or the black hole solution, after fine tuning one parameter. This fine tuning is expected because both solutions have a negative mode. We have not analyzed the full flow. However, we have classically integrated out the massive modes to land on a non-linear sigma model whose geometry has the same topology as either the \HP or black hole solutions. These manifolds are not Ricci flat and are expected to flow to the actual solutions (this is the step we have not verified). Nevertheless, we can see interesting features already in this description. These linear sigma models involve free fields with an appropriate superpotential. For generic values of the parameters the supersymmetric vacua describe a manifold that has the topology of the \HP solution or the black hole solution. We get one or the other depending on the detailed values of the parameters. At the transition point, in the type II case, there is a new branch that classically opens up. Quantum mechanically, this branch leads to extra massive vacua. But nothing special happens in the heterotic case. So, in the heterotic case, this suggests   a continuous transition between the two. In the type II case, there seems to be a special point where the two massive vacua merge with the non-linear sigma model branch.  In the type II case, the two branches (the \HP and black hole solutions) have a different spectrum of D-branes and at the transition point, some of the D-branes should become tensionless. This difference between the spectrum of D-branes can also be viewed as resulting from a difference between K-theory invariants on the two sides.

In the type II case, even though they  are not continuously connected as CFTs, we could still have a continuous connection but with a special point at which the    conformal field theory is singular and the classical limit of the thermodynamic quantities might not be analytic at that point. In other words, we could have a higher order phase transition. This is just one possibility, but others, such as first order phase transitions are also possible in principle. We will not give evidence for or against any of these possibilities in this paper. 

We also provide a solution generating procedure that produces a solution with momentum and winding string charges along an internal circle starting from the neutral solution. This is a generalization of the procedure used to generate such solutions in the gravity approximation \cite{Horowitz:1992jp,Sen:1994eb}. We provide a simple formula relating the thermodynamic properties of the charged solution to those of the uncharged solution. This formula is based on two observations. First, the classical on shell action for string theory is given purely by a boundary term which can be computed at infinity in terms of the mass and the asymptotic property of the dilaton. Second, the solution generating transformation is a simple $O(2,2)$ transformation at infinity and can be used to determine the asymptotics of the new solution. 
Of course, this was well known for the black hole case \cite{Horowitz:1992jp,Sen:1994eb}. We have simply extended this to include all $\alpha'$ corrections. In particular, we also generate a new solution which is the charged version of the \HP solution. 

The charged version of the \HP solution can describe near extremal configurations with momentum and winding charges in four dimensions. In other words, instead of a {\it singular} black hole solution we get a {\it non-singular}  fuzzball-like solution that describes an oscillating string with momentum and winding along the internal circle. 
This solution exactly reproduces the entropy of an oscillating string with momentum and winding. 

 We work out these solutions both for the type II and heterotic strings. For the heterotic string we reproduce the expected answers both for BPS and non-BPS ``extremal'' charged configurations.\footnote{The ``extremal'' is in quotation marks because for the non-BPS case there are other configurations with lower energy.}  
 
 We also discuss an open string analog of the two possible configurations. This arises when we consider a brane/anti-brane system separated by some distance $L$. When $L$ is very large there is a classical solution that connects the brane and the anti-brane. While for small $L$ of order the string scale there is a solution with a condensate of the open string mode stretching between the brane and the antibrane that is about to become tachyonic, but is not yet tachyonic. This is a solution that follows from the same logic as that of the \HP solution. In this case, the supersymmetric index is the same on the two sides. And a linear sigma model argument also suggests that the two configurations could be continuously connected. 

This paper is organized as follows.

 In section \ref{sgsol} we review the \HP solution, slightly extending the work of \cite{Horowitz:1997jc} by providing the explicit form of the solution, giving the form of the solution in dimensions $D=4,5,6$ and presenting a formula for the classical entropy. We also interpret the \HP solution as a solution producing the decay of the Kaluza-Klein vacuum, analogous to \cite{Witten:1981gj},  which was expected from the observation in \cite{Atick:1988si} that the Hagedorn transition should be first order. 

In section \ref{BhHP} we present the linear sigma model construction for the heterotic and type II cases. We also discuss the computation of the supersymmetric index for both cases, and describe the reasons why, for the type II case, they cannnot be continuously connected as CFTs.   

In section \ref{ActionTree} we recall that  the on shell classical action of string theory is a boundary term, and we show it can be expressed in terms of the asymptotic data of the metric and dilaton.  

In section \ref{GenCh} we describe the procedure to generate the charged solution starting from the uncharged one. We treat both the type II and heterotic cases, which involve slightly different transformations.

In section \ref{OpenString} we discuss an open string analog of the Horowitz-Polchinski solution and the black hole solution. 

In section \ref{concl} we conclude with a few remarks.

 \section{The self gravitating string solution} 
 \la{sgsol}

 In \cite{Horowitz:1997jc} Horowitz and Polchinski constructed a solution describing a self gravitating string. In this section we review their work while adding a few additional results and comments. See \cite{Damour:1999aw,Khuri:1999ez,Kawamoto:2015zha} for further perspectives on this solution. 
 
 The solution is most simply viewed as a solution with a localized winding condensate, which arises as follows.  A free string theory at finite temperature in thermal equilibrium can be described in terms of a Euclidean theory on a circle of length $\beta$, or radius $R= \beta/2\pi$ (we will work in terms of the ``radius'' of the circle instead of the length just to eliminate factors of $2\pi$ in some formulas). String theory has a limiting temperature called the Hagedorn temperature \cite{Hagedorn:1965st}. At this special value of the radius, $R_H$, a winding mode becomes massless \cite{Kogan,Sathiapalan}.  This mode would be tachyonic for $R< R_H$ (or temperatures $T> T_H$). In general,  the mass of this winding mode, viewed as a $D-1$ dimensional field, is  
\be \la{MassII}
 m^2 = { R^2 - R_H^2 \over \alpha'^2 } ~,~~~~~~   R_H^{\rm Bosonic} =2 l_s ~,~~~~~~~~~~R_H^{\rm Type ~II} = \sqrt{2}l_s ~,~~~~l_s \equiv \sqrt{\alpha'} 
 \ee 
for the bosonic or type II strings or   
 \be   \la{MassHet}  m^2 =  { R^2 \over \alpha'^2}  + { 1 \over 4 R^2}- { R^2_H \over \alpha'^2}  - { 1 \over 4 R^2_H} ~,~~~~~~~~~~~R_H^{\rm Heterotic} =\left(1 + { 1 \over \sqrt{2} } \right)l_s
 \ee 
 for the heterotic string where the thermal mode has also a half unit of momentum
   \cite{OBrien,McGuigan,Atick:1988si}, in addition to the unit of winding. We can understand the need for this extra half unit of momentum as follows  (see also
   section \ref{QuanNum} below). 
   We consider a heterotic string wound on a circle with antiperiodic boundary conditions for the Green-Schwarz fermions\footnote{The same answer can be obtained by looking at the string in the NS sector (of the RNS formalism) and imposing a GSO projection with an additional factor of $(-1)^w$, where $w$ is the winding number on the thermal circle.}. Parametrizing the momentum on the circle as $p =n/R$ we find the conditions 
   \be 
  L_0^L = - {\alpha' m^2 \over 4 } +  {\alpha' \over 4 }  \left( { R \over \alpha'} +{ n \over R} \right)^2 =   1    ~,~~~~~~~L_0^R = - {\alpha' m^2 \over 4 } +  {\alpha' \over 4 }  \left( { R \over \alpha'} -{ n \over R} \right)^2 =   \half.
   \ee 
From the difference between the two equations we find that $n=\half $.  
   
  We will now consider the effective theory for temperatures $T\sim T_H$ or $R - R_H \ll l_s $. In this regime the winding mode is a light field in the $D-1$ dimensional theory that results by Kaluza Klein reduction on the circle.  We denote the winding mode by $\chi$ and we write the radius of the circle as $Re^\varphi$ where $R$ is the asymptotic value of the radius. 
  Then the $d=D-1$ dimensional theory contains the following terms (omitting others that will be zero on the solution):
  \be \la{Action}
  I_d = { 1 \over 16 \pi G_N} \int d^d x \sqrt{g} 
  e^{ - 2 \phi_d } \left[ - {\cal R} - 4 (\nabla \phi_d)^2 +  (\nabla \varphi)^2 + |\nabla \chi|^2 + m^2(\varphi) |\chi|^2 \right] 
  \ee 
  where we wrote the action in terms of the $d$ dimensional dilaton. The field $\chi$ is complex because the string can wind the circle in either direction. However, for the solutions we will discuss,  the phase of the field is constant and we can view it as a real field. 
 The mass can be expanded as 
 \be \la{massphi}
 m^2(\varphi)  = m^2_\infty + { \kappa \over \alpha'} \varphi   + o(\varphi^2)  ~,~~~~~~~~~~~m^2_{\infty} \sim  { \kappa (R-R_H) \over \alpha' R_H} 
 \ee 
 where we expanded the mass in \nref{MassII} or \nref{MassHet} in powers of $\varphi$ after the replacement $R \to e^\varphi R$ in those formulas. The first term is the asymptotic value given by \nref{MassII} or \nref{MassHet},  expanded to first order in $R-R_H$. $\kappa$ is the first derivative with respect to $\varphi$, or $\kappa \equiv \alpha' R\partial_R m^2$. In other words, 
 \be 
  \kappa^{\rm Bosonic} = 8 ~,~~~~~~~
  \kappa^{\rm Type~II}= 4 ~,~~~~~~~
  \kappa^{\rm Heterotic} = 4 \sqrt{2}
 \ee 
  where we have set $R\to R_H$ after taking the derivative, since we are expanding for small $R-R_H$. 
  
    It is important in this discussion that $\kappa$ is very large compared to $\alpha' m_\infty^2$, 
and therefore it leads to the most important interaction term in the Lagrangian, which couples $|\chi|^2$ and $\varphi$. For this reason we can neglect all other interaction terms in the Lagrangian and we can set the $d$ dimensional metric to the identity and $\phi_d$ to a constant. The equations of motion for $\varphi$ and $\chi$ then become 
  \bea \la{chieq} 
  0 &=& -\nabla^2 \chi + \left( m_\infty^2 + { \kappa \over \alpha'} \varphi \right) \chi    ,
   \\
   0 &=& - 2 \nabla^2 \varphi + {\kappa \over \alpha'} |\chi|^2 .
  \eea
  
  We can solve the second equation and obtain 
  \be \la{PotExp}
  \varphi(x) =  - \frac{\kappa }{2 (d-2) \omega_{d-1} \alpha' } \int d^d y { |\chi(y)|^2 \over |\vec x - \vec y|^{d-2} } 
  \ee 
 where $\omega_{d-1}$ is the area of the unit $(d-1)$-sphere. Inserting this in \nref{chieq} we get a single integro-differential equation for $\chi$:
\begin{equation}\la{EOM}
 - \nabla^2 \chi (x) - \frac{\kappa^2 }{2 (d-2) \omega_{d-1} \alpha'^2 } \int d^d y { |\chi(y)|^2 \over |\vec x - \vec y|^{d-2} } \chi(x) = - m_{\infty}^2 \chi (x) .
\end{equation}
It is useful to use rescaled variables\footnote{The case of $d=4$ needs to be treated slightly differently; see sec. \ref{sec:d=4}.}
   \be \la{Rescale}
   \hat x = x  \frac{m_\infty}{\sqrt{\zeta}} ~,~~~~~~
   \chi(x) =\frac{\alpha' \sqrt{2(d-2) \omega_{d-1}}}{\kappa} \frac{m_{\infty}^2}{\zeta } \hat \chi( \hat x ) 
   \ee 
where $\zeta $ is a numerical constant to be determined. 
   Then the equation takes the form 
   \be \la{EqID}
   -\hat \nabla^2 \hat \chi(x) - 
   \int d^d \hat y { | \hat \chi (\hat y) |^2 \over |\hat x - \hat y|^{d-2} }\hat{\chi} (\hat x)  = - \zeta \hat \chi(\hat x) .
   \ee 
   In terms of the rescaled variables, $\varphi$ is given by
   \begin{equation} \la{RescalePhi}
   \varphi =  \frac{m_{\infty}^2 \alpha'}{\kappa \zeta}  \hat{\varphi}(\hat{x})~, \quad ~~~~~~~~\hat{\varphi}(\hat{x}) \equiv  - \int d^d \hat{y} { |\hat{\chi} (\hat y)|^2 \over |\hat{ x} - \hat y|^{d-2} } ~.
   \end{equation}
So far, $\zeta$ has been an arbitrary constant. We now  impose the additional condition 
  \be \la{IntChi}
  \int d^d \hat{x} |\hat \chi(\hat x) |^2 =1  ~~\longrightarrow ~~~\int d^d x |\chi(x)|^2 = \frac{2\alpha'^2 (d-2) \omega_{d-1}}{\kappa^2} \zeta^{{d \over 2 } -2} m_\infty^{4-d} .
 \ee 
  With this extra condition on $\hat \chi$, the equation  \nref{EqID} becomes an eigenvalue equation for $\zeta$, in the sense that there exists a solution satisfying (\ref{IntChi}) only for discrete values of $\zeta$ (for $d<6$).   In the ``ground state,'' a solution with $\chi$ everywhere positive, $\zeta $ has some numerical value independent of the value of $R-R_H$. We will discuss special cases below, give the explicit values for $\zeta$ for $d=3,4,5$,  and explain that there is no normalizable solution for $d\geq 6$.

  Note that we have scaled out completely the dependence on $R-R_H$ which comes in through $m_\infty$. In particular, from (\ref{Rescale}) we see that the size  of the solution scales as 
  \be \la{SizeGen}
  \ell \propto  \frac{1}{m_\infty} \propto \sqrt{ l_s \over R -R_H} .
  \ee 
  So as the temperature approaches the Hagedorn temperature the solution becomes larger.
  
   A posteriori, we can check that the approximations were correct in the regime of validity of the \HP solution, which is $\alpha' m^2_\infty \propto (R-R_H)/l_s  \ll 1 $. 
First,  note that $\varphi$ has an explicit factor of $m^2_\infty$ in \nref{RescalePhi}, ensuring that $\varphi \ll 1$. In addition, 
  we find that $\chi \ll 1$ from \nref{Rescale}. This ensures we can neglect higher order terms in the Lagrangian that involve winding. These include terms of order $|\chi|^4 $ \cite{Dine:2003ca,Schulgin:2011zb}, as well as interactions of the form $ \chi^2 \bar \chi_{2}$ that would create fields with winding number two.\footnote{The winding number two mode associated with the tachyon is projected out by the GSO projection in the Type II case, but we can consider other modes with winding number two. }   These interactions are present but their effects are parametrically smaller than the terms that were kept.   
    
  We are also interested in computing the mass and the entropy of the solution. 
  It is easiest to compute the entropy first. For this purpose we need to recall that when we take the derivative of the action with respect to $\beta$ we want to keep the $D$ dimensional dilaton fixed, so that at infinity we have $e^{ - 2 \phi_d} = e^{ -2 \phi_D} \beta $ with $\phi_D$ fixed. This explicit factor of $\beta$ doesn't contribute when we compute the entropy, so
  \be 
  S = (1 - \beta \partial_\beta) \log Z = 
 (1 - \beta \partial_\beta)  (-I_d) = \int d^d x \, \sqrt{g} e^{-2 \phi_d} \beta  \partial_\beta {\cal L}_d 
  \ee 
  where we consider only the explicit $\beta$ dependence of the $d$-dimensional Lagrangian, since the implicit dependence vanishes due to the equations of motion (we are evaluating the action on a solution). This explicit dependence comes from the dependence of $m(R)$ and gives 
  \be 
  S = { 1 \over 16 \pi G_N} \int d^dx\, \sqrt{g} e^{-2 \phi_d} (R \partial_R m^2) |\chi|^2  \sim  {\kappa \over \alpha'} { \beta_H \over 16 \pi G_N} \int d^dx  |\chi|^2.   \la{EntHP}
  \ee 
  where we expanded the entropy to leading order in $R-R_H$.  
  We see that this is a classical solution which has a non-zero entropy! It shares this feature with the black hole solution. Here it is arising because the winding mode has a mass that depends on a non-local feature, namely the total size of the circle. Otherwise,  a completely local $D$ dimensional Lagrangian on a spacetime where the time circle never shrinks would lead to zero entropy \cite{Gibbons:1976ue}. 
  Since the answer depends on the integral of $|\chi|^2$, we can use \nref{IntChi} to determine its value.

 Once we have the entropy, we can estimate the mass from thermodynamics, since the temperature is close to the Hagedorn temperature we expect that 
  $M \sim T_H S  $, so to first order in $R -R_H$,  
  \begin{equation} \la{MassThA}
	M = \frac{2 \alpha' (d-2) \omega_{d-1} }{\kappa} \frac{1 }{16\pi G_N} \zeta^{\frac{d}{2}-2} m_{\infty}^{4-d} \propto \left( R - R_H \right)^{\frac{4-d}{2}}.
\end{equation}
 We see that it has a behavior that is sensitive to the dimension $d$. We discuss some cases below in more detail. 
We will also check \nref{MassThA}  more explicitly by a direct computation of the mass. In writing this formula,  we have assumed that the asymptotic value of the $D$ dimensional dilaton has been absorbed into $G_N$. In other words, we  set this asymptotic value to zero,  $\phi_{D , \infty} =0$.

It is tempting to identify the field $\varphi$ with the Newtonian potential. Actually, the field $\varphi$ leads to both a Newtonian potential and a non-trivial 
dilaton in $D$ dimensions. This is physically related to the fact that different parts of a highly excited string attract each other through both gravity 
and the dilaton force, since in Einstein frame the tension of the string scales as $T \propto e^{4 \phi_D /(D-2)}$ and so that the string sources the dilaton field.
More precisely, the $D$ dimensional dilaton field is obtained from
\be \la{HPDil}
e^{  -2\phi_d} = e^{ - 2 \phi_D } \beta e^\varphi ~~~~ \longrightarrow ~~~~~\phi_D =   \half \varphi 
\ee 
 where we assume that $\phi_d$ is a constant, since its coupling to the 
winding mode involves $\alpha' m_\infty^2$, which is of higher order, and we fix an additive constant in $\phi_D$ by assuming that $\varphi$ and $\phi_D$ vanish at infinity.
 Similarly, one can extract the mass by writing the string metric in the large $\rho$ region  as  
\bea \la{HPasymp}
ds^2 &=& e^{ 2 \varphi} dt^2 + d\vec x^2 \sim   e^{ 4 \phi_D \over D-2 } \left(  f dt^2 + {d\rho^2 \over f } + \rho^2 
d\Omega_{D-2}^2 \right) ~,~~~~~{\rm as } ~~~~~ \rho \to \infty \\ 
& ~& f \sim  1 - { \mu \over \rho^{D-3} } = 1 + 2 { D-3 \over D-2} \varphi  ~,~~~~~~~ \notag
\eea 
where all functions are expanded to the first order as $\rho\to \infty$. 
To put the solution in this form, we introduced a radial coordinate $\rho$ that
 is not simply $|\vec x|$, but satisfies   $|\vec x|^2 \sim  e^{ 4 \phi_D\over D-2} \rho^2$ for large $|\vec x|$.   
 The factor in the parenthesis in \nref{HPasymp} is the Einstein frame metric, which could then be used to extract the mass
   \be \la{MassUs}
   M =  { (D-2) \omega_{D-2} \over 16 \pi G_N} \mu .
   \ee 
   We find agreement with \nref{MassThA} once we use the large $|x|$ limit of \nref{PotExp} together with \nref{IntChi}.

\subsection{Breakdown of the solution for very small $R-R_H$ }

   \begin{figure}[t]
    \begin{center}
    \includegraphics[scale=.25]{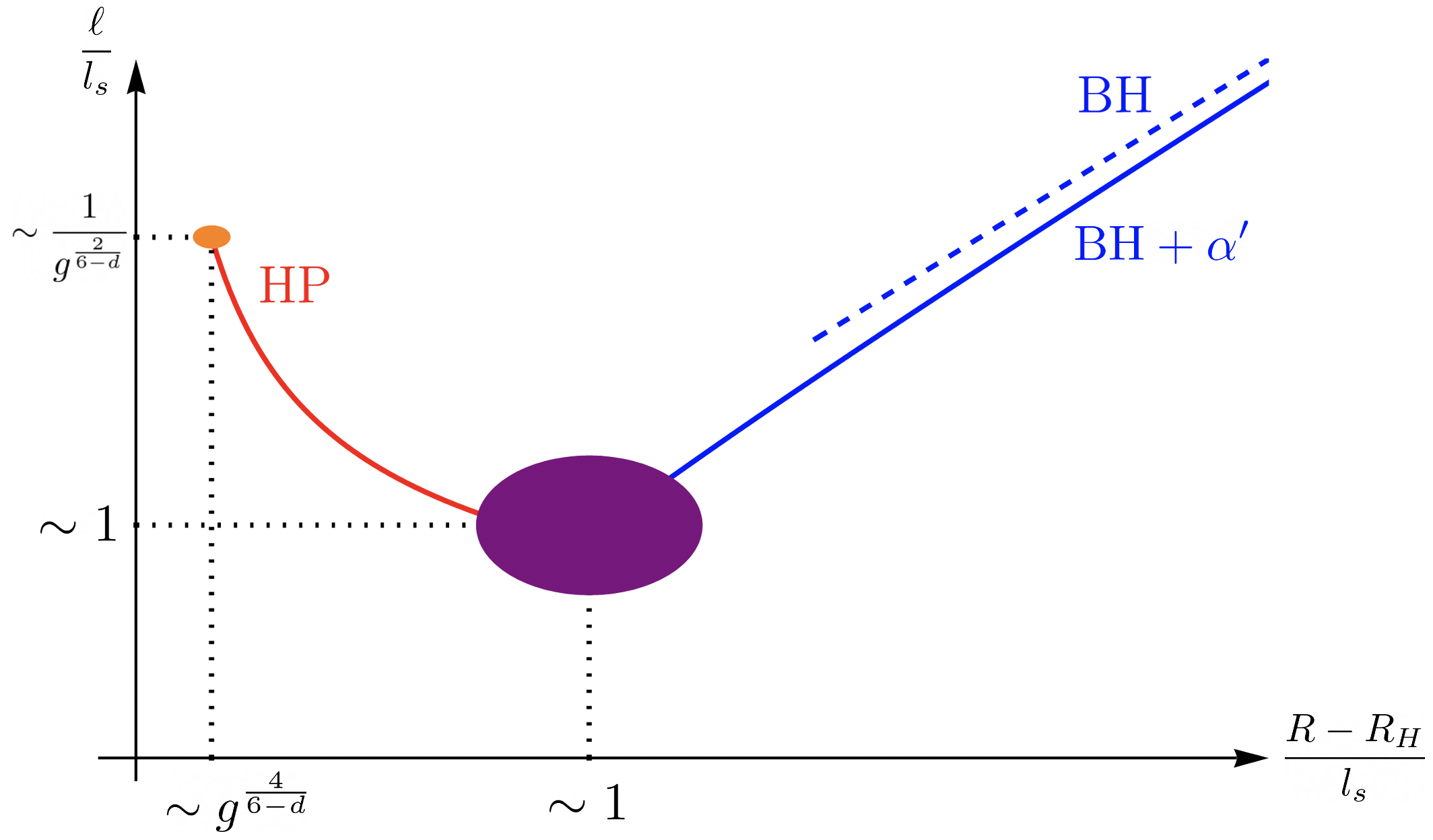}
    \end{center}
    \caption{In red, the size of the Horowitz-Polchinski (HP) solution as a function of the temperature, or $R-R_H$, for $d=3,4,5$. The HP description breaks down at $R-R_H \sim g^{\frac{4}{6-d}} l_s$, as denoted by the orange blob, where we should transit to the free string picture. In dashed blue, we also plot the curve for a black hole (BH), applicable when $R\gg l_s$. We also sketched the known leading $\alpha'$ correction for the black hole \cite{Callan:1988hs,Myers:1987qx,Chen:2021qrz} in solid blue, which lowers the temperature for a fixed size (the leading correction vanishes for $d=3$ in the bosonic or heterotic case). 
 The intermediate regime between the HP and BH descriptions, being the focus of section \ref{BhHP},  is represented by the purple blob and is not analytically tractable at present. The plot, as well as fig. \ref{fig:phase3d}, \ref{fig:phase4d}, \ref{fig:phase5d}, are only qualitative.}
   \label{fig:size}
\end{figure}

When we get too close to the Hagedorn temperature these solutions cease to be valid \cite{Horowitz:1997jc}. In this derivation we were treating them as classical solutions, meaning that we were thinking of the string coupling as being smaller than anything else. However, for a fixed and small string coupling, the solutions cease to be valid when  $R-R_H$ becomes a certain positive power of the coupling. This arises as follows. We can estimate the size of quantum fluctuations by looking at the change of the action under an order one uniform rescaling of $\chi$, $\chi \to \lambda \chi$. Of course, we could consider other fluctuations, but this will suffice for our estimate. This gives 
\be 
\delta I_d  \propto  {  \lambda^2  \over g^2 } m^2_{\infty} \int d^d x\, |\chi|^2 \propto \lambda^2  { m_{\infty}^{6 -d}   \over g^2 } .
\ee 
We want the coefficient of $\lambda^2$ to be much larger than one in order for the classical solution to be a good approximation. This happens when
\be  \la{ClassVal}
  { R-R_H \over l_s } \gtrsim g^{ 4 \over 6 - d} .
  \ee 
 
  We can wonder what happens for smaller values of $R-R_H$. The idea is that we go over to a free string description \cite{Horowitz:1997jc}. In the free string description the entropy and the size of the string are given by 
  \be 
  S \sim \beta_H M ~,~~~~~~~ \ell \sim l_s \sqrt{ l_s M} ,
  \ee 
  where the size is given by that of a random walk with $l_s M$ steps, each of size $l_s$ \cite{Mitchell:1987hr,Mitchell:1987th}. 
  Note, in particular that the size grows as $M$ increases.

 We now discuss some special cases in more detail.
 
 \subsection{$d=3$}\la{sec:d=3}


\begin{figure}[t]
    \begin{center}
    \includegraphics[scale=.25]{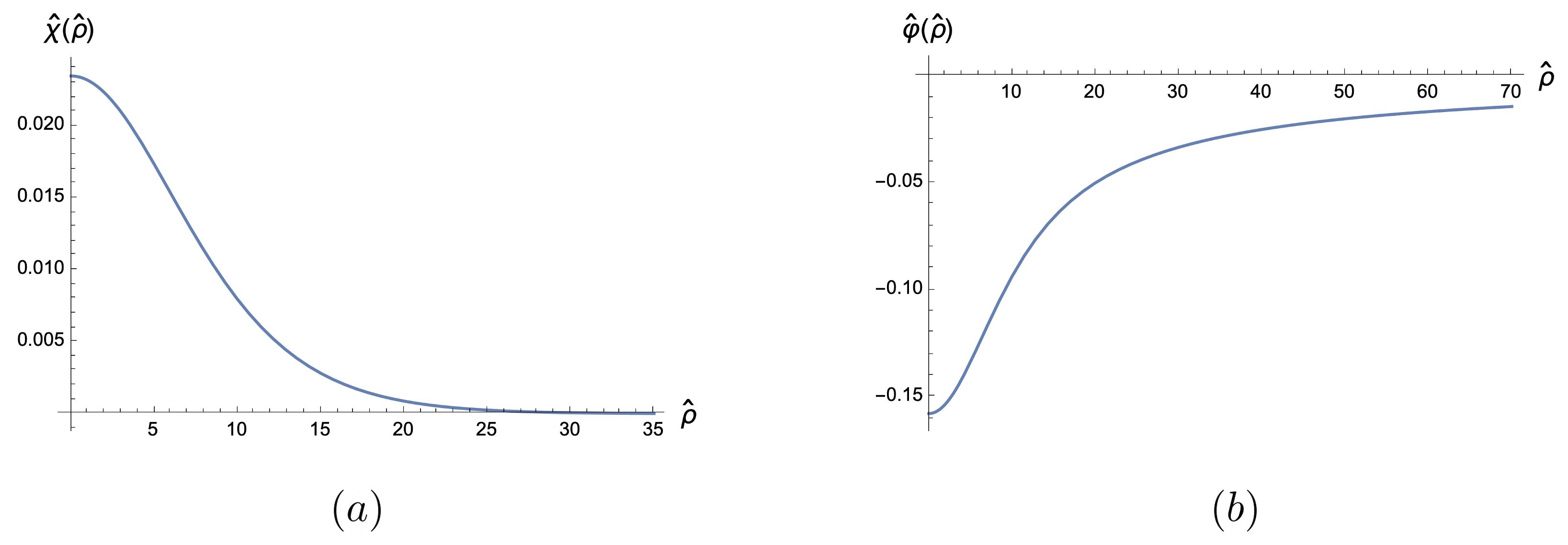}
    \end{center}
    \caption{Solution for $d=3$. (a) Plot of the rescaled profile $\hat \chi(\hat \rho)$. (b) Plot of the rescaled gravitational potential $\hat \varphi (\hat \rho) \equiv -\int d^3 \hat{y}\, \frac{|\hat{\chi} (\hat y)|^2}{|\hat{x}-\hat{y}|}$. The true values of $\chi$ and $\varphi$ are related to the rescaled ones through (\ref{Rescale}) and (\ref{RescalePhi}). 
}
    \label{fig:HP3d}
\end{figure}

The eigenvalue problem (\ref{EqID}) has a lowest energy solution, which is spherical symmetric and decays exponentially towards infinity. One can find this solution by numerically solving the equation (\ref{EqID}) under (\ref{IntChi}), and get $\zeta = 0.0813\pm 0.0001 $.\footnote{In practice, it is easier to solve the coupled differential equations (\ref{chieq}) using the shooting method than solving (\ref{EqID}) directly.} In fact, the equation (\ref{EqID}) for the case of $d=3$ has appeared before in the study of gravitating Bose condensates  \cite{PhysRev.187.1767,PhysRevD.35.3640,Bernstein}. We plot the radial profile of the rescaled solution $\hat{\chi}(\hat{\rho})$ and $\hat{\varphi}(\hat{\rho})$ in fig. \ref{fig:HP3d}. We have $\varphi(0) \approx - 1.9 \frac{m_{\infty}^2 \alpha'}{\kappa}$, so when $m_{\infty}^2 \alpha' \ll 1$ we indeed have $\varphi (0) \ll 1$, i.e. the gravitational backreaction is small.

\begin{figure}[t]
    \begin{center}
    \includegraphics[scale=.26]{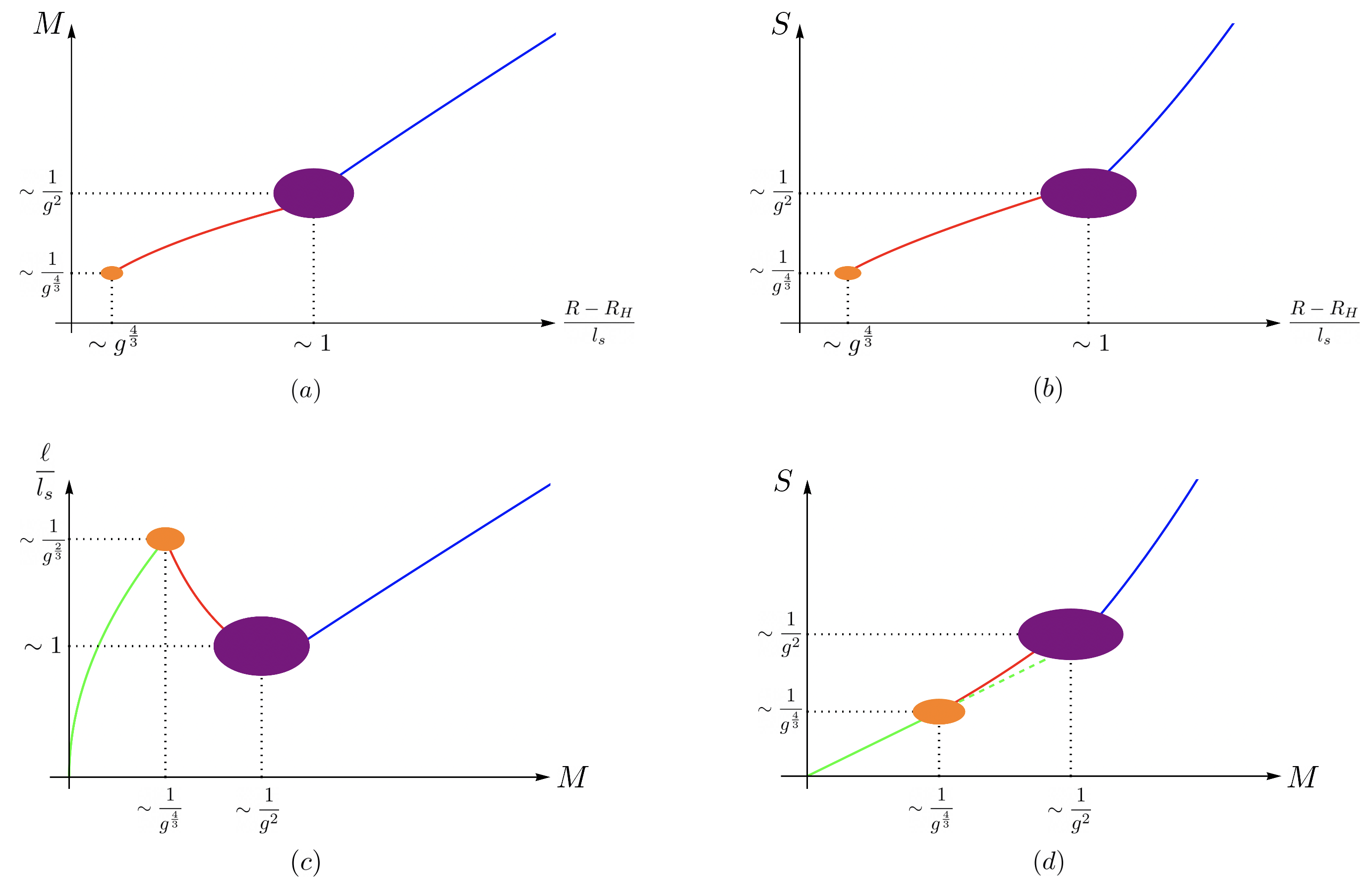}
    \end{center}
    \caption{The phase diagrams for $d=3$. In the plots, the blue, red, green lines represent the black hole, Horowitz-Polchinski solution and free strings, respectively. The orange and purple blobs denote the transition regions between the solutions. In (d), we note that the entropy of the Horowitz-Polchinski solution increases relative to the extrapolation of the free string result (represented by the green dashed line), see \nref{Scorr}. }
    \label{fig:phase3d}
\end{figure}
 
For $d=3$ the mass decreases as we approach the Hagedorn temperature as
 \begin{equation}\label{Mass3d}
 	M = \frac{ \alpha' }{\kappa} \frac{1}{2 G_N}  \frac{m_{\infty}}{\sqrt{\zeta}} = { 1 \over 2 G_N} \sqrt{ \alpha' (R-R_H) \over \kappa R_H \zeta}  . 
 \end{equation}
where we used \nref{massphi}. 
 The entropy also decreases as it is simply linear in the mass to the leading order of $R-R_H$. We plot the dependence of the mass and the entropy on the temperature in fig. \ref{fig:phase3d} (a), (b). For comparison, we plot the behavior of the black hole in the same figure.

In fact, using thermodynamics we can also calculate the next correction to the relation between the mass and the entropy. We note that
  \be \label{dSdM}
   { d S \over d M} = \beta = \beta_H + (\beta - \beta_H) = \beta_H  +  \xi M^2 , \quad \xi \equiv \zeta \frac{8\pi G_N^2 \kappa R_H}{\alpha'},
  \ee 
where we used (\ref{Mass3d}) in the last equality.
Integrating this relation, we get
\begin{equation}
 S = \beta_H  M + \frac{\xi}{3} M^3 . \la{Scorr}
\end{equation}
The second term is the leading correction away from the free string behavior due to self gravitation. The behavior of the size and the entropy as functions of the mass are depicted in fig. \ref{fig:phase3d} (c) and (d). These diagrams are particularly useful when we want to interpret the solutions as solutions in the microcanonical ensemble.


 \subsection{$d=4$}\la{sec:d=4}
 
 \begin{figure}[t]
    \begin{center}
    \includegraphics[scale=.25]{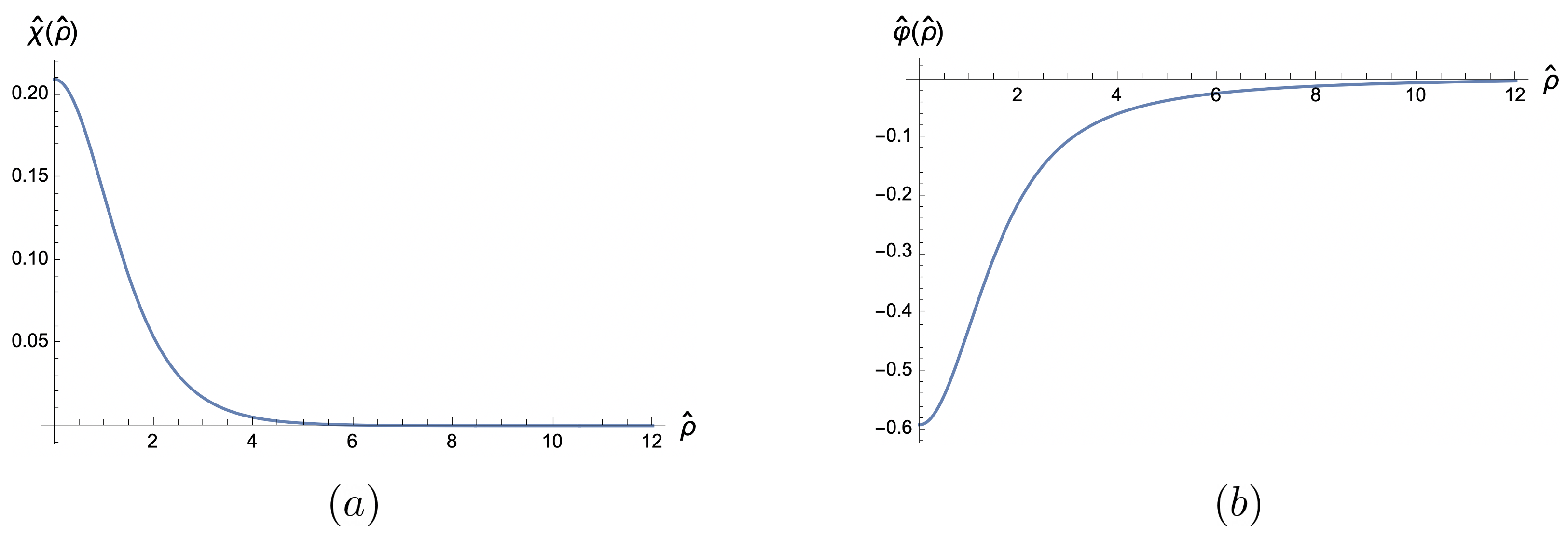}
    \end{center}
    \caption{Solution for $d=4$. (a) Plot of the rescaled profile $\hat \chi(\hat \rho)$. (b) Plot of the rescaled gravitational potential. 
    The true values of $\chi$ and $\varphi$ are related to the rescaled ones through (\ref{rescale4d}) and (\ref{RescalePhi4d}). }
    \label{fig:HP4d}
\end{figure}

\begin{figure}[t]
    \begin{center}
    \includegraphics[scale=.26]{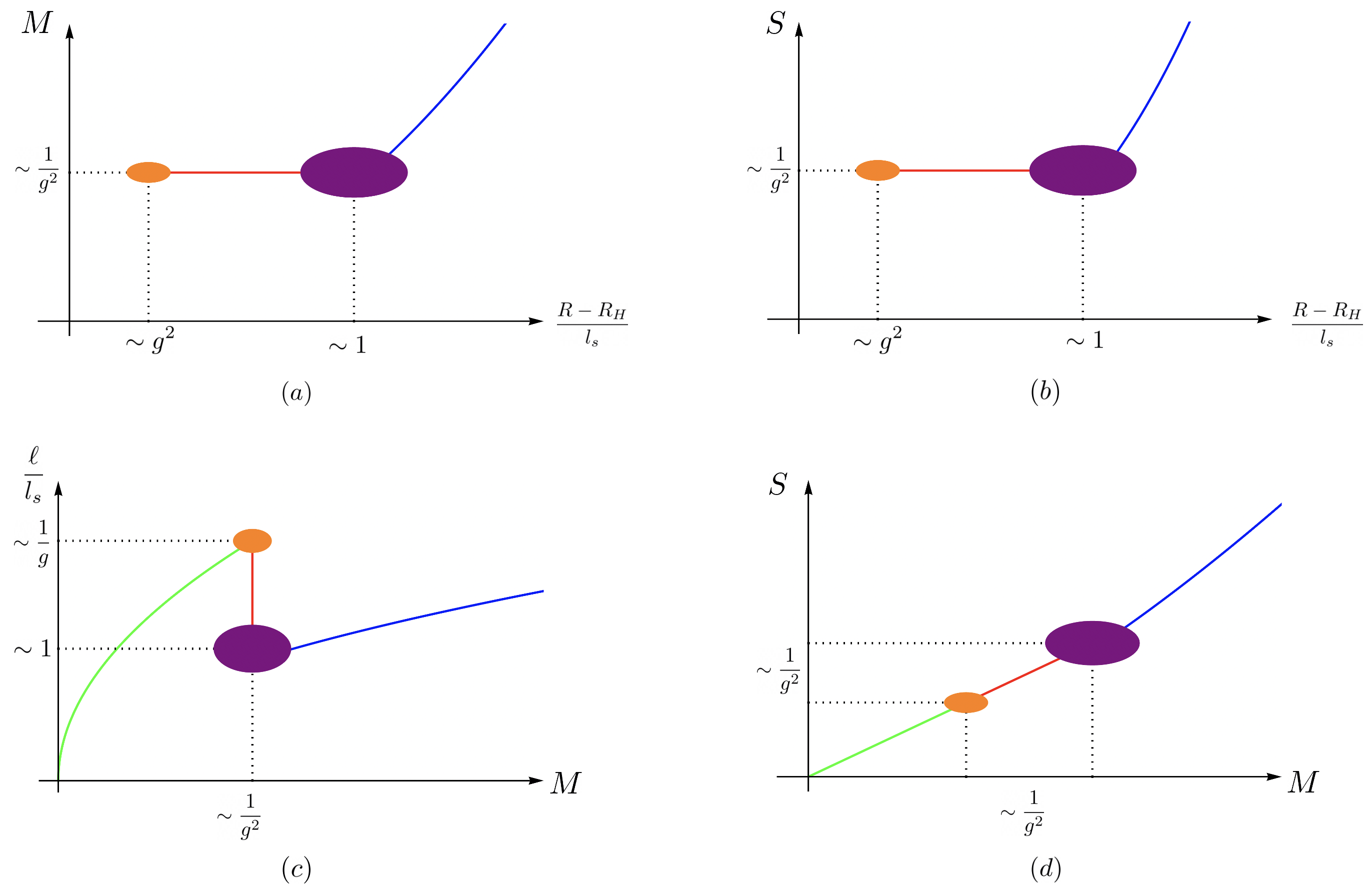}
    \end{center}
    \caption{The phase diagrams for $d=4$. In the plots, the blue, red, green lines represent the black hole, Horowitz-Polchinski solution and free strings, respectively. The orange and purple blobs denote the transition regions between the solutions.  }
    \label{fig:phase4d}
\end{figure}

The case of $d=4$ needs to be treated slightly differently, as the rescaling introduced in (\ref{Rescale}) cannot be used to fix the normalization of $\hat{\chi}$ to be one. Therefore we apply a different rescaling:
\begin{equation}\la{rescale4d}
	\hat{x} =m_{\infty} x , \quad \chi(x) = \frac{\alpha' \sqrt{2(d-2) \omega_{d-1}}}{\kappa} \frac{m_{\infty}^2}{ \zeta } \hat{\chi}(\hat{x}),
\end{equation}
under which the equation (\ref{EOM}) takes the form
\begin{equation}\label{eomd=4}
	- \hat{\nabla}^2 \hat{\chi} (x) - \frac{1}{\zeta^2}  \int d^4 \hat y { | \hat \chi (\hat y) |^2 \over |\hat x - \hat y|^{2} }\hat{\chi} (\hat x) = - \hat{\chi} (\hat{x}).
\end{equation}
In this case, the potential $\varphi$   is given in terms of the rescaled variables as
\begin{equation} \la{RescalePhi4d}
 \varphi = \frac{m_{\infty}^2 \alpha'}{\zeta^2 \kappa} \hat{\varphi} (\hat{x}) ,\quad \hat{\varphi}(\hat{x}) \equiv - \int d^4 \hat{y} { |\hat{\chi} (\hat y)|^2 \over |\hat{ x} - \hat y|^{2} } .
\end{equation}
Now we can further use the freedom of $\zeta$ to normalize $\hat{\chi}$
\begin{equation}
	\int d^4 \hat{x} |\hat \chi(\hat x) |^2 =1  \longrightarrow \int d^4 x |\chi(x)|^2 =  \frac{8\pi^2 \alpha'^2 }{ \zeta^2 \kappa^2}. 
\end{equation}
The mass $M$ of the solution in this case is given by
\begin{equation}\la{M4d}
	M \approx \frac{1}{2\pi R_H} S \approx \frac{\kappa}{\alpha'} \frac{1}{16\pi G_N} \int d^4 x\, |\chi(x)|^2 = \frac{\pi\alpha'}{2 \kappa G_N} \frac{1}{\zeta^2}.
\end{equation}
Therefore we can also express (\ref{eomd=4}) using the mass as
\begin{equation}
  - \hat{\nabla}^2 \hat{\chi} (x) - \frac{2 \kappa G_N M}{\pi\alpha'}  \int d^4 \hat y { | \hat \chi (\hat y) |^2 \over |\hat x - \hat y|^{2} }\hat{\chi} (\hat x) = - \hat{\chi} (\hat{x}).
\end{equation}
The normalizable solution we are looking for only exists for a particular value of the mass $M$. In other words, to the leading order in $R-R_H$, the mass is independent of the temperature and takes the value in (\ref{M4d}). One can find the explicit solution numerically, which gives $\zeta = 0.361 \pm 0.001$. The explicit solution is shown in fig. \ref{fig:HP4d}. The gravitational potential at the center is $\varphi(0) \approx -4.5 \frac{m_{\infty}^2 \alpha'}{\kappa}$.
The entropy is also independent of the temperature,  to the leading order in $R-R_H$, since it is proportional to the mass \nref{M4d}.  We plot the mass and the entropy as functions of the temperature in fig. \ref{fig:phase4d} (a), (b).

Since the mass is independent of the temperature to the order we are working, in this case we cannot use the method in (\ref{dSdM}) to work out the correction to the $S$ vs $M$ relation. 
 Similarly, we cannot determine how the size $\ell$ of the solution varies with the mass $M$. Therefore the plots in the microcanonical ensemble (fig. \ref{fig:phase4d} (c), (d)) are a bit more sketchy compared to the $d=3$ case.

 \subsection{$d=5$}\la{sec:d=5}

 \begin{figure}[t]
    \begin{center}
    \includegraphics[scale=.25]{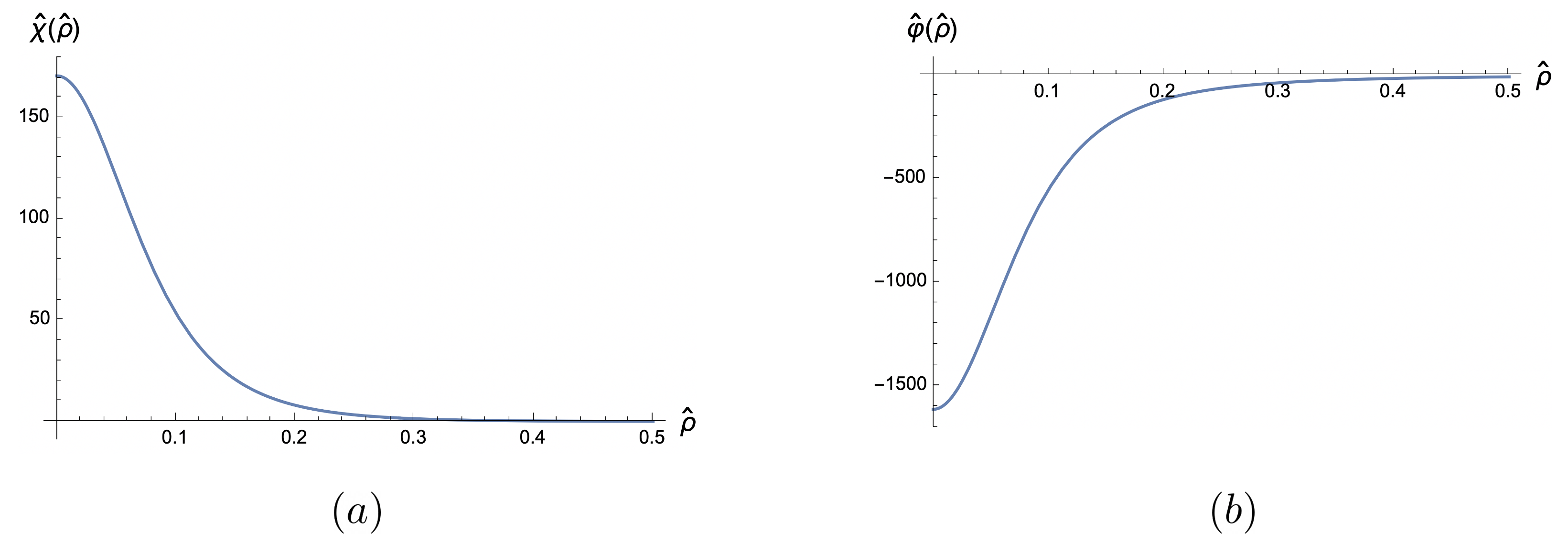}
    \end{center}
    \caption{Solution for $d=5$. (a) Plot of the rescaled profile $\hat \chi(\hat \rho)$. (b) Plot of the rescaled gravitational potential $\hat \varphi (\hat \rho) \equiv -\int d^5 \hat{y}\, \frac{|\hat{\chi} (\hat y)|^2}{|\hat{x}-\hat{y}|^3}$. The true values of $\chi$ and $\varphi$ are related to the rescaled ones through (\ref{Rescale}) and (\ref{RescalePhi}). }
    \label{fig:HP5d}
\end{figure}

By numerically solving the equation (\ref{EqID}) under the condition (\ref{IntChi}), one gets $\zeta \approx 108 \pm 1 $. The explicit solution is shown in fig. \ref{fig:HP5d}. The gravitational potential at the center is $\varphi (0) \approx - 14. \frac{m_{\infty}^2 \alpha'}{\kappa}$. The mass increases as we approach the Hagedorn temperature as 
\begin{equation}
	M = \frac{ \pi \alpha' }{\kappa G_N} \frac{\sqrt{\zeta}}{m_{\infty}} = \frac{\pi}{G_N} \left( \frac{\alpha'}{\kappa}\right)^{\frac{3}{2}} \sqrt{\frac{ \zeta R_H }{R-R_H}},
\end{equation}
where we used (\ref{massphi}).

\begin{figure}[t]
    \begin{center}
    \includegraphics[scale=.28]{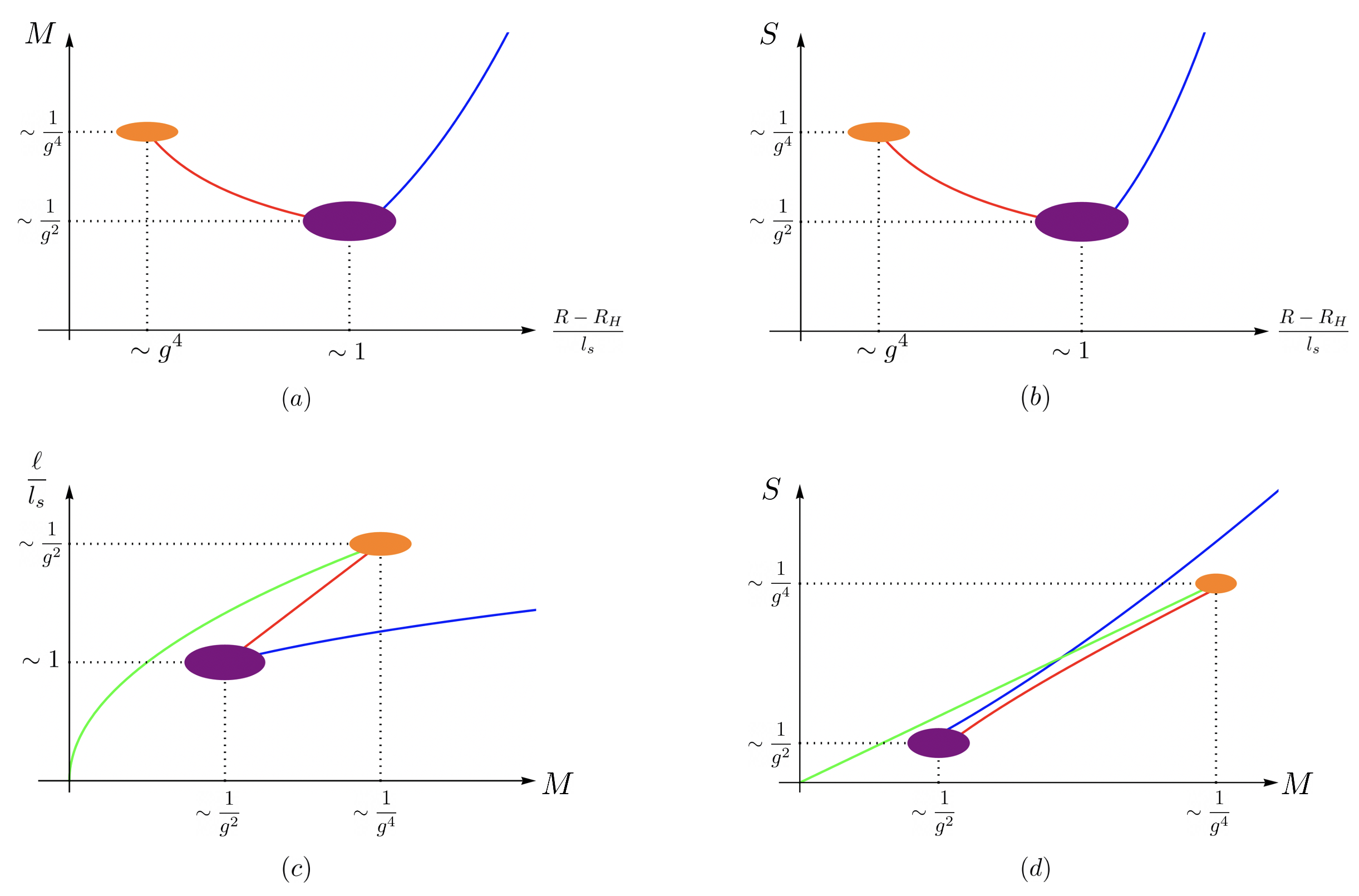}
    \end{center}
    \caption{The phase diagrams for $d=5$. In the plots, the blue, red, green lines represent the black hole, Horowitz-Polchinski solution and free strings, respectively. The orange and purple blobs denote the transition regions between the solutions. }
    \label{fig:phase5d}
\end{figure}

As before, by using thermodynamics we can also calculate the next correction to the relation between the mass and the entropy.  
 \be 
   { d S \over d M} =  \beta = \beta_H + (\beta - \beta_H) = \beta_H   + \frac{\xi}{M^2},\quad \xi \equiv \frac{2 \zeta R_H}{G_N^2} \left( \frac{\pi \alpha'}{\kappa} \right)^3 .
  \ee 
Integrating this we get
\begin{equation}\la{Scorr5d}
 S = \beta_H  M - \frac{\xi}{M}.
\end{equation}
We plot the mass and the entropy as functions of the temperature in fig. \ref{fig:phase5d} (a), (b).

In the canonical ensemble, the Horowitz-Polchinski solution and the black hole apply for separate regimes of the temperature, just as the cases for $d=3,4$. However, the behavior in the microcanonical ensemble is particularly interesting in $d=5$, as can be seen from fig. \ref{fig:phase5d} (c), (d). There exists a range of mass $\frac{1}{g^2} \ll \ell_s M \ll \frac{1}{g^4}$ where the Horowitz-Polchinski solution overlaps with the black hole and the free string. The Horowitz-Polchinski solution has the intermediate size, while it has the smallest entropy. As we will show in sec. \ref{sec:nmodes}, the Horowitz-Polchinski solution in $d=5$ is unstable in the microcanonical ensemble. This suggests that in the Lorentzian picture, the self-gravitating string solution is unstable towards either collapsing into a black hole, or expanding and forming a more dilute string gas.

   \subsection{$d\geq 6$}\la{sec:dgeq6}
   
  For $d\geq 6$, it can be shown that there do not exist normalizable solutions to (\ref{EqID}) and (\ref{IntChi}).  One way to see this is to consider the effective action for the winding mode, which can be derived by integrating out $\varphi$ from the original action (\ref{Action}):
\begin{equation}\la{Ieff}
\begin{aligned}
 I_{\rm eff} & = \frac{1}{16 \pi G_N}  \left[ \int d^d x \,  |\nabla \chi|^2  - \frac{\kappa^2}{4(d-2) \omega_{d-1} \alpha'^2} \int d^d x \int d^d y \frac{|\chi (x)|^2 |\chi (y)|^2}{ |\vec{x} - \vec{y}|^{d-2}} +    \int d^d x \, m^2_{\infty} | \chi|^2 \right]  \\
& \equiv I_1 - I_2 + I_3,
\end{aligned}
\end{equation}
where $I_1, I_2, I_3 > 0$ correspond to the three terms on the first line, respectively. Here we are only keeping the expression to the leading order in $(R-R_H)$, so that we can set the metric to be flat and the dilaton to be zero. The equation of motion (\ref{EOM}) follows from this action. Assuming that there exists a normalizable solution $\chi_*$, we can derive the ratios of the on-shell values $I_{1,*},I_{2,*},I_{3,*}$ by considering variations $\chi (x) = \lambda \chi_* (x/\gamma)$. It is easy to see that under such variations,
\begin{equation}
	I_{\rm eff} = \lambda^2 \gamma^{d-2} I_{1,*}  - \lambda^4\gamma^{d+2} I_{2,*}  + \lambda^2\gamma^d I_{3,*},
\end{equation}
Since $\chi_*$ is a solution, we have
\begin{equation}
 \left.\frac{\partial I_{\rm eff}}{\partial \lambda} \right|_{\lambda , \gamma =1} =  \left.\frac{\partial I_{\rm eff}}{\partial \gamma} \right|_{\lambda , \gamma =1} = 0,
\end{equation}
which gives
\begin{equation}\la{I2I3inI1}
	I_{2,*} = \frac{2 I_{1,*}}{d-2}, \quad ~~~~~~~~~~~~I_{3,*} = \frac{(6-d) I_{1,*}}{d-2}.
\end{equation}
From the second equation, we immediately see that consistency with $I_{1,*},\,I_{3,*}>0$ requires $d<6$, which shows that there are no normalizable solutions in $d\geq 6$.
Also, note that (\ref{ClassVal}) suggests, even if classical solutions were found in $d\geq 6$, there does not exist a parameter regime where we can satisfy both $R-R_H \ll l_s$ and $g\ll 1$ so that the solution is trustworthy.


As a side remark, taking (\ref{I2I3inI1}) back into (\ref{Ieff}), we can find an explicit expression for the free energy $F$ of the Horowitz-Polchinski solution
\begin{equation}
	F = I_{\rm eff} = \frac{2  I_{3,*} }{6-d} = \frac{2}{6-d} \frac{1}{16\pi G_N} \int d^dx\, m_{\infty}^2 |\chi|^2  = \frac{2}{6-d} \frac{1}{16\pi G_N} \frac{2\alpha'^2 (d-2) \omega_{d-1}}{ \kappa^2} \zeta^{\frac{d}{2} - 2} m_{\infty}^{6-d}
\end{equation}
where we used (\ref{IntChi}) in the last equality.\footnote{The expression should be modified in $d=4$ as $I_{\rm eff} = \frac{1}{16\pi G_N} \frac{8\pi^2 \alpha'^2}{\zeta^2 \kappa^2} m_\infty^2$ since we defined $\zeta$ differently.} We see that the free energy is positive, and it decreases as $(R - R_H)^{\frac{6-d}{2}}$ as we approach the Hagedorn temperature. 

We could check that the action is consistent with the entropy and the mass in (\ref{EntHP}) and (\ref{MassThA}). For the mass, we have
\begin{equation}
\begin{aligned}
M & = \beta\frac{d F}{ d\beta} + F \approx   \frac{2}{16\pi G_N} \frac{2\alpha'^2 (d-2) \omega_{d-1}}{ \kappa^2} \zeta^{\frac{d}{2} - 2} m_{\infty}^{5-d} \beta \frac{d m_{\infty}}{ d\beta}  \\
& =  \frac{2 \alpha' (d-2) \omega_{d-1} }{\kappa} \frac{1 }{16\pi G_N} \zeta^{\frac{d}{2}-2} m_{\infty}^{4-d}
 \end{aligned}
\end{equation}
which agrees with (\ref{MassThA}). Notice that we've dropped the $+F$ term on the first line since it is higher order in $(R-R_H)$ than the $\beta\frac{dF}{d\beta}$ term and we cannot trust it since we also need to take into account of the corrections to the action. As a result, for the entropy, we simply have that
\begin{equation}
	S = \beta^2 \frac{dF}{d\beta} \approx \beta_H M
\end{equation}
which agrees with (\ref{EntHP}).

 \subsection{Interpretation as a bubble nucleating the decay of the Kaluza-Klein vacuum}

\begin{figure}[t]
    \begin{center}
    \includegraphics[scale=.15]{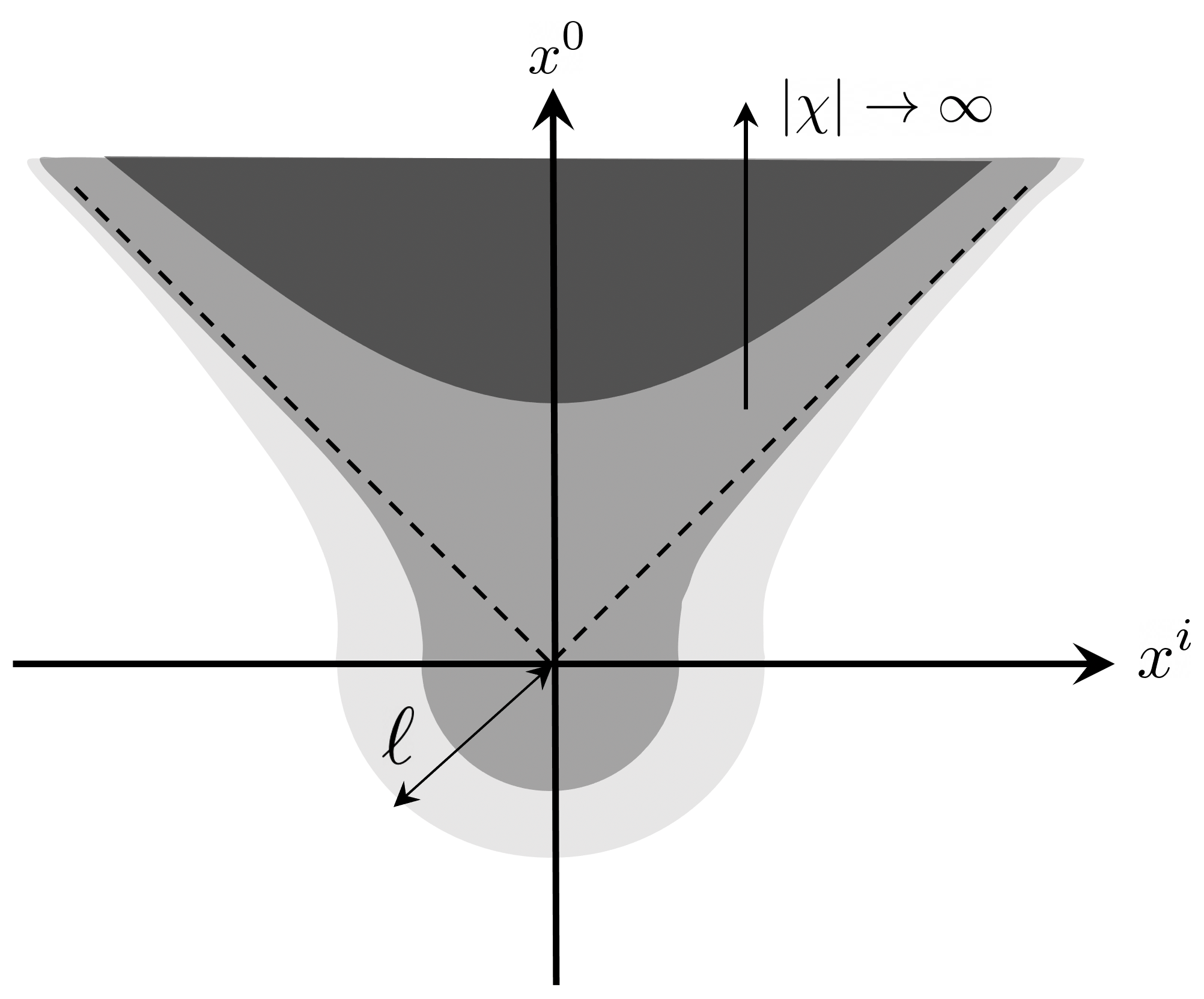}
    \end{center}
    \caption{In the lower part, we illustrate the Euclidean solution, which has an $SO(d)$ symmetry in $d$ dimensional Euclidean space. We can then continue the solution into Lorentzian signature, as shown in the upper half of the figure. The analytic continuation of the winding mode becomes large in the region where   the color gets darkest,  and we cease to trust the solution there. In the figure, $x^0$ is the time direction of the $d$ dimensional spacetime, while $x^i, i =1,...,d-1$ represent  the spatial directions. 
 }
    \label{fig:bounce}
\end{figure}

Though we are mainly interested in
giving a thermal interpretation to the circle, we could also think of the circle as an ordinary spatial dimension which has been compactified with anti-periodic boundary conditions for the fermions. In this setup we have $d$ spacetime dimensions. The localized solution in Euclidean space can be viewed as a type of bounce solution that mediates the decay of this circle compactification. In other words, we can analytically continue the solution by choosing time to be one of the $d$ dimensions. As is usual with spherically symmetric Euclidean bounces, the Lorenzian solution describes an expanding bubble where the winding mode is condensing. Unfortunately, even though the Euclidean solution is under control, the Lorenzian solution is not under control because the winding mode becomes  large in the region of the Lorentzian solution that is to the future of the center of the bubble (see fig. \ref{fig:bounce}). This means that higher order terms in $\chi$ in the action become important. 

This bubble solution is related to a feature discussed in \cite{Atick:1988si}. There it was noticed that the massless fields imply that the effective potential for the winding mode should have a negative quartic term. This negative quartic term is precisely the one arising from integrating out the field $\varphi$ above which gives the term 
\be 
 - \int d^d x d^d y {|\chi(x)|^2 |\chi(y)|^2 \over |\vec x - \vec y|^{d-2 } }
 \ee 
 in the effective potential. 
 The bubble solution computes the probability for the nucleation of a phase with larger values of $\chi$. 
 
 A relevant observation is that the effective mass squared for the field $\chi$ is negative at the center of the solution, despite being positive far away. This is because the size of the circle becomes smaller near the center. Recall that the size of the circle is changed by the field $\varphi$. We can see that it is negative by multiplying the equation for $\chi$ by $\chi^* $ and integrating. After an integration by parts we  get 
 \be 
   - \int |\nabla \chi|^2 = \int m^2(\varphi) |\chi|^2 
   \ee 
   We see that the left hand side is manifestly negative, which is possible only if $m^2(\varphi) < 0$ in some region. Since $m^2(\varphi)$ acquires its lowest value at the center, we see that the $\chi$ field has negative mass squared at that point. 
   As we continue to the Lorentzian solution we see that $m^2$ becomes even more negative in the region to the future of the center of the bubble. 
  
  When the radius is large $R\gg R_H$, the Euclidean black hole solution gives a somewhat similar bubble producing the decay of flat space \cite{Witten:1981gj}.  
   
\subsection{Negative modes of the solutions} \la{sec:nmodes}

In sections \ref{sec:d=3} - \ref{sec:d=5}, we have shown that normalizable saddle point solutions exist for (\ref{Action}) in $d=3,4,5$. However, we have not shown that these solutions minimize the action. In the case of the black hole solution, there is a well-known negative mode which lowers the action \cite{Gross:1982cv}. We will now show that a similar negative mode also exists  for  the \HP solutions. In addition, the interpretation of the solution as nucleating the decay of flat space suggests that there should be a negative mode to produce the requisite $i$ multiplying the full amplitude \cite{Coleman:1985rnk}. 

 Instead of examining all possible variations around a solution $\chi_*$, here we focus on the ones that are given by simple rescalings. In other words, we vary $\chi$ around the solution $\chi_*$ through $\chi (x) = \lambda \chi_* (x/\gamma)$, which was considered in sec. \ref{sec:dgeq6}. We could now expand $I_{\rm eff}$ in (\ref{Ieff}) to second order of $\delta \lambda= \lambda - 1,\delta\gamma = \gamma - 1,$ and get
\begin{equation}
 I_{\rm eff} = \frac{2 I_{1,*} }{d-2} + I_{1,*} \begin{pmatrix}
     \delta \gamma & \delta \lambda
    \end{pmatrix} 
  H
    \begin{pmatrix}  
     \delta \gamma \\ \delta \lambda
    \end{pmatrix} ,\quad ~~~~ H \equiv   \begin{pmatrix}
    - \frac{d^2 - 2d + 8}{d-2}  & -\frac{2 (d+2)}{d-2} \\
    -\frac{2 (d+2)}{d-2}  & - \frac{8}{d-2}
    \end{pmatrix} .
\end{equation}
It is straightforward to check that the Hessian matrix $H$ has one positive and one negative eigenvalue, for $d=3,4,5$. Therefore we identified a single negative mode for the solutions. This negative mode is similar to the one for the Euclidean black hole, as it can be intuitively interpreted as increasing or decreasing the mass away from the saddle point value. We have not attempted to prove that it is the \emph{only} negative mode of the solutions.

So far we have been interpreting the self gravitating string solutions as solutions in the canonical ensemble, namely we are fixing the temperature. We could also interpreting them as solutions in the microcanonical ensemble, where we fix the total mass $M$ instead. In this case we should view the $I_3$ term in (\ref{Ieff}) as coming from demanding that $\int d^d x\, |\chi (x)|^2$ stays a constant, with $m_{\infty}^2$ being a Lagrange multiplier. In other words, the appropriate effective action for the microcanonical ensemble is
\begin{equation}
	I_{\rm eff}^{\rm micro} = I_1 - I_2,
\end{equation}
subject to the constraint that $\int d^d x\, |\chi (x)|^2$ stays unchanged. This is the question examined in \cite{Horowitz:1997jc}, and we repeat their argument here using our notations. We can still consider the variation $\chi (x) = \lambda \chi_* (x/\gamma)$ while subjecting to the constraint $\lambda^2 \gamma^d = 1$. Therefore
\begin{equation}\label{actionmicro}
\begin{aligned}
    I_{\rm eff}^{\rm micro} & = \lambda^2 \gamma^{d-2} I_{1,*} - \lambda^4 \gamma^{d+2} I_{2,*}  \\
    & = \gamma^{-2 }I_{1,*} - \gamma^{-d+2} I_{2,*}. 
\end{aligned}
\end{equation}
Expanding around $\gamma =1$ and using (\ref{I2I3inI1}), we get
\begin{equation}\la{microexpand}
	I_{\rm eff}^{\rm micro} = \frac{d-4}{d-2} I_{1,*} - (d-4) I_{1,*} (\gamma-1)^2 + \mathcal{O}((\gamma - 1)^3).
\end{equation}
Therefore we find a difference between $d=3$ and $d=4,5$. In $d=3$, the variation we considered is not a negative mode. In fact, it was proven in \cite{LiebProof} that the solution $\chi_*$ minimizes the action (\ref{actionmicro}) globally. However, for $d=5$ we do find a negative mode. The case of $d=4$ is inconclusive as the higher order terms in (\ref{microexpand}) also vanish. To resolve the fate of the $d=4$ case, we need to go to higher orders in $(R - R_H)$.

    \section{The connection between the \HP  and black hole solutions} 
    
 \la{BhHP}
 \subsection{General Remarks}
 
 We have reviewed in detail the properties of the \HP solution because it is natural to conjecture that, at least for small values of $d$,
  it might be continuously connected to the black hole regime.  The usual black hole solution is trustworthy for $R\gg R_H$,  while the \HP solution is valid for $R-R_H \ll l_s$. However both have the following properties in common 
  
\begin{figure}[h]
\begin{center}
\includegraphics[scale=.2]{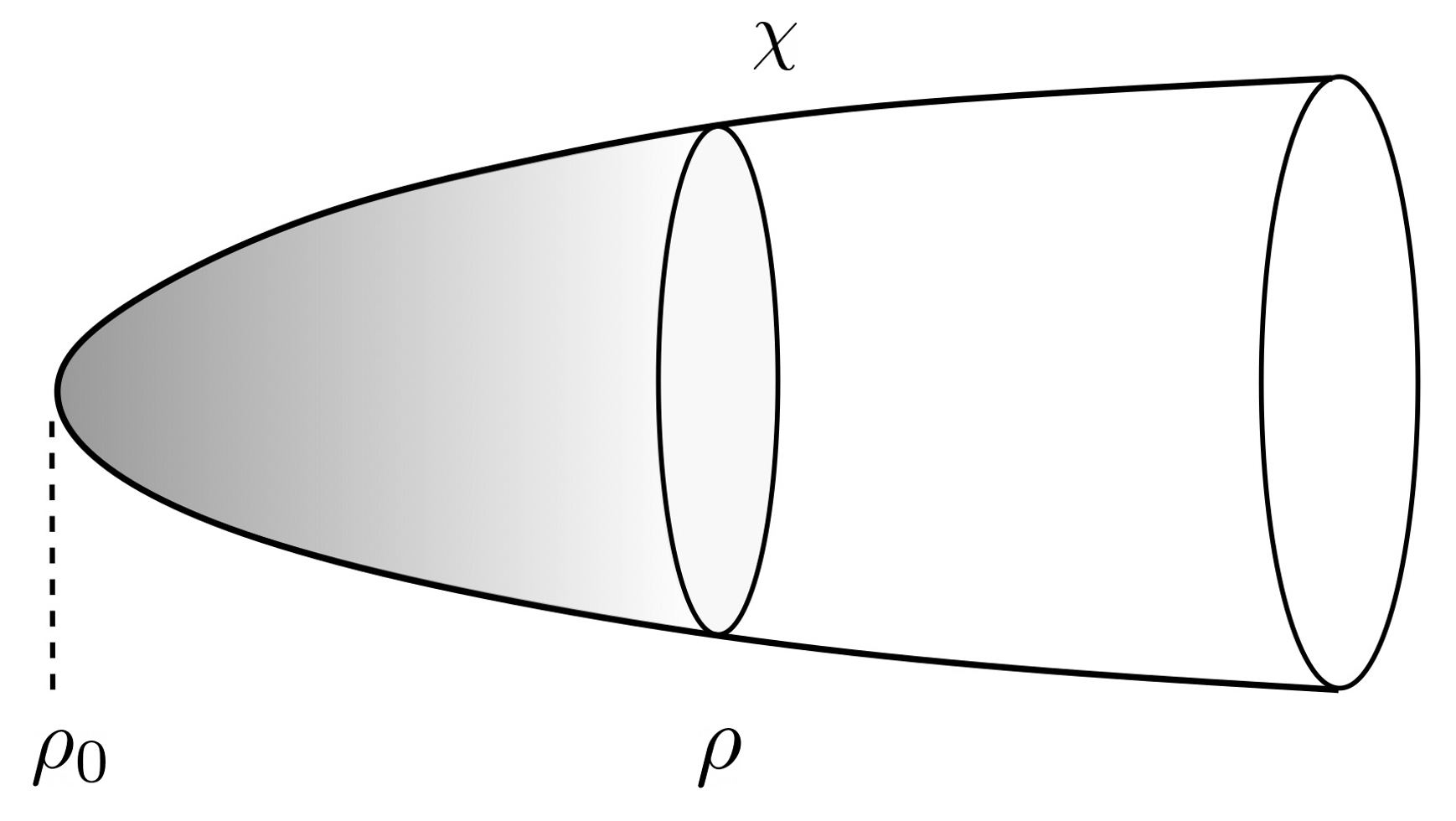}
\caption{A Euclidean string worldsheet wrapping the cigar gives rise to the expectation value of the winding mode at position $\rho$, $\chi(\rho) \propto e^{ - { 1 \over 2 \pi \alpha' } ({\rm area})} \propto e^{ - (\rho-\rho_0) R/\alpha' }$ in Schwarzschild coordinates, for $\rho\gg \rho_0$, and $R\gg l_s$.  }
\label{StringWinding}
\end{center}
\end{figure}

  \begin{itemize}
  	\item Spontaneous breaking of the winding symmetry at the classical level.\footnote{The symmetry is broken classically;  in the full string theory,
	  after integrating over a zero-mode, the symmetry is restored.} This is obvious for the \HP solution, since $\chi$ is charged under the winding gauge symmetry. For the black hole case this is due to the fact that a string that is wound on the thermal circle can go to the horizon and become unwound.  
 More precisely, one can argue that the winding mode has a vacuum expectation value in the presence of a black hole. This can be estimated as follows. The expectation value of $\chi$ at some position $\rho$ can be computed by considering a classical string worldsheet that wraps the Euclidean cigar and ends at $\rho$, see figure \ref{StringWinding}. Of course, this gives something very small when $R\gg l_s$, but it shows that this vacuum expectation value is non-zero. One can improve this computation by considering the small fluctuations of the worldsheet, etc. Nevertheless we do not know how to describe this field near the horizon, where it would vary over string scales. In addition, we also expect to have all higher winding modes contributing near the horizon. For a review and further references see \cite{Mertens:2015ola}.\footnote{The winding condensate can be viewed as a gas of strings,  and there have been
   a number of papers which suggest that it might account for all of the black hole entropy, see e.g. \cite{Dabholkar:2001if}. 
  However, these do not seem to rely on controlled approximations, in the sense that the gas of strings would be strongly coupled. Recent attempts include  
 \cite{Jafferis:2021ywg,Brustein:2021cza,Giveon:2021gsc}. 
  Based on our current understanding, the most reasonable statement would be that the winding mode contributes to a part of the total entropy of the black hole, and its contribution can becomes large at a special temperature for some black holes, see for example \cite{Chen:2021emg}. This is in spirit similar to quantum corrections to the black hole entropy, with the main difference being that the winding mode contributes at the classical level. }   
 This spontaneous breaking has an associated zero mode. For the \HP solution it is the phase of $\chi$, for the black hole it is the integral of the $B$ field on the cigar geometry.

 \item
 Both the \HP solution and the black hole have a non-zero classical entropy. 
\end{itemize}

These similarities would lead us to conjecture that the black hole and the \HP solution could be continuously connected as classical string theory solutions.\footnote{As a side remark, we notice that the two coupled SYK model introduced in \cite{Maldacena:2018lmt} at finite temperature has a somewhat similar transition where a  hot wormhole like phase (analogous to the \HP solution) transitions into two separate black holes, analogous to the black hole phase. In that case, there is evidence that the transition is smooth \cite{Maldacena:2018lmt,Maldacena:2019ufo}. }
 
A classical string theory solution is a certain CFT.   Then the conjecture is that by changing a parameter of this solution, 
namely the radius of the circle at infinity,  we can interpolate between the black hole and the \HP solution. Of course, we can only 
make this conjecture for the dimensions where the \HP solution exists. We will be considering solutions which are the product of an internal CFT times a CFT that describes the \HP or black hole solutions in $D$ dimensions. Our discussion centers on the CFT that describes the \HP or black hole solutions.  

Unfortunately, it is too difficult to decide whether it is possible to smoothly interpolate between the black hole and the \HP solution via a family of classical
solutions -- two-dimensional conformal field theories.   For one thing, one does not have a very concrete construction of either the black hole or the \HP solution  as a CFT away from the asymptotic regimes of $R\gg l_s$ or $R-R_H \ll l_s$ that we have discussed. These descriptions are not valid near the hypothetical transition
region.     However, for either Type II or the heterotic string,\footnote{Type I would be similar to Type II, since the linear sigma models that
we consider for Type II are invariant under worldsheet orientation reversal (and could be extended by adding space-filling branes), and therefore have simple analogs for Type I.    We do not consider
the bosonic string since it has a tachyon and it is not clear that we can generally prevent its condensation on general backgrounds.} one can construct linear sigma models that plausibly flow in the infrared to the black hole or the \HP solution.   Then we can ask whether, in the framework
of linear sigma models,  one can interpolate smoothly between a linear sigma model that looks like it could flow to the black hole and one that looks
like it could flow to the \HP solution.        If so, this roughly means that off-shell, it is possible to make a smooth interpolation between the black hole and the \HP
solution, in the classical limit of string theory.

We describe the linear sigma models for the heterotic string in section \ref{heterotic}, and those for Type II in section \ref{typetwo}. 
Perhaps the most surprising thing to come out of this study is that there is a difference between the heterotic and Type II superstring models.   For the heterotic
string, one can interpolate smoothly, in the sense of linear sigma models, between the black hole and the \HP solution.  For Type II, this is not possible.  
  In section \ref{analysis},
we discuss the difference between the two cases in a more general way.

The results that we will find do not necessarily mean that the two phases are not continuously connected for Type II.   We only learn that these two phases
cannot be smoothly connected as classical solutions of string theory. It is possible that they are connected at a critical point at which quantum effects are important.
One possible mechanism would be to have a sigma model in which  the dilaton becomes infinite in some region of the target space.  This happens in some 
simple non-critical strings \cite{Douglas:2003up}, or when we approach a conifold transition in string theory \cite{Ooguri:1995wj}.

 \subsection{Linear sigma models for the heterotic string}\label{heterotic}
 
\subsubsection{Construction of the model}\label{heteconstruction}
 
 A worldsheet theory for the heterotic string should have $(0,1)$ supersymmetry, which can be realized in a superspace with bosonic coordinates $x^-,x^+$
 and a single fermionic coordinate $\theta=\theta^+$.   The supersymmetry generator is $Q=\partial_\theta+i \theta\partial_{x_+}$, and commutes with
 $D=\partial_\theta-i\theta\partial_{x^+}$, which can be used in constructing supersymmetric actions.   For more detail, see for instance \cite{Brooks,GaiottoW}.
 
 We will use two types of superfields in constructing linear sigma models, namely a scalar superfield $\Phi(x,\theta)=\phi(x)+i \theta \psi_+(x)$, and
 a fermi superfield $\Lambda(x,\theta)=\lambda_-+\theta F$.    Here $\phi$ is a (real) scalar field, $\psi_+$ and $\lambda_-$ are fermion fields of the indicated chirality,
 and $F$ is an auxiliary field.   Vector multiplets are also possible, but we will not make use of them.
 
 We consider a model with $n$ scalar superfields $\Phi_i$ and $m$ fermi superfields $\Lambda_\alpha$.
To generate an ordinary potential energy for the scalar fields, one starts with a superspace interaction of the form
\be\label{superaction} \int d^2x d\theta \sum_\alpha \Lambda_\alpha W_\alpha(\Phi_i), \ee
with some functions $W_\alpha$.  After integrating over $\theta$ and integrating out the auxiliary fields, this leads to an ordinary potential of the form
\be\label{ordpotl} V(\phi_i)=\frac{1}{2} \sum_\alpha W_\alpha(\phi_i)^2.  \ee
We see that if $m<n$, it is natural for a model of this kind to lead at low energies to a nonlinear sigma model with a target space of dimension $n-m$.
The target space is defined by the vanishing of the potential:
\be\label{targsp} W_1(\phi_i)=\cdots = W_m(\phi_i)=0. \ee
The superspace action (\ref{superaction}) also leads to a Yukawa coupling
\be\label{yukawa}i \int d^2x\,\sum_{\alpha,i} \frac{\partial W_\alpha}{\partial\phi_i}\lambda_{\alpha,-}\psi_{i,+}, \ee
which will be relevant later.

For our application, to study the black hole or the HP solution in $D=d+1$ dimensions, we choose $n=D+1$, $m=1$.    We take the scalar superfields to consist
of a $d$-plet $\vec \YY=(\YY_1,\cdots, \YY_d)$  and a 2-component vector  $\vec \XX=(\XX_1,\XX_2)$.  We write $\vec Y$, $\vec X$ for 
the bottom components of $\vec\YY$, $\vec\XX$.    We assume these fields have canonical kinetic energy
\be\label{canonical} \int d^2x\, \frac{1}{2}\left(\partial_\alpha \vec X\cdot \partial^\alpha \vec X+ \partial_\alpha \vec Y\cdot \partial^\alpha \vec Y\right), \ee
after integrating over $\theta$.   We also introduce a single fermi superfield $\Lambda$, and an interaction (\ref{superaction}) with 
 $W=m\left( (\vec\YY^2+a)(\vec \XX^2-b)+c \right)$.   Here  $a,b,c$
are dimensionless constants and $m$ is a constant with dimensions of mass.  Classically, the model leads to a nonlinear sigma model supported on the locus 
$W=0$, namely
\be\label{classloc} (\vec Y^2+a)(\vec X^2-b) +c=0 \ee
or
\be\label{eqform} \vec X ^2=b-\frac{c}{\vec Y^2+a}. \ee
At the classical level, the target space metric of the nonlinear sigma model, assuming the fields are normalized canonically as in eqn. (\ref{canonical}), is the flat metric
$d s^2= d\vec X^2+ d\vec Y^2$, restricted to the locus of eqn. (\ref{classloc}) or eqn. (\ref{eqform}).   This metric does not depend on $m$, but it does depend on
$a,b,c$.     The region in which the nonlinear sigma model is weakly coupled can be reached by scaling up the target space metric by a large factor.
We can reach this regime by scaling $a,b,c\to ta,tb,t^2c$ with $t\gg 1$.  In other words,  we take $a,b,c$ large with $a/b$ and  $ab/c$ fixed.

We are mainly interested in the case $b,c>0$.   Classically, for $|\vec Y|\gg 1$, $\vec X$ parametrizes a circle of radius $b^{1/2}$, corresponding to an inverse
temperature $\beta =2\pi b^{1/2}$.   As $\vec Y$ decreases, the circle parametrized by $\vec X$
becomes smaller, as expected for both the black hole and the \HP solution.
Classically, whether we get something more like the black hole or more like the HP solution depends on the value of $b-c/a$.  If $b-c/a<0$,  then $\vec Y$ is restricted
to $\vec Y^2\geq c/b-a$.   At $\vec Y^2=c/b -a$, the $\vec X$ circle shrinks to a point and the space ends.  This is qualitatively similar to the Euclidean
Schwarzschild solution; the topology is $S^{d-1}\times B$, where $B$ is a two-dimensional disc.   If instead $b-c/a>0$, then arbitrary values of $\vec Y$ are possible
and the topology is $\RR^{d}\times S^1$, like that of the \HP solution.  
  In summary, 
\be \la{BHHPca}
c - a b> 0 ~\sim ~{\rm Black~hole} ~,~~~~~~~~~~~c-ab< 0 ~\sim ~{\rm Horowitz~Polchinski~solution}
\ee
in terms of their topological nature. 

Although the space of classical ground states is singular at $ab-c=0$, one expects the two-dimensional supersymmetric field theory to vary smoothly 
with the parameters $a,b,c$.   That is because no new flat direction in field space  opens up at $ab-c=0$.   Accordingly, there is no way for
new low energy states to appear or disappear at or near $ab-c=0$; there is nowhere for them to go.  The situation for Type II is different, as we will see in 
section \ref{typetwo}.

Therefore, we have found a smooth off-shell continuation from something resembling a black hole to something resembling the \HP solution.   In particular, it must be impossible to distinguish them by the supersymmetric index $\Tr\,(-1)^F$ or any other invariant of a two-dimensional theory
with $(0,1)$ supersymmetry.   In fact, for a $(0,1)$ sigma model with target $M$,  $\Tr\,(-1)^F$ is the index of the Dirac operator on $M$.   This is odd under
parity, so it vanishes for both the black hole and the \HP spacetime.  We explain in another way in section \ref{analysis} why no invariant of a $(0,1)$ supersymmetric
model can distinguish the black hole from the \HP spacetime.

 In the usual description of the \HP solution, conservation of string winding number is explicitly broken by a condensate of strings that carry winding number.
 In the linear sigma model description, winding number is not a conserved quantity because the circle $|\vec X|^2=b$ is  contractible in the full
 field space, regardless of the values of the parameters.     Accordingly, 
for suitable values of $a,b,c$, the linear sigma model has the potential to spontaneously generate the effects of the winding condensate and recover the \HP
solution.

Though we have found a smooth off-shell continuation between the two solutions,
  the region in which the nonlinear sigma model is under good control (large $a,b,c$ with fixed $a/b$ and $ab/c$, as explained above)
is very far from the region of the possible
crossover between the black hole and the \HP solution, which is expected to occur for $b\sim 1$.

The construction has a few further limitations.      One is that, at least at this level, it is not sensitive to the value of the spatial dimension 
$d$, while the \HP solution depends very much
on $d$.   Perhaps the renormalization group (RG) running of the model to the infrared is sensitive to $d$, but the semiclassical picture in the ultraviolet is not.

A second point is that the linear sigma model has more parameters than one might wish.   
The Euclidean black hole and the \HP solution    depend on a single parameter,
the inverse temperature $\beta$, which determines the mass or energy.   Another parameter is the asymptotic value of the $D$-dimensional dilaton field $\phi_D$, but
the linear sigma model as we have formulated it does not see this parameter.\footnote{Shifting the asymptotic value of $\phi_D$ by a constant can be accomplished
by adding to the worldsheet action a multiple of the worldsheet Euler characteristic  $\int d^2 x\sqrt g R$.    We could generalize the linear sigma model to a
curved worldsheet and add such a coupling.    Then we would say that the linear sigma model depends on four dimensionless parameters ($a,b,c$ and the
asymptotic value of $\phi_D$), while  two, $\beta$ and $\phi_D$,  suffice for a description of the black hole or the \HP solution.}   The Euclidean black hole and the \HP solution
are both unstable, with a single unstable mode we identified in sec. \ref{sec:nmodes}.  Therefore, we expect a CFT description of the black hole or the \HP solution to have   one relevant operator. Under the influence of this relevant operator, we expect that a generic RG trajectory will flow to either an empty flat space, or to a configuration that   expands under the flow and ``eats up" the spacetime.  
For the Euclidean black hole, this was demonstrated in \cite{Wiseman:2011by,Headrick:2006ti}. 
Hence a linear sigma model description of the black hole or the HP solution needs at least two dimensionless parameters -- one relevant parameter must be adjusted
to get an  RG
 flow to the black hole or the HP solution, and a second will become the inverse  temperature.   The linear sigma model that we have described
actually has three dimensionless parameters.   There is no contradiction here, since the third parameter certainly might become irrelevant in the infrared in the
RG sense.  But a linear sigma model would be more useful if it had only the necessary dimensionless parameters, or at least, if we knew more about which
parameters are the important ones.

In the particular case of $D=4$ there appears to be another parameter in the black hole side, which corresponds to turning on an NS $B$ field on the $S^2$.  This is a marginal coupling at the level of the classical sigma model. However, we expect that worldsheet instanton corrections can generate a superpotential that fixes it to zero.\footnote{Perhaps $\int B =\half$ can also be considered, but deformations away from it are expected to be relevant, so it is a more fine tuned value than $\int B =0$. } This parameter is not present on the \HP side. Therefore we do not expect to have an extra continuous parameter in either case.

Another issue is the following.   We can think of the RG running from the linear sigma model to the infrared in two steps.  First we integrate
out the massive modes of the sigma model to get a nonlinear sigma model, and then we run the nonlinear sigma model to the infrared.   In the above
discussion, we carried out the first step classically, but it is actually necessary to be more careful.    The magnitude $|\vec X|$  of the field $\vec X$ is a massive
scalar of mass $M_{|\vec X|}(\vec Y)\sim 2m \sqrt{b}{\vec Y}^2$ for large ${\vec Y}^2$. Quantum fluctuations of $|\vec X|$ give a one-loop contribution to the vacuum expectation value $\langle \vec X^2\rangle$
that is
\be\label{oneloop}\int \frac{d ^2 k}{(2\pi)^2}\frac{1}{k^2+M_{|\vec X|}(\vec Y)^2}=\frac{1}{4\pi}\log (\Lambda^2 / M^2_{|\vec X|}(\vec Y)),\ee
where $\Lambda$ is an ultraviolet cutoff.   Hence instead of describing the target space of the nonlinear sigma model by the equation (\ref{eqform}), it would
be more accurate to write
\be\label{neqform} \vec X^2= b_r -\frac{1}{4\pi}\log (\mu^2/M^2_{|\vec X|}(\vec Y)) -\frac{c}{\vec Y^2+a},\ee
where $\mu$ is a renormalization point and $b_r$ is a renormalized parameter $b_r=b-\frac{1}{2\pi}\log(\Lambda/\mu).$  For large $\vec Y$, the right hand
side of eqn. (\ref{neqform}) grows as $\frac{1}{\pi}\log |\vec Y|.$ 
Hence in this approximation, the inverse temperature grows as $\log^{1/2} |\vec Y|$ for large $\vec Y$, rather than approaching a constant. 

\EW{I made a couple of minor changes of wording above, but substantial changes in this section begin here, with one exception: I moved the remarks
on string winding number in the linear model to the bottom of p. 27, and changed the wording a lilttle.}
This is a significant drawback of the model, because with the radius of the circle growing logarithmically at infinity, it is doubtful that the RG flow will have the 
desired behavior.  The radius of the circle is a scalar field in $d$ dimensions, and under RG flow it tends to relax to its average value.   Since the average of
$\log |\vec Y|$ is divergent, the RG flow might bring the model to a  low temperature limit.

However, it is possible to slightly complicate the model and avoid this issue.  We add another scalar superfield $ \XX'$, and a second fermi multiplet $\Upsilon$.   
We take the superspace coupling to be
\be\label{extendedaction} \int d^2 x\,d\theta \left(m \Lambda \left( (\vec\YY^2+a)(\vec \XX^2-\XX'{}^2-b)+c \right)+m' \Upsilon (1+f \vec \YY^2)\XX'\right), \ee
with new constants $m'$, $f$.
Classically, the new fields $\XX'$ and $\Upsilon$ are massive, for any value of $\vec Y$, and  $\XX'$ vanishes in any supersymmetric state. So adding these fields
does not affect  the description of the classical ground states.
But quantum mechanically, the problematic logarithm is absent in this more complicated model.  For large $\vec Y$, both
$\XX$ and $\XX'$ acquire masses proportional to $|\vec Y|^2$.
Thus eqn. (\ref{neqform}) is replaced with
\be\label{repform}\vec X^2= b_r -\frac{1}{4\pi} \log (M^2_{X'}(\vec Y)/M^2_{|\vec X|}(\vec Y)) - \frac{c}{\vec{Y}^2 + a}.\ee
Now the argument of the logarithm has a constant limit for $\vec Y\to\infty$, so likewise the radius
 of the $\vec X$ circle  has a constant limit.  
  Note that since $\XX'$ has only one component while $\vec \XX$ has two, this modification of the model does not eliminate the existence
of a logarithmic renormalization of $b$ due to ``normal ordering'', that is, vacuum fluctuations in $\vec X^2-X'{}^2$.   However, at large $\vec Y$,
only one of the two components of $\vec\XX$ is massive, and adding $\XX'$ does cancel the logarithmically growing
$\vec Y$-dependent part of the renormalization of $b$.   This cancellation depends on the precise relative factor $-1$ between the $\vec\XX^2$ and $\XX'{}^2$ terms
in eqn. (\ref{extendedaction}).   That factor is not affected by normal ordering, which is the only ultraviolet divergent renormalization in a supersymmetric theory 
 in two  dimensions with polynomial couplings.  If we slightly change the coefficient $-1$ in the action so that a $\log |\vec Y|$ term appears in eqn. (\ref{repform})
 with a small coefficient,
 then it becomes important at exponentially large values of $|\vec Y|$, presumably not affecting the RG behavior in the interior of the spacetime.
 
 By further
 elaborations of the model, one can eliminate the logarithmic renormalization of $b$ if this is desired.  One way to do this is to take $\Upsilon$ and $\XX'$ to
 be two-component fields, but with couplings such that only one component of $\XX'$ gets a mass proportional to $\vec Y^2$ for large $\vec Y$.   For
 this, one can replace the coupling in (\ref{extendedaction}) with
  \be\label{furtherelab}\int d^2 x\,d\theta \left(m \Lambda \left( (\vec\YY^2+a)(\vec \XX^2-\vec \XX'{}^2-b)+c \right)+m' (\Upsilon_1 \XX'_1+\Upsilon_2(1+f \vec \YY^2)\XX'_2)\right).\ee
As these examples illustrate, there are many ways to add additional massive fields without changing the fact that classically, the model leads to a nonlinear sigma model of something similar
to the black hole or the \HP solution, depending on the sign of $ab-c$.

Much of what we have said has an analog for the Type II problem to which we come in section \ref{typetwo}.   Before turning to that problem,
however, we consider one more question for the heterotic string.

\subsubsection{Deriving the quantum numbers of the heterotic string thermal winding mode}\la{QuanNum} 

\EW{In this section, I made only minor changes in wording.}

 An important subtlety of the thermal ensemble for the
heterotic string is that the ground state of a string wrapping around the thermal circle has unusual quantum numbers  \cite{Rohm:1983aq,OBrien,McGuigan,Atick:1988si}.  It is odd under $(-1)^F$, the operator that distinguishes worldsheet bosons and fermions and appears in the GSO projection.   And it has a half-unit of $\SP$, the momentum around the thermal circle.
Let us see how these properties arise in the linear sigma model.

The basic idea is to determine the quantum numbers of the fermion ground state in the presence of a string that winds around the thermal circle. 
    In the presence of a  background $\vec X$,
it makes sense to ask how the fermion ground state transforms under those symmetries that preserve the background.   
The relevant symmetries are $(-1)^F$ and rotation of $\vec X$, which ultimately is interpreted as translation along the thermal circle.
$(-1)^F$ remains
a symmetry in the presence of any background, but rotation of $\vec X$ is only a symmetry for special choices of background.

The field $\vec \YY$ plays no role in the analysis and we can just set it equal to a constant.  Thus we can consider a slightly simpler problem
with scalar superfields $\XX_1,\XX_2$, a fermi superfield $\Lambda$, and a coupling (\ref{superaction}) with $W=m(\vec\XX^2-b)$.  The scalar multiplets contain
positive chirality fermions $\psi_{1,+}, \psi_{2,+}$, and the fermi multiplet contains a negative chirality fermion $\lambda_-$. 
We consider the theory on a circle of circumference $2\pi $, with metric $dx^2$, $0\leq x\leq 2\pi $. Since we are fixing the worldsheet metric, the limit in which the
linear sigma model may reduce to a CFT is $m\to \infty$.  
Since we are interested in states in the NS sector, we put antiperiodic boundary conditions for   
the fermions  under $x\to x+2\pi $.    The kinetic energy for $\psi_{1,+},\psi_{2,+}$ is
$-i d/d x$, and the kinetic energy for the opposite chirality fermion field $\lambda_-$ is $i d/dx$.    Because the fermions are antiperiodic, the eigenvalues of
$-i d/dx$ are arbitrary elements $n$ of $ \frac{1}{2}+\ZZ$. Including the effects of the Yukawa coupling (\ref{yukawa}), the single-particle Hamiltonian in a basis
$\begin{pmatrix}\psi_{1,+}\cr \psi_{2,+}\cr \lambda_-\end{pmatrix}$ is
\be\label{hamton} H=\begin{pmatrix}-i\frac{d}{dx} &0&0\cr 0&- i\frac{d}{dx} & 0\cr 0&0& i\frac{d}{dx} \end{pmatrix}+im\begin{pmatrix}
0 & 0& X_1 \cr 0 & 0 & X_2 \cr -X_1 & -X_2 & 0 \end{pmatrix}.\ee
First suppose that $\vec X=0$.  There are no single-particle fermion states of zero energy, since $n$ takes half-integer values.
 As usual the fermion ground state is found by filling all the negative energy states.   The fermion ground state at $\vec X=0$ corresponds
 to the identity operator of a free fermion CFT, and has vacuum quantum numbers -- it is invariant under $(-1)^F$ and under rotation of $\vec X$.  
 Now as a warmup,
 consider the case that $\vec X$ is a non-zero constant; this preserves the $(-1)^F$ symmetry but not, of course, the symmetry of rotation of $\vec X$.   
Regardless of $\vec X$, as long as it is constant,
$H$ has no zero eigenvalues.   In fact, as long as $\vec X$ is constant, $H$ is equivalent to a free  Dirac Hamiltonian for three fermion modes
 of which two are massive; regardless of the mass, there are no zero-modes, since the fermions are antiperiodic.   Absence of single-particle zero-modes
means that as we vary $\vec X$, the filled fermion states remain filled and the empty ones remain empty, so the ground state quantum numbers 
(under symmetries preserved by the background) do not change, and hence the ground state continues to be invariant under $(-1)^F$.      Now let us consider instead a winding state, with $(X_1,X_2)=|\vec X|(\cos x ,\sin x )$.   This is invariant under rotation of $\vec X$, combined with a translation of $x$.
  That combined operation is what ``translation along the thermal circle'' means in a winding sector.  
It is possible to solve explicitly for the eigenvalues of the single-particle Hamiltonian $H$ in the presence of a background of this kind.   The answer is most
simply described if one first makes a change of variables\footnote{We are going to use this change of variables only as a shortcut to determine the eigenvalues of the
single-particle Hamiltonian.   We are not making a quantum change of variables that might have an anomaly.}
\be\label{conjugating}\begin{pmatrix}\psi_{1,+}'\cr \psi_{2,+}'\cr \lambda_-'\end{pmatrix}=  \begin{pmatrix} \cos x & \sin x &0\cr -\sin x &\cos x &0\cr 0&0&1\end{pmatrix}
\begin{pmatrix}\psi_{1,+}\cr \psi_{2,+}\cr \lambda_-\end{pmatrix}. \ee 
After the change of variables, the operator that generates rotation of $\vec X$ (together with translation of $x$) becomes just $\SP=-i d/dx$.
The single-particle Hamiltonian becomes
\be\label{newhamton} H'=\begin{pmatrix}-i\frac{d}{dx} &0&0\cr 0&- i\frac{d}{dx} & 0\cr 0&0& i\frac{d}{dx} \end{pmatrix} +i\begin{pmatrix}0&1&0\cr
-1&0&0\cr 0&0&0\end{pmatrix}+im|X|\begin{pmatrix}
0 & 0& 1 \cr 0 & 0 & 0 \cr -1 & 0 & 0 \end{pmatrix}.\ee   
In a sector with $-id/dx$ acting as $n\in \frac{1}{2}+\ZZ$, this is
\be\label{fteham} H'_n=
 \begin{pmatrix}n &0&0\cr 0&n & 0\cr 0&0& -n \end{pmatrix} +i\begin{pmatrix}0&1&0\cr
-1&0&0\cr 0&0&0\end{pmatrix}+im|X|\begin{pmatrix}0 & 0& 1 \cr 0 & 0 & 0 \cr -1 & 0 & 0 \end{pmatrix}.\ee
We see that 
\be \la{DeHn}
\det H'_n=-n(n^2-1+m^2 |X|^2)   
\ee    Since $n$ (being valued in $\frac{1}{2}+\ZZ$) never vanishes, for $H'_n$ to have a zero-mode,
we need $n^2-1+ m^2 | X|^2=0$.   This occurs precisely if $n=\pm 1/2$ and  $|X|=\frac{\sqrt 3}{2 m  }$.

At $|X|=0$, the eigenvalues of $H_{1/2}$ are $1/2,-1/2,-1/2$.   For $|X|\not=0$ and any sufficiently large $m$, two eigenvalues of $H_{1/2}$ are positive and one is negative.
Thus, at $n=1/2$, there is one single-particle state whose energy goes from negative to positive as $|X|$ is turned on.    The complex conjugate of this
state\footnote{Note that $H'_{-n}$ is minus the complex conjugate of $H'_n$.}
is a single particle state  at $n=-1/2$ whose energy goes from positive to negative as $|X|$ is turned on. Since the IR limit involves $m\to \infty$, we always cross this zero mode if $|X|\not=0$.   Quantum mechanically, these states correspond to operators
$a,a^\dagger$ obeying the usual fermion relations $a^2=a^\dagger{}^2=0$, $\{a,a^\dagger\}=1$.   The part of the Hamiltonian that depends on
$a,a^\dagger$ is $f(|X|) a^\dagger a$, where $f(0)>0$, but $f(|X|)<0$ if $|X|>\frac{\sqrt 3}{2m  }$.   When $f(|X|)$ becomes negative, the fermion ground state jumps from being
annihilated by $a$ to being annihilated by $a^\dagger$.   A single fermion state of momentum $\SP=-1/2$ that formerly was unoccupied becomes occupied.
Therefore, the fermion ground state becomes odd under $(-1)^F$, and carries momentum $\SP=-1/2$.    These are the standard
CFT results for a winding state of the heterotic string at nonzero temperature, showing that at least in this respect, the linear sigma model does
reproduce the results of the CFT.

This discussion has been for the mode with winding number one. 
For winding number $k$,  we multiply the middle term in \nref{fteham} by $k$ and the  $-1$ in \nref{DeHn} becomes $- k^2$. This means that $k$ modes cross zero and that the fermion number becomes $(-1)^k$ and the worldsheet momentum becomes $- k^2/2$, which implies that $\SP$ is half-integral when $k$ is odd, but integral  for $k$ even, as in \cite{Rohm:1983aq,OBrien,McGuigan,Atick:1988si}.

  \subsection{Linear sigma models for Type II superstrings}\label{typetwo} 
  
  \subsubsection{Construction of the model}\label{typetwoconstruction}
  
  In Type II superstring theory, we want a worldsheet theory that has $(1,1)$ supersymmetry, and also a $\ZZ_2$ chiral $R$-symmetry,   {which is needed for the GSO projection.  $(1,1)$ supersymmetry
  can be realized in a superspace with bosonic coordinates $x^+,x^-$, and corresponding fermionic coordinates $\theta^+,\theta^-$.     The only
  type of superfield that we will consider is a scalar superfield $\Phi(x,\theta)= \phi(x)+i\theta^+\psi_+(x) +i \theta^-\psi_-(x)+i\theta^+\theta^- F$,
  where $\phi$ is an ordinary (real) 
  scalar field, $\psi_\pm$ are fermi fields of the indicated chirality, and $F$ is an auxiliary field.    Such a  scalar superfield of $(1,1)$ supersymmetry
  decomposes under $(0,1)$ supersymmetry as the direct sum of a scalar superfield and a fermi superfield.
  
  Let us first discuss what happens if we do not impose a $\ZZ_2$ $R$-symmetry.    Consider a system of $N$ chiral superfields $\Phi_i$.   To generate a nontrivial
  potential for the ordinary scalar fields $\phi_i$, we introduce a real-valued function $W(\Phi_i)$ known as the superpotential, and include in the action a term
  \be\label{superac}\int d^2x\,d^2\theta \,\,W(\Phi_k).  \ee
  After integrating over $\theta^\pm$ and integrating out the auxiliary fields, this leads to an ordinary potential energy
  \be\label{ordpot} V(\phi_k)=\frac{1}{2}\sum_{i=1}^N \left(\frac{\partial W(\phi_1,\cdots,\phi_N)}{\partial \phi_i}\right)^2. \ee
  To find a classical state with zero energy and unbroken supersymmetry, we need to solve the equations $\partial_k W=0$, $k=1,\cdots, N$.   
  These are $N$ equations for $N$ unknowns, so generically the solutions are isolated.   It is not natural to get at low energies a nonlinear sigma model with
  a target space of positive dimension.
  
Ordinary global symmetries (as opposed to $R$-symmetries) make it possible to generate some sigma models, but not of sufficient generality to study the
black hole and the \HP solution.   If one assumes a group $H$ of ordinary global symmetries and requires $W$ to be invariant under an $H$ action on the $\Phi$'s,
then it is natural to get a space of supersymmetric states whose connected components are homogeneous spaces for $H$.  Degeneracy beyond that is not natural.
The black hole and \HP spaces are not homogeneous
spaces for any global symmetry group, so we cannot get them by imposing ordinary symmetries. 

What does work is to assume a chiral $\ZZ_2$  $R$ symmetry $\tau$, which acts on the fermionic coordinates by $\theta^\pm\to \pm\theta^\pm$.   (Combining this
with the universal $(-1)^F$ symmetry that acts by $\theta^\pm\to -\theta^\pm$, we get an opposite chirality $\ZZ_2$ $R$ symmetry $\theta^\pm\to \mp\theta^\pm$.) 
In any event, we want to assume such a symmetry because it is part of the structure of Type II superstring theory.
Since the fermionic measure $d^2\theta$ is odd under $\tau$, the superpotential $W$ must also be odd to ensure invariance of $\int d^2\theta\,W$.   
To make it possible for the superpotential to be odd under $\tau$,
we need superfields that are $\tau$-odd.

In general, we introduce $n$ $\tau$-even superfields $\Phi_i$, transforming by
\be\label{eventrans}\Phi_i(x,\theta^+,\theta^-)\to \Phi_i(x,\theta^+,-\theta^-), \ee
and $m$ $\tau$-odd ones $\PP_\alpha$, transforming as
\be\label{oddtrans}\PP_\alpha(x,\theta^+,\theta^-)\to - \PP_\alpha(x,\theta^+,-\theta^-).\ee
We introduce a superpotential of the general form
\be\label{superform} W(\PP_\alpha,\Phi_i)=\sum_\alpha \PP_\alpha W_\alpha(\Phi_i).\ee
This is certainly odd under the $\ZZ_2$ $R$-symmetry, since it is homogeneous and linear in the odd superfields $\PP_\alpha$.
Let us denote the ordinary scalar fields that are the bottom components of $\PP_\alpha$ and $\Phi_i$ as $P_\alpha$ and $\phi_i$.
Then by evaluating eqn. (\ref{ordpotl}), one finds  the potential 
\be\label{speccase} V(P_\alpha,\phi_k)=\frac{1}{2} \sum_\alpha W_\alpha(\phi_k)^2+\frac{1}{2}\sum_i \left(\sum_\alpha P_\alpha\partial_i W_\alpha( \phi_k)\right)^2.\ee

As in section \ref{heterotic}, assuming that $n>m$, the equations $W_\alpha(\phi_k)=0,~~\alpha=1,\cdots,m$ generically define a manifold $M$ of dimension $n-m$.
In contrast to section \ref{heterotic}, to find a supersymmetric classical state, we now have the additional equations
\be\label{addeqs} \sum_\alpha P_\alpha\partial_i W_\alpha( \phi_k)=0,~~i=1,\cdots, n.\ee
These equations are clearly  satisfied if the $P_\alpha$ all vanish. 
A solution with nonzero $P_\alpha$ arises at and only at a singularity of\footnote{$M$ is smooth at a  point $\phi_k=\phi_{k,0}$  if and only if the
equations $W_\alpha=0$ place $m$ independent constraints on a first order variation $\delta \phi_k$ of the $\phi_k$ at that point.   Saying that eqns. (\ref{addeqs})
are satisfied with some nonzero $P_\alpha$ is equivalent to saying that
 a linear combination $\sum_\alpha P_\alpha W_\alpha=0$ of the equations places no constraint
on the $\delta\phi_k$ in first order. This is the condition for a singularity.}
$M$.
In particular,  if $M$ is smooth,  the equations are only satisfied for $P_\alpha=0$,
and the low energy physics, at the classical level, will be a nonlinear sigma model with target $M$.   But if 
$M$ is singular,    the equations have solutions with $P_\alpha\not=0$.  Solutions with $P_\alpha\not=0$ are precisely the ones that are not invariant
under the chiral $R$-symmetry.
The equations (\ref{addeqs}) are homogeneous and linear in the variables
$P_\alpha$,  so if there are solutions with $P_\alpha\not=0$, then the space of those solutions is a cone and in particular is not compact.

For our purposes, just as in section \ref{heterotic}, we take $n=D+1$, $m=1$.   We take the $\tau$-even superfields to be a $d$-plet $\vec\YY$ and a pair
$\vec \XX$, and we introduce a single $\tau$-odd superfield $\PP$.  We denote
 the bottom components of $\vec \YY,\, \vec \XX$ and $\PP$ as $\vec Y, \vec X$, and $P$. For the superpotential, we pick 
\be\label{superp}W=m \PP\left((\vec\YY^2+a)(\vec \XX^2-b)+c\right).   \ee
 The  equations $W_\alpha=0$ become 
 \be\label{classloc2} (\vec Y^2+a)(\vec X^2-b) +c=0 ,\ee
 familiar from section \ref{heterotic}.    For generic values of $a,b,c$, this describes a smooth manifold $M$, which is qualitatively similar to the black hole
 or the spacetime underlying the \HP solution depending on the sign of $ab-c$.    When $M$ is smooth, $P$ vanishes in all classical supersymmetric states.
 
 What is different from the heterotic string is that the superfield $\PP$ contains an ordinary scalar field $P$, not just a bosonic auxiliary field.  To minimize the
 energy, there are additional conditions  (\ref{addeqs})  involving $P$.  These conditions now  become
 \be\label{peqns} P\vec Y(\vec X^2-b)= P(\vec Y^2+a)\vec X=0.\ee
 For $a>0$, $b\not=0$, 
and assuming $P\not=0$, these conditions are only satisfied at $\vec X=\vec Y=0$, which is consistent with eqn. (\ref{classloc2}) if and only if $ab=c$.  This
 just reflects the fact that $M$ is singular at $ab=c$ and its singularity is at $\vec X=\vec Y=0$.       Thus, precisely
 when we try to make a transition at the classical level from the black hole to the \HP spacetime, a new branch opens up in the space of classical ground states,
 parameterized by $P$.   This is a branch on which the $\ZZ_2$ chiral $R$-symmetry is spontaneously broken.   However, the picture is rather different quantum
 mechanically, as we explain in section \ref{effectivesuper}.   
 
We ultimately do not know how the model based on the superpotential $W$ of eqn.  (\ref{superp}), and variants of this model that we describe momentarily, behave in the
crossover between the Horowitz-Polchinski model and the black hole.   However, we can describe what in a sense is the minimal model in which the puzzle
arises.    The model based on $W$ has some sort of singular behavior, at the classical level, at $ab=c$.   The singularity occurs at $\PP=\XX=\YY=0$.
If we expand around this point, the leading terms in the superpotential are
\be\label{wnew}\widetilde W=m\PP\left(-b \vec\YY^2+a \vec \XX^2+c-ab\right). \ee
We do not know how this model behaves when $ab-c$ is changing sign, but whatever the answer is, we suspect that the behavior of the more complete
model (\ref{superp}) is similar. A linear sigma model similar to \nref{wnew} also appeared in  appendix A of 
\cite{Adams:2005rb} as part of a related topology change discussion.

 Many remarks in section \ref{heteconstruction} have obvious analogs.   One  point that merits some discussion is the quantum correction to the
 radius of the $\vec X$ circle at large $\vec Y$.   In the model as presented so far, this radius grows as $\log^{1/2} |\vec Y|$,  as in the analogous
 heterotic string model of section \ref{heteconstruction}.   This again can be avoided by adding additional massive fields.  As in the discussion of the heterotic
 string, a minimal choice is to add another $\tau$-even superfield $\XX'$, and another $\tau$-odd one $\TT$, and take the superpotential to be
 \be\label{typetwosecond} W=m \PP\left((\vec\YY^2+a)(\vec \XX^2-\XX'^2-b)+c\right)+m' \TT(1+f \vec \YY^2)\XX'.\ee
As in the case of the heterotic string, a cancellation between the $\vec Y$ dependence of $\langle \vec X^2\rangle$  and of $\langle X'^2\rangle$
ensures that the $\vec X$ circle has a constant radius for $\vec Y\to\infty$.   If one also wishes to avoid the need for
 a logarithmic renormalization of $b$, one can do this, as in  eqn. (\ref{furtherelab}), by taking $\TT$ and $\XX'$ to be two-component fields and
 choosing
 \be\label{typetwothird}W=m \PP\left((\vec\YY^2+a)(\vec \XX^2-\vec \XX'^2-b)+c\right)+m' (\TT_1\XX_1'+\TT_2(1+f\vec \YY^2)\XX'_2).\ee
There are many other ways to add additional massive fields without changing the low energy behavior at the classical level.
  
 It is straightforward to repeat for Type II the analysis given in section \ref{QuanNum} of the quantum numbers of the ground state of 
a string that wraps around the thermal circle.   One simply gets two copies of the previous computation, one copy for modes of positive chirality and one
for modes of negative chirality (more precisely, one copy for modes even under the chiral $R$-symmetry and one for  modes that are odd).   The half unit of momentum
cancels between the two copies.   However, the ground state in this winding sector is odd under the chiral $R$-symmetry $\tau$, and also odd under the opposite
chirality $R$-symmetry $\tau (-1)^F$.   It is even under $(-1)^F$.   This agrees with the standard string theory result \cite{Rohm:1983aq}.

 \subsubsection{The effective superpotential}\label{effectivesuper}
 
 \EW{The following two paragraphs have been changed.  No other changes in this section.}
 
 At the quantum level, in contrast to somewhat similar problems with more supersymmetry, what happens at $ab=c$ is not the opening up of a new branch of vacua
 parametrized by $P$.    
 Rather, because of the spontaneous breaking of the chiral $R$-symmetry on the branch with $P\not=0$, there is no obstruction to generating an effective superpotential
 $W_\eff(\PP)$ on this branch, and such a superpotential is in fact generated.  Such a superpotential lifts the degeneracy of the $P$ branch, generically leaving
 a finite number of massive vacua.
 
 The details depend on precisely which model one considers.   We will make the analysis for the ``original'' model with superpotential presented in eqn. (\ref{superp}).   
 Other models such as the ones described in eqns. (\ref{typetwosecond}) and (\ref{typetwothird}) can be treated in a similar fashion.   Though the details
 are model-dependent, some properties are general.   A nontrivial superpotential for $\PP$ is always generated.   This leads to new massive vacua at large
 values\footnote{Analogs of massive vacua at large $P$ have been found previously in two-dimensional models with $(2,2)$ supersymmetry.  See 
\cite{Harvey:2001wm} and section 13.6 of  \cite{Deligne:1999qp}. }
 are of $P$.   The number of such vacua depends on the coupling parameters.   The last statement is unavoidable, for a topological reason that will be 
 explained in section \ref{analysis}.  
 
 A convenient way to compute the effective superpotential for $\PP$ is as follows.
 Let $F$ be the auxiliary field in the multiplet $\PP$.
   In the free field theory of the superfield $\PP$, perturbed by a superpotential
 $W_\eff(\PP)$, $F$ is related to $P$ by $F=\partial_P W_\eff(P)$.   Hence we can determine $\partial_P W_\eff(P)$ -- and therefore $W_\eff(P)$ by integration --
 by computing $\langle F\rangle$ as a function of $\langle P\rangle$, on the branch of classical ground states
 with $\langle P\rangle \not=0$.    Going back to the linear sigma model with canonical kinetic energy
 and  superpotential 
 $W=m \PP\left((\vec\YY^2+a)(\vec \XX^2-b)+c\right),$ we see that the formula $F=\partial_P W$ (which holds by the equations of motion of
 the linear sigma model) gives  
 \be\label{auxfield}F= m\left((Y^2+a)(X^2-b)+c\right).\ee
 So to determine $\langle F\rangle$, 
 we just have to compute the expectation value of the operator on the right hand side of eqn. (\ref{auxfield}) in a vacuum with nonzero $P$.
 This can be done reliably in perturbation theory only in the region where $P$ is large.   
 At large $P$, the energy is minimized at $\vec X=\vec Y=0$ and the superfields $\vec\XX$, $\vec\YY$ have masses
 proportional to $P$.  We can integrate out $\vec \XX$, $\vec \YY$ in perturbation theory by evaluating Feynman diagrams with $\XX$ and $\YY$ propagators.
 Because $\vec \XX$ and $\vec \YY$ have masses of order $P$
 and  the linear sigma model is superrenormalizable, most  such Feynman diagrams generate contributions proportional to negative powers of $P$.   
 The only relevant exceptions are the one-loop ``bubble'' contributions to $\langle \vec X^2\rangle $
 and $\langle\vec Y^2\rangle$.   
 To evaluate the expectation value of $m\left((Y^2+a)(X^2-b)+c\right)$ in a state with large $P$,  we just replace
 $X^2$ and $Y^2$  with their  expectation values $\langle\vec X^2\rangle$ and $\langle \vec Y^2\rangle$, computed
 from the bubble diagram.   The computation is the same as the one in eqn. (\ref{oneloop}), except that now $\vec X$ has two massive components and $\vec Y$
 has $d$ of them.    The result is 
 \be\label{effective}F=\frac{\partial W_\eff (P)}{\partial P}= \left(  -\frac{d}{4\pi}\log (P^2/\mu^2)+a_r     \right)    
 \left(-\frac{1}{2\pi} \log (P^2/\mu^2)-b_r\right) +c,\ee where $a_r $ and $b_r$ are renormalized versions of $a$ and $b$.
 It is possible to integrate this with respect to $P$ to get the effective superpotential (and this function is an odd function of $P$, as expected).
 However, what we actually want is to identify possible vacuum states in the region of large $P$.  The condition for a vacuum state is $\partial W_\eff(P)/\partial P=0$,
 the vanishing of the right hand side of eqn. (\ref{effective}).   
In terms of $S=\frac{1}{4\pi}\log( P^2/\mu^2)$, the equation for a vacuum state becomes a quadratic equation for $S$:
 \be\label{quadeq} \left(dS-a_r\right)\left(2S +b_r\right)+c=0, \ee
 leading to
\be\label{soln}S=\frac{-(db_r-2a_r)\pm\sqrt{(db_r-2a_r)^2-8d(c-a_r b_r )} }{4d}   .\ee
We recall from section \ref{heterotic} that the nonlinear sigma model is weakly coupled for large $a_r,b_r,$ and $c$ with $a_r/b_r$ and $c/a_r b_r$ fixed.\footnote{Because the renormalization from $a$ and $b$ to $a_r$ and $b_r$ is by an additive constant, this assertion is unaffected by replacing $a$ and $b$ with
$a_r$ and $b_r$.}   In this regime, the solutions for $S$ are generically large but not necessarily real or positive.   Solutions with ${\mathrm{Im}}\,S\not=0$ do not correspond
to vacuum states, since the field $P$ is real.   Solutions with $S$ real and negative correspond to small values of $P$ at which the computation is not reliable (small $P$ is
the region of the nonlinear sigma model).    
But solutions with $S$ real and positive (and sufficiently large) correspond to a pair of massive supersymmetric vacua, at $P=\pm \mu\exp(2\pi S)$.  Observe that if $S$ is
large, $P$ is exponentially large.


%
%

From eqn. (\ref{soln}), we find the following three cases, where we label HP or BH in terms of the sign of $c -ab$, as in \nref{BHHPca}, see also figure \ref{fig:vacua}.

(HP$_2$): When $c-a_r b_r<0$. In this case there is one positive and one negative solution of \nref{soln}, but we can trust only the positive one, which leads to two vacua with opposite values of $\langle P \rangle $. 
 
(BH$_0$): This is when there are no positive solutions of \nref{soln}, which happens under two circumstances, when  $c-a_r b_r>0$ and  $db_r-2a_r>0$ or, alternatively, when $d b_r - 2 a_r < 0$ and $c-a_r b_r$ is sufficiently positive so that there is a pair of complex solutions.  In this case there are no vacua   in the large $|P|$ region. 

(BH$_4$): If $db_r-2a_r<0$ and $c-a_r b_r > 0$ but not too positive, then we have two positive real roots in \nref{soln}. For each of them we can have two opposite values of $\langle P \rangle$, so that in total we have four vacua.

\begin{figure}[t]
    \begin{center}
    \includegraphics[scale=.2]{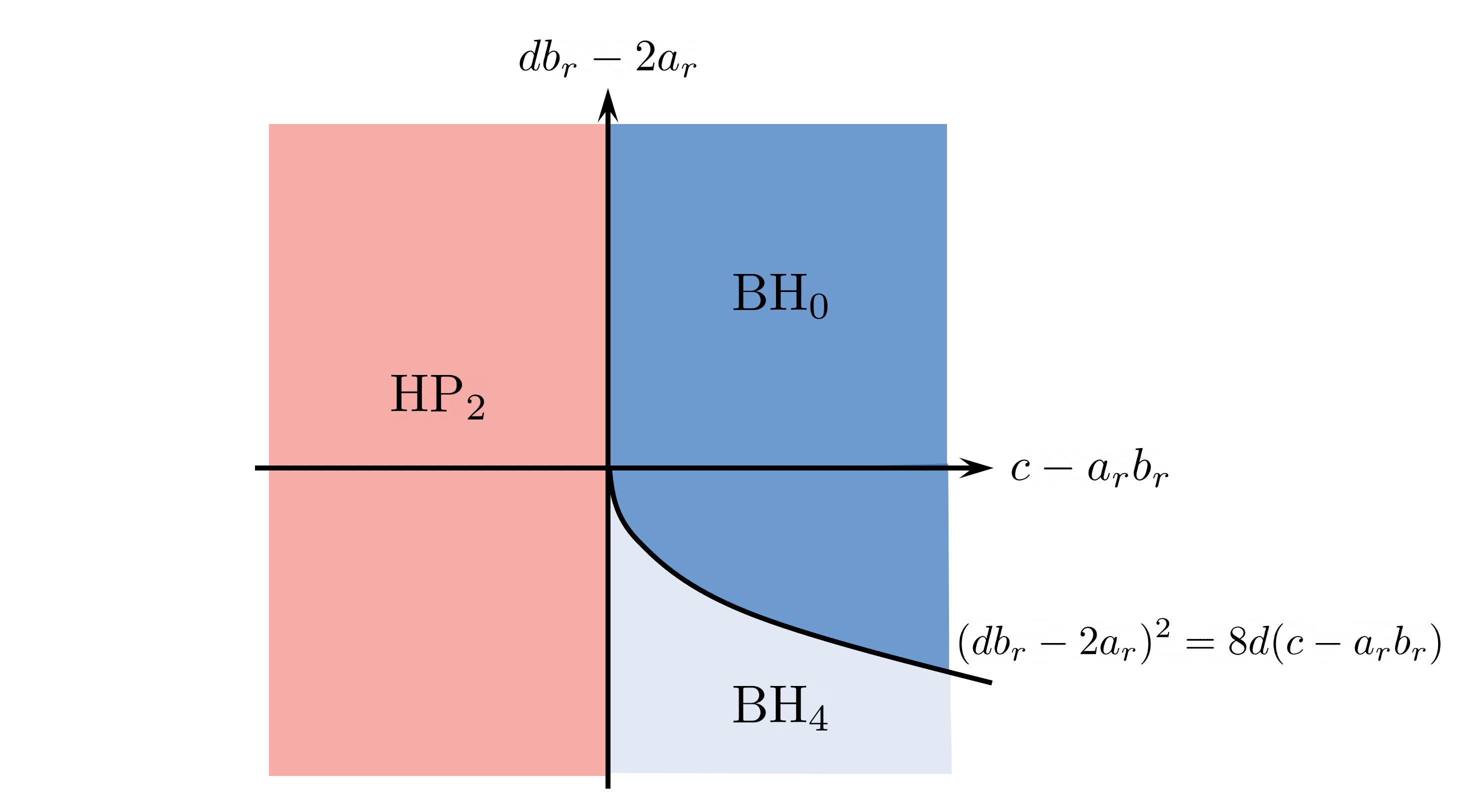}
    \end{center}
    \caption{  We  find three different possibilities for the massive vacua, depending on the values of $db_r - 2a_r$ and $c- a_r b_r$. The subscripts $0,2,4$ label how many massive vacua there are in each case. The analysis is not reliable when we are very close to the vertical axis in the plot, where $|P|$ becomes small.
}
   \label{fig:vacua}
\end{figure}

These massive vacua exist in addition to the non-compact, non-linear sigma model branches that have the topology of the \HP solution or the black hole solution. Note that
within the region of parameter space that we have the \HP solution, we also have two massive vacua. Surprisingly, in the region of parameter space corresponding to the black hole we have a region with no massive vacua and a region with four massive vacua.  
 At least in the linear sigma model, there cannot be a smooth crossover from the black hole to the \HP solution, because somewhere along
the way a pair of massive vacua appears or disappears.   But we do not know what this phenomenon corresponds to in the infrared,
after hypothetically flowing from the linear sigma model to a family of CFT's and moving to the crossover region with $c-a_r b_r$ small. 

In section \ref{analysis}, we will explain in terms of the supersymmetric index $\Tr\,(-1)^F$ the appearance or disappearance of massive vacua in interpolating between
the \HP solution and the black hole.   We will also explain what is happening in the transition from region (BH$_0$) to region (BH$_4$), where the number of massive vacua
changes even though the nonlinear sigma model can be weakly coupled and nothing in particular is happening to it.

    \subsection{The topological obstruction}\label{analysis}

    Consider a supersymmetric field theory with Hamiltonian $H$, Ramond sector Hilbert space $\HH$, and a space $\HH_0$ of supersymmetric ground states.
    In a family of such theories with discrete spectrum, the index
    \be\label{thindex}I=\Tr_{\HH_0}\,(-1)^F=\Tr_\HH \,(-1)^F e^{-\hat\beta H}\ee
    is a constant.    Here $\hat \beta$ is an arbitrary positive number; the last expression does not depend on $\hat\beta$ because states of positive energy occur in
    bose-fermi pairs and cancel out of the trace.\footnote{$\hat \beta$ is completely unrelated to the spacetime inverse temperature $\beta$.}
    In the case of a nonlinear sigma model with $(1,1)$ supersymmetry and with 
    target a $D$-dimensional 
    compact manifold $M$, the space $\HH_0$ is the direct sum of the real cohomology group $H^q(M)$, whose dimension is the Betti number $b_q$.   
    The operator $(-1)^F$ acts on
    $H^q(M)$ as $(-1)^q$.  The index is therefore
    \be\label{eulchar} I=\sum_{q=0}^D \Tr_{H^q(M)}(-1)^F =\sum_{q=0}^D (-1)^q b_q = \chi(M), \ee
    where $\chi(M)$ is the Euler characteristic of $M$.
    
    A slight generalization of this is to consider a supersymmetric theory with a global symmetry $T$ that commutes with supersymmetry.
    In a family of such theories with discrete spectrum, the quantity
    \be\label{lefquantity} I_T=\Tr_{\HH_0}\,(-1)^F T = \Tr_\HH \,(-1)^F T e^{-\hat{\beta} H} \ee
    is a constant.   In the case of a nonlinear sigma model with target $M$, if $T$ is a symmetry of $M$, then
    \be\label{lefdef} I_T=\sum_{q=0}^D (-1)^q\Tr_{H^q(M)} T,\ee
    a quantity that is known as the Lefschetz number $L_T$ of $T$.
    
    We are not in such a simple situation, because the black hole and the spacetime $\RR^d\times S^1$ that underlie the \HP solution are not compact.
    However, the two spacetimes are equivalent at infinity.   One would expect that any obstruction to interpolating smoothly from the black hole to the \HP solution
    is local, and depends on what is happening in  the interior of the spacetime, not on the geometry at infinity, where nothing is changing.   Without changing anything in the interior, we can ``cap off''
    the black hole and \HP solutions at spatial infinity and compactify them, making the discussion of $\Tr\,(-1)^F$ more straightforward.   We can do this, for example,
    by gluing in at infinity a copy of $B\times S^{D-2}$, where $B$ is a two-dimensional disc.   The black hole then becomes $S^2\times S^{D-2}$, and the \HP
    solution becomes $S^D$.   We have
    \begin{align}\label{eulchartwo} \chi(S^2\times S^{D-2} )& = 2(1+(-1)^D) \cr   \chi(S^D)&=1+(-1)^D. \end{align}
    Thus the two differ by 2 if $D$ is even, but agree if $D$ is odd.
    
To explore the case of odd $D$, it is convenient to let $T$ be a reflection symmetry that reverses the orientation of $S^D$ or $S^{D-2}$.  
In the linear sigma model, this is the symmetry that reverses the sign of one component of $\YY$, say $\YY_1\to -\YY_1$.
With this choice of $T$, one has
\begin{align}\label{lefsphere} L_T(S^2\times S^{D-2}) & = 2(1+(-1)^{D-1}) \cr 
                                               L_T(S^D)&=1+(-1)^{D-1}. \end{align}
     These differ by 2 if $D$ is odd, but agree if $D$ is even.
  
  \EW{Following paragraph added.}   
In the case of a nonlinear sigma model with noncompact target space, in general  invariants such as $\Tr\,(-1)^F$ and $I_T$ cannot be computed just by counting
normalizable ground states in the theory formulated on a circle, because  there may be boundary contributions to these invariants.
However, when we compare the black hole to the \HP solution, a possible boundary contribution cancels out and so the comparison can be made
by counting ground states.   For a nonlinear sigma model with target space $M$ formulated on a circle, the normalizable ground states correspond
to normalizable harmonic forms on $M$.   
  The \HP spacetime is $\RR^{d-1}\times S^1$
with a flat metric; it has no normalizable harmonic forms.   However, the black hole spacetime has two of them.   The black hole metric is
\be\label{bhmetirc} ds^2=\left(1-\frac{\mu}{\rho^{D-3}}\right)dt ^2 +\frac{1}{1-\frac{\mu}{\rho^{D-3}}}d \rho^2 +\rho^2 d\Omega_{D-2}^2. \ee
In this spacetime, there are two normalizable harmonic forms.   One is  $\theta=d \rho\, d t/\rho^{D-2}$, which is easily seen to be normalizable 
and to satisfy $d\theta=d\star \theta=0$ (where $\star$ is the Hodge star operator\footnote{In its action on the supersymmetric ground states, the chiral
$R$ symmetry $\tau$ is equivalent to $\star$.  The anomalous commutatation relations (\ref{anomcomm}) that we discuss momentarily can therefore
be seen in the action of $(-1)^F$, $T$, and $\star$ on the states $\theta$ and $\star\theta$. Since $\star$ exchanges these two states, which carry
different quantum numbers under $(-1)^F$ and $T$, clearly these operators do not commute.}),
and therefore to be harmonic.    The second is $\star \theta$, which satisfies the same conditions.  Here $\theta$ is a two-form and $\star\theta$ is a $D-2$-form,
so they contribute $1+(-1)^D$ to $\Tr\,(-1)^F$.     Since $\theta$ is even under $T$ and $\star\theta$ is odd, these two states contribute $1+(-1)^{D-1}$ to $L_T$.
For a proof that there are no other normalizable harmonic forms in the Schwarzschild spacetime, see \cite{Etesi:2000mv}.
Thus, the contributions of $\theta$ and $\star\theta$ account for the difference between the values of $\Tr\,(-1)^F$ and $L_T$ for the black hole and  the \HP spacetime.

     The importance of this matter for our purposes is that the 
     different  values  of $\chi$ and $L_T$ are an obstruction to a smooth interpolation between the black hole and the \HP solution; depending on $D$,
     either $\chi$ or $L_T$ differs by 2 between the black hole and the \HP solution.   With this in mind, the results found in section \ref{typetwo} are natural.
     In that analysis, we found that in interpolating from the black hole region to the \HP region, the number of massive vacua at large $P$ changes by $\pm 2$
     (see fig. \ref{fig:vacua}).   With appropriate assumptions about the quantum numbers of the massive
     vacua under $(-1)^F$ and $T$, this can account for the jumping of $\chi$ or $L_T$ by 2.

         \EW{Some changes in this paragraph}
     The necessary statements about the quantum numbers of the massive vacua can be largely explained as follows.  We recall that the chiral $R$-symmetry $\tau$
     exchanges pairs of massive vacua at positive and negative $P$.  Classically, $\tau$ commutes with $(-1)^F$ and $T$, but quantum mechanically, in the
     Ramond sector, there is an anomaly
     \be\label{anomcomm}\tau (-1)^F =(-1)^D (-1)^F \tau,~~~~   \tau T = -T\tau.\ee
     The anomaly can be found by turning off the Yukawa couplings and computing explicitly the action of $\tau$, $(-1)^F$, and $T$ on the Ramond-Ramond ground states of the free theory. Acting on these ground states, the left and right moving    fermion zero modes obey the same algebra as gamma matrices. The operators $\tau$, $(-1)^F$ and $T$ are products of gamma matrices (or fermion zero modes) that are uniquely determined by requiring that they transform
     the fermion zero modes in the expected way.   Given these facts, it is
      straightforward to identify the operators representing $\tau$, $(-1)^F$, and $T$ in the space of ground states and to obtain \nref{anomcomm}.   Eqn. (\ref{anomcomm}) shows that pairs of massive vacua related by $\tau$ have opposite eigenvalues of $T$,
     and have $(-1)^F $ differing by a factor $(-1)^D$.  Therefore,  a pair of massive vacua exchanged by $\tau$ always have opposite values of $T$,
     and have opposite values of $(-1)^F$ if and only if $D$ is odd.  These statements  make the pattern of jumping of massive
     vacua found in section \ref{typetwo} consistent  with the jumping of $\chi$ and $L_T$ in the transition between the black hole and the \HP solution.    
     
Finally, in section \ref{typetwo}, we found that it is possible for two massive vacua to annihilate at large $P$. 
This occurs (both at positive $P$ and at negative $P$)
in the transition between the ${\mathrm{BH}}_0$ and ${\mathrm{BH}}_4$ regions of fig. \ref{fig:vacua}.   The linear sigma model
remains weakly coupled and reliable through this transition.  There is a simple interpretation:
the two vacua that appear or disappear have opposite values of $(-1)^F$ and the same value of $T$, so they make
no net contribution to $\Tr\,(-1)^F$ or $I_T$.     A simple example in which a pair of massive vacua annihilate
is a theory with one scalar superfield $\PP$ and a superpotential $\PP^3-a\PP$, with parameter $a$.  For $a>0$, there is a pair of massive vacua
at $\PP=\pm \sqrt{a/3}$; for $a<0$, there are none.

\subsection{D-Branes}\label{dbranes}

\EW{Various relatively minor changes in this section.}

The invariance of $\Tr\,(-1)^F$ and $I_T$ is not the whole story concerning the obstruction to interpolating smoothly between the black hole and the \HP solution.
There is a mismatch between the $D$-branes of the black hole and of the \HP spacetime.
In describing this mismatch, we will, except at the end of this section, consider the black hole or \HP CFT, together with its $D$-branes, as opposed to the full string theory 
(in which, in addition to the $D$-branes,  one would  also consider Ramond-Ramond fields
and associated collective coordinates).    For a basic illustration of the mismatch, observe that in the 
black hole spacetime, which is topologically $B\times S^{D-2}$, where $B$ is a two-dimensional ball,
it is possible to have a $D$-brane wrapped on $S^{D-2}$ (times a point in $B$).
Such a $D$-brane has no analog in the \HP apacetime.   Conversely, in the \HP spacetime $\RR^{D-1}\times S^1$, one can have a brane wrapped on $S^1$;
this has no direct analog in the black hole spacetime.

To express more systematically the difference between the black hole and the \HP spacetime, we observe the following.  In any CFT with (1,1) supersymmetry, one  can define a group of conserved charges of the $D$-branes (boundary states that preserve one worldsheet supercharge).
These charges are invariant under continuous deformation of a $D$-brane, are additive if one takes the direct sum of two $D$-branes, and vanish for any
$D$-brane system that can annihilate by tachyon condensation.  
In the case of a supersymmetric nonlinear sigma model with target $M$, the conserved charges make up what is called the $K$-theory of $M$.  $K$-theory is similar
to cohomology, but  unlike cohomology, which is $\ZZ$-graded, $K$-theory is only $\ZZ_2$-graded.    Thus there are two $K$-theory groups, $K^0(X)$ 
(often called just $K(X)$) and $K^1(X)$.   If $M$ is a ten-manifold that is a target space of Type II superstring theory, then $K^0(M)$ and $K^1(M)$ are 
associated to $D$-branes of Type IIB and Type IIA superstring theory on $M$, respectively \cite{Witten:1998cd}.  

$K^0(M)$ and $K^1(M)$ are finitely generated abelian groups.     In general, they have torsion subgroups, which correspond to possible discrete charges carried by
$D$-branes; if we ignore the torsion subgroups, then $K^0(M)$ is a lattice, isomorphic to $\ZZ^c$ for some integer $c$, and similarly $K^1(M)$ is a lattice,
isomorphic to some $\ZZ^{c'}$.   In fact, if we define the sum of the even Betti numbers of $M$, $b_+=\sum_{i\in2\ZZ} b_i$, and the sum of the
odd Betti numbers, $b_-=\sum_{i\in 2\ZZ+1} b_i$, then\footnote{The lattice $K^0(M)$ has the same rank as the direct sum of the even
cohomology groups, but in general the lattice structure is different. The same applies for the relation between $K^1(M)$ and the odd cohomology groups.
The discrete charges are also different in K theory from what they would be in cohomology.} $c=b_+$, $c'=b_-$.    

Since $K^0(M)$ and $K^1(M)$ are both invariants in a family of theories with $(1,1)$ supersymmetry, we conclude that $b_+$ and $b_-$ are both invariants.
This is a stronger statement than invariance of $\Tr\,(-1)^F=b_+-b_-$.

To sum up, in section \ref{typetwo}, we found that a Type II  linear sigma model does not make a smooth interpolation between the black hole and \HP phases,
and here we have explained why this might be expected.   By contrast, in section \ref{heterotic}, we learned that    a linear sigma model with $(0,1)$
supersymmetry, appropriate for the heterotic string, can make a smooth interpolation between the two phases.   In fact, it is believed that properties of a $(0,1)$
supersymmetric model in two dimensions that are invariant under arbitrary continuous deformations (including arbitrary RG flows) are very restrictive.
In the case of a sigma model with target $M$, it is believed that all such invariants are cobordism invariants\footnote{   
For example, in a $(0,1)$ nonlinear sigma model
with target $M$ (and no fermi multiplets), $\Tr\,(-1)^F$ is equal to the index of the Dirac operator on $M$, which is a cobordism invariant (this compares to
$(1,1)$ supersymmetry, where $\Tr \,(-1)^F$ is the Euler characteristic, not a cobordism invariant).   More generally,
the elliptic genus of $M$ is a cobordism invariant.}    of $M$.    The black hole and the \HP spacetime
are cobordant; a  cobordism between them is described
by the familiar equation
\be\label{cobordism} (\vec Y^2+a)(\vec X^2-b)+c =0, \ee
which we view now as describing a smooth manifold of dimension $D+1$ parametrized by $\vec X$, $\vec Y$, and $c$ (with $a$ and $b$ viewed as constants).
This manifold has the topology of the black hole if restricted to a value of $c$ greater than $ab$, or of the \HP solution if restricted to a value of $c$ less than $ab$.
So it establishes a cobordism between the two spaces.   The invariants that we have been discussing, such as $\Tr\,(-1)^F$, $K^0$, and $K^1$,
are not cobordism invariants and can distinguish the black hole from the \HP spacetime in the context of Type II superstring theory. 
In other words, the fact that these quantities do not have the same values for the black hole and the \HP solution is an obstruction to the existence of a smooth
transition between them at the level of two-dimensional supersymmetric field theory. 
 
 \EW{Some changes in this paragraph}
 
So far we have considered D-branes purely from the standpoint of conformal field theory.  In the full string theory, the picture is somewhat different.  
In string theory, a D-brane is the source of some RR flux; if the D-brane carries a conserved charge, this flux can be measured at infinity.   However,
depending on the bulk topology, a given RR flux at infinity may not need a D-brane source.
 For example, the \HP solution in Type IIA superstring theory has a D0 brane wrapping the Euclidean time circle. On the black hole side, there
 is no wrapped D0-brane, but  we can obtain a configuration with the same RR flux at infinity by simply taking the RR two-form field $F_2$ to
 be a multiple of the harmonic two-form $\theta$ 
 that was described in section \ref{analysis}.   
This gives a Euclidean description of a black hole carrying D0 brane charge.   The difference is that on  one side the flux is sourced by an explicit D-brane while on the other it can exist as an everywhere regular flux.   
In the IIB theory, we can have a similar discussion involving branes which are also wrapped along some spacetime dimensions other than those
that parametrize the black hole or the \HP solution.

  \section{Action of a classical string theory solution } \la{ActionTree}
  
  Let us consider a classical lagrangian of the form  
  \be \la{CalLa}
  I = \int d^Dx\, \sqrt{g} e^{ - 2 \phi } {\cal L}( \partial_\mu \phi   , \eta  ) 
  \ee 
  where $\phi$ is the dilaton  and $\eta$ denotes any other fields, including the metric. The important property is that the only $\phi$-dependent term that does not involve derivatives of $\phi$ is the overall factor of $e^{ - 2 \phi}$ in the action. 
  Then the equation of motion for $\phi$ implies that 
  \be \la{EOMdil}
  -2 e^{ - 2 \phi } {\cal L } = \partial_\mu \left( e^{-2\phi}{ \delta {\cal L} \over \delta \partial_\mu \phi } \right).
   \ee
   This implies that, on shell, the action density is a total derivative and therefore the full action reduces to a boundary term \cite{Tseytlin:1988tv}. This argument also works if the action involves higher derivatives of $\phi$. The general form of the Lagrangian \nref{CalLa} is correct to all orders in the $\alpha'$ expansion,\footnote{We might worry about the RR fields, which in the usual presentation do not involve a factor of $e^{ - 2\phi}$ in their action. However, as discussed in section 12.1 of \cite{Polchinski:1998rr},  we can pull out a factor of $e^{ -2\phi}$ at the cost of complicating the expressions for the field strength and gauge transformations, which now contain terms proportional to the gradients of $\phi$. In fact, this is the form that comes out naturally from  worldsheet perturbation theory. The  $\alpha'$ corrections should continue to contain this overall factor of $e^{ -2\phi}$ once we choose this other definition of the fields. Of course, in this case, the boundary term can involve the RR potentials and gauge fields at the boundary. }
though only  at tree level in the string coupling. 

If we consider an asymptotically flat solution then the curvatures becomes small as we go to infinity. This means that the boundary term can be evaluated at infinity using the lowest order terms in the action 
   \be 
   I \sim { 1 \over 16 \pi G_N} \left[ \int_{\cal M} d^D x\, \sqrt{g}  e^{ - 2 \phi } ( - {\cal R } - 4 (\nabla \phi)^2  + \cdots ) -  2 \int_{\partial \mathcal{M}  }  d^{D-1} x \,\sqrt{h} e^{ - 2 \phi } K \right]
   \ee 
   where we have also added the Gibbons-Hawking-York boundary term and $h$ is the induced metric on the boundary.    Using the equation of motion for the dilaton \nref{EOMdil} and integrating by parts we find that the total action becomes \cite{Kazakov:2001pj}   
      \be \la{FinOsh}
   -I =  { 1\over 8 \pi G_N}   \int_{\partial \mathcal{M} }  d^{D-1} y \, e^{ - 2 \phi} \sqrt{h} (K - 2 \partial_n \phi) = { 1\over 8 \pi G_N}  \int_{\partial \mathcal{M}  }  d^{D-1} y \, \partial_n( e^{ - 2 \phi } \sqrt{h }   ).
   \ee 
   This is the final form for the on shell action for any classical string theory solution. 
  The action is purely determined in terms of the asymptotic values of the metric and the dilaton at large radius $\rho$ 
   \bea 
  \phi & \sim &    { C_\phi \over \rho^{D-3} }  
     \cr 
   ds^2 & \sim & e^{ 4 \phi \over D-2} \left[ f dt^2 + { d\rho^2 \over f } + \rho^2 d\Omega_{D-2}^2 \right] ~,~~~~~~~ f \sim 1 - { \mu \over \rho^{D-3} } \la{StFm}
   \eea
 where we have given the form of the string frame metric, which acquires the usual form in Einstein frame, whose expression is the term in the brackets above. $C_\phi$ and $\mu$ are some constants that depend on the details of the solution. We have also absorbed the asymptotic value of the dilaton into $G_N$ so that $\phi_{D , \infty}$ is set to zero.  Of course, this asymptotic value is free parameter of the classical solution. 
   
   We can now evaluate the Euclidean action and after subtracting appropriate counterterms we find 
   \bea 
   \log Z \sim - I &=& { \omega_{D-2} \beta \over 8 \pi G_N}   \lim_{\rho\to \infty} \left[ \sqrt{f} e^{ - { 2 \phi \over D-2} } \partial_\rho \left( \sqrt{f} e^{  {2 \phi \over D-2} } \rho^{D-2} \right) - \sqrt{f} \partial_\rho \rho^{D-2} \right] \la{RegExp}
   \cr\log Z  &=& - { \omega_{D-2} \over 16 \pi G_N}  \beta \left[ \mu +  4{ D-3 \over D -2} C_\phi   \right]  \la{FreeEn} 
   \eea 
   where $\beta$ is the period of the Euclidean time direction. Note that in  \nref{RegExp} we subtracted the flat space answer with a circle with the Einstein frame length $\beta \sqrt{f} $ at the cutoff $\rho$. Note also that the factors in front of $\partial_\rho$ in the first term comes from the relation between $\partial_n$ and $\partial_\rho$ in the string frame, \nref{StFm}.  $C_\phi$ and $\mu$  are functions of $\beta$ which  depend on the details of the solution. In fact, all the $\alpha'$ corrections appear through the  detailed dependence of $\beta $ in  $\mu$ and $C_\phi$. In deriving this formula we have assumed that the solution does not carry any gauge charges, although those can be also incorporated.

   This argument is only valid to all orders in $\alpha'$ expansion. But we suspect that the result is also true exactly in $\alpha'$, 
     since the naive value of the classical action from the CFT point of view is zero, due to the division by the volume of $SL(2,{\mathbb C})$. If we interpret this zero as a manifestation that the action is a total derivative, then we would obtain the above result, \nref{FreeEn}. It would be nice to find out whether this is true or not for the full CFT solution. 
   
   From this classical action we can compute the mass as 
   \be \la{Massth}
   M = - \partial_\beta \log Z  =   { \omega_{D-2} \over 16 \pi G_N}    \left[   \partial_\beta (\beta \mu )+  4{ D-3 \over D -2} \partial_\beta (\beta C_\phi)     \right].
   \ee 
   On the other hand we also expect that the mass should be given in terms of $\mu$ by \nref{MassUs}.
   Demanding consistency between \nref{Massth} and \nref{MassUs} we get a consistency condition relating $\mu$ and $C_\phi$
   \be \la{EqnCmu}
  0 = \beta^{D-2} \partial_\beta \left( { \mu \over \beta^{D-3} } \right) + 4{ D-3 \over D -2} \partial_\beta (\beta C_\phi) .
   \ee 
  The equivalence of the two methods for computing the mass comes from the fact that we can think of changing the temperature as either varying $\beta$, or varying the asymptotic value of the $g_{tt}$ metric component. The first way gives us (\ref{Massth}), while the second way leads to the ADM mass in (\ref{MassUs}). This works in general, while the special property of classical string theory is that through (\ref{FreeEn}) the first way gives an expression involving only $\mu$ and $C_\phi$.  
    
 In the case that $C_\phi=0$, \nref{EqnCmu} is automatically obeyed by the Schwarzschild black hole, which has $\mu \propto \beta^{D-3}$. Stringy corrections imply some deviation from this relation and \nref{EqnCmu} implies that necessarily a non-zero dilaton profile is generated. We have checked that indeed \nref{EqnCmu} is obeyed for the known $\alpha'$ corrections to a black hole \cite{Callan:1988hs,Myers:1987qx,Chen:2021qrz}. 
   
   We can also compute $C_\phi$ and $\mu$ for the \HP solution, and from (\ref{MassThA}) to (\ref{HPasymp}) we get
   \be 
   \mu = - 4 \frac{D-3}{D-2} C_\phi \propto (\beta - \beta_H)^{\frac{5-D}{2}}. 
   \ee 
   The precise relative numerical coefficient implies that the equation \nref{EqnCmu} is obeyed to leading order in the $\beta-\beta_H$ expansion. By adding suitable higher order terms to $\mu$ and $C_\phi$, we could, in principle, obey \nref{EqnCmu} at higher orders.   

  
  The formula \nref{FreeEn} also implies that the entropy of a classical string theory solution can be written in terms of boundary quantities\footnote{It is curious to note that the combination in the parenthesis corresponds to the coefficient in the asymptotic fall-off of the $g_{tt}$ component of the \emph{string} frame metric, as can be seen from (\ref{StFm}).}  
    \be \label{EntBdy}
  S = {\omega_{D-2} (D-3) \over 16 \pi G_N} \beta \left( \mu - { 4 \over D-2 } C_\phi \right)
  \ee 
  Of course, for the black hole case, the equations of motion also imply that it can be written as the usual area of the horizon, and also the corresponding Wald corrections \cite{Wald:1993nt} if we include $\alpha'$ corrections to the effective action.    For the \HP solution case, it can be written as 
  \nref{EntHP}. We should emphasize that \nref{EntBdy} is true only for classical string theory, and therefore classical gravity. However, it is not true for general theories of gravity. In particular, it is not true when we include higher curvature corrections in M-theory.\footnote{One way to see that it cannot be true is that in M-theory we have no dilaton. Without the dilaton the equation \nref{EqnCmu} imlies that $\mu \propto \beta^{D-3}$, which is the classical answer. However, higher curvature corrections modify this relation \cite{Chen:2021qrz}.}
  
   \section{Generating charged solutions } 
 \la{GenCh}

 The solutions we have discussed so far are neutral -- they do not carry any gauge charges.  In this section,  we show that these solutions can be used to produce charged solutions. More specifically, consider a compactification which contains an internal spatial circle. 
 We can then generate solutions that carry momentum and fundamental string winding charges along this circle. This can be done by a sequence of solution generating transformations which map one solution into another. These have been used previously in order to generate charged black hole solutions  starting from uncharged solutions \cite{Horne:1991cn,Sen:1994eb}.     We want to emphasize  that these transformations also make sense at the level of the worldsheet CFT, transforming one worldsheet CFT into another. These transformations can be applied to the  \HP solution, to the stringy corrected black hole, or to any other interpolating solution we find in the future. We will further see that we can concentrate on the action of those transformations at infinity and use them to calculate the thermodynamic properties of the transformed solution in terms of those of the seed solution. 
 
 Let us first discuss the solution generating transformations in the asymptotic region, when $\rho$ is very large. In the Euclidean theory we have two circles: the original Euclidean time circle and the extra internal circle we have introduced. The original theory is $D+1$ dimensional and after reduction on the two circles the theory is $d=D-1$ dimensional. This $d$ dimensional theory has an $O(2,2)$ symmetry group which acts on the scalars that determine the geometry of the two torus spanned by the Euclidean time circle and the extra spatial circle    \cite{Narain:1985jj,Narain:1986am}.
 
We follow the formalism in \cite{Maharana:1992my} to organize the $O(2,2)$ symmetry. The scalars can be combined into a matrix (we set $\alpha'=1$ in this section) 
   \begin{equation}
  M = \begin{pmatrix}
   G^{-1}  & - G^{-1} B\\
   B G^{-1} & G - B G^{-1} B   
  \end{pmatrix} ~,~~~~~~~~~\alpha'=1
 \end{equation}
where $G$ and $B$ are the metric and the $B$ field on the two torus, viewed as $2\times 2$ matrices. Under an $O(2,2)$ transformation $\Omega$, namely a transformation that satisfies
 \begin{equation}
  \Omega^T \eta \Omega = \eta, \quad    \eta = \begin{pmatrix}
    0 & 1_{2\times 2} \\
    1_{2\times 2} & 0
    \end{pmatrix},
 \end{equation}
$M$ transforms as
\begin{equation}\la{Mtransf}
  M \rightarrow \Omega M \Omega^T. 
\end{equation}

Note that in principle the above explicit transformation is only accurate near the boundary of the spacetime, where the gradients of the fields are negligible. When the gradients are large, one can have $\alpha'$ corrections to the form of the transformation \cite{Bergshoeff:1995cg,Kaloper:1997ux,Bedoya:2014pma}. This is related to a subtlety in the heterotic case, as we will discuss in sec. \ref{sec:SolnGeneHet}. Luckily, for our main purpose of computing thermodynamic quantities, the transformation at infinity is good enough. 

Another useful fact is that a string state which carries winding $w_{t},w_y$ and momentum $n_t, n_y$ has mass
\begin{equation}\label{mstr}
    m^2 =  \begin{pmatrix}
    w_t & w_y & n_t & n_y
    \end{pmatrix} M^{-1} \begin{pmatrix}
    w_t \\ w_y \\ n_t \\ n_y
    \end{pmatrix}.
\end{equation}

In \cite{Sen:1991zi}  it was argued that these solution generating transformations should extend to all orders in $\alpha'$ for situations where the seed solution has no momentum and winding. 
However, as we have argued above, both the \HP solution as well as the black hole solution lead to an expectation value for a winding mode. In the heterotic case, this winding mode also has a half unit of momentum,  a property we have encountered in sec. \ref{QuanNum}.  

Nevertheless, we can still consider transformations $\Omega$ that do not ``touch'' the winding mode. In particular we require that if we have a mode with the quantum numbers of the winding mode excited in these solutions, then $\Omega$ should leave it invariant 
\be \la{condition}
\Omega_{\rm type~II}   \begin{pmatrix}
    1 \\0 \\ 0\\ 0
    \end{pmatrix} = \begin{pmatrix}
    1 \\0 \\ 0\\ 0
    \end{pmatrix} ~,~~~~~~~~~~~~
\Omega_{\rm heterotic}  \begin{pmatrix}
    1 \\0 \\ \half \\ 0
    \end{pmatrix} = \begin{pmatrix}
    1 \\0 \\ \half \\ 0
    \end{pmatrix}    
\ee 

This produces a set of allowed transformations that we can use in each case to generate new solutions. Since the transformations are different in each of the two cases we will treat each of them separately. There is a three parameter set of allowed transformations. One of the parameters is trivial because it just changes the radius of the internal circle. The remaining two parameters will generate momentum and winding charges. 

So far, we have defined the solution generating transformations in the asymptotic region. However, we think that these can be extended to the full CFT solution. For the type II case we will present a more detailed argument in section \ref{CFTarg}. 

First we will discuss some common features. Denoting the parameters of the seed solution in terms of tilde quantities and the parameters of the transformed solution without tildes we have 
\be \la{ParRel}
M = \Omega \tilde   M \Omega^T  ~,~~~~~~\tilde G = \begin{pmatrix}
    \tilde R^2 & 0  \\0 & \tilde r^2    \end{pmatrix} ~,~~~~~~~G =\begin{pmatrix}
     R^2 + r^2 a^2  & r^2 a  \\  r^2 a &  r^2    \end{pmatrix} ~,~~~~~B_{12} = b
\ee 
where we are writing the final metric in a Kaluza-Klein form 
$ ds^2 = R^2 dt^2 + r^2 (dy + a dt )^2$ and we take both $t$ and $y$ to be circles of size $2\pi$. $a$ and $b$ are then related to the chemical potentials for the associated charges 
\be \la{ChemPot}
a = i R \mu_p ~,~~~~~~~ b = i 
R \mu_w ~,
\ee 
 where $a,b,R$ are values at infinity.  The relation between the tilde and untilde quantities is determined by $\Omega$ and depends on those parameters. We will give the explicit expressions below for the type II and the heterotic cases. 

As we discussed in section \ref{ActionTree}, the action is given by equation \nref{FinOsh} which depends only on the metric and dilaton at the boundary and can be written in terms of 
\be 
\log Z=- I = { 1 \over 8 \pi G_N} \int d^{d-1}y\, \partial_n ( e^{ - 2 \phi_d} \sqrt{h} ) ,
\ee 
where $\phi_d$ is the $d$ dimensional dilaton ($d=D-1$) and $h$ is the metric on the $d-1$ dimensional boundary. Written in this form it is manifest that the action is {\it independent} of $\Omega$ and it has the same value before and after the transformation. 
 In other words, the new action $\log Z$, as a function of $R,r,\mu_p,\mu_w,\phi_D$ where all the quantities are evaluated at infinity, has the same value as the original action $\log \tilde{Z}$, which is a function of $\tilde{R}, \tilde{\phi}_D$,\footnote{The dependence on $\tilde{r}$ is implicitly included in $\tilde{\phi}_D$.}
\begin{equation}\la{Transca}
\log Z(R,r,\mu_p,\mu_w,\phi_D) = \log \widetilde Z(\tilde R , \tilde \phi_D).
\end{equation}
The dependence of the action on the asymptotic value of the dilaton is particularly simple, so it is convenient to strip it out completely by defining
\begin{equation}\la{Zprime}
	e^{-2\phi_D} \log Z'(R,r,\mu_p,\mu_w) \equiv \log Z(R,r,\mu_p,\mu_w,\phi_D),\quad e^{-2\tilde{\phi}_D} \log \tilde{Z}' (\tilde{R}) \equiv \log  \widetilde Z(\tilde R , \tilde \phi_D),
\end{equation}
where the primed quantities no longer depend on the asymptotic value of the dilaton.
We can further relate $\phi_D$ and $\tilde{\phi}_D$ by utilizing the fact that $\phi_d$ is invariant under the transformation, which means
\be \la{Dilrel}
e^{ - 2 \tilde \phi_D}\tilde{R} = e^{ -2 \phi_D} R. 
\ee 
Combining (\ref{Transca}), (\ref{Zprime}), (\ref{Dilrel}), we have 
\be 
\log{ Z }'(R,r,\mu_p,\mu_w) = { R \over \tilde R }  \log {\widetilde Z }'(\tilde R).  \la{Partfinx}
\ee 
This is a key relation that relates the new action and the old action with the dilaton being taken care of, and we can now use it to relate the thermodynamic quantities. We first recall that with the chemical potentials defined in (\ref{ChemPot}), the action has the form
\begin{equation}\la{Grand}
	\log Z = e^{-2\phi_D} \log Z' =  S - 2\pi R E + 2\pi R \mu_p Q_p + 2\pi R \mu_w Q_w.
\end{equation}
So the various thermodynamic quantities can be computed as
\bea  \la{ThQu}
 2\pi  R Q_p &=& e^{ - 2 \phi_D} \partial_{\mu_p} \log Z' ~,~~~~~~~~ 2\pi  R Q_w = e^{ - 2 \phi_D} \partial_{\mu_w} \log Z'   , \\ \notag
 2\pi E &=&  e^{ - 2 \phi_D} \left( - \partial_R + { \mu_p \over R } \partial_{\mu_p} +{ \mu_w \over R } \partial_{\mu_w} \right)\log Z' , \\ \notag
S &=& e^{ - 2 \phi_D} ( 1- R\partial_R) \log Z' ,
\eea 
where the partial derivative with respect to $R$ are taken for fixed values of the chemical potentials $\mu_p$, $\mu_w$ (and also fixed $\phi_D$ and $r$).  With these definitions $Q_p$ and $Q_w$ are normalized so that they have integer eigenvalues.     After using \nref{Partfinx} we can express \nref{ThQu} in terms of the energy and entropy of the original solution. More explicitly, denote the original entropy and energy as $\tilde{S}$, $\tilde{E}$, and similarly strip out the dilaton dependence by defining
\be \la{tildep}
 e^{ - 2 \tilde \phi_D} \tilde S' \equiv \tilde{S} ~,~~~~~~~~~~~~~~~~
 e^{ - 2 \tilde \phi_D } \tilde E'\equiv \tilde{E} ,
\ee 
we then obtain 
\bea \notag
2\pi Q_p &=&- e^{ - 2 \phi_D}  { 1 \over \tilde R^2 }{ \partial \tilde R \over \partial \mu_p }  \tilde S' ~,~~~~~~~~~
2\pi Q_w =-e^{ - 2 \phi_D}   { 1 \over \tilde R^2 }{ \partial \tilde R \over \partial \mu_w } \tilde S' ~,
\\ \la{NewOldF}
S &=& e^{ -2 \phi_D}  {R^2 \over \tilde R^2}  { \partial \tilde R \over \partial R }\tilde S'  ~,
\\\notag
2\pi E &=& e^{ -  2 \phi_D} \left[ 2\pi \tilde E' + \left( -1  + {R \over \tilde R } { \partial \tilde R \over \partial R } - { \mu_p \over \tilde R} {\partial \tilde R \over \partial \mu_p} -{ \mu_w \over \tilde R} {\partial \tilde R \over \partial \mu_w}  \right) { \tilde S' \over  \tilde R }\right].
\eea 
We derive these formulas in detail in appendix  \ref{app:rel}. What's been done here is that we've expressed the thermodynamic quantities $S,E,Q_p,Q_w$ of the new solution in terms of $\tilde{S}',\tilde{E}'$ of the old one, up to relations between parameters $R,\mu_p,\mu_w$ and $\tilde{R}$ that will be worked out in sec. \ref{sec:SolnGeneBos} and \ref{sec:SolnGeneHet}.  

We emphasize that these formulas are valid both for the black hole and the \HP solution as well as any other interpolating solution we might find in the future. 

One interesting aspect is that we can take a limit of the chemical potentials  so that we approach the extremal limit where $E  - (Q_p/r \pm Q_w r)  $ becomes small. We will find that as we approach this limit the original radius $\tilde R$ becomes small and we should transition from the black hole to the \HP solution. So the black hole to string transition for the neutral black hole can be lifted into a similar transition as we approach extremality for a black hole with momentum and winding charges. We will see that the entropy has the expected extremal value based on the corresponding oscillating string (type II or heterotic). This agreement is not a miracle, it is related to the fact that the \HP solution has an entropy that matches that of a highly oscillating string, plus further corrections due to self attraction.  We will discuss more on the approach to extremality in sec. \ref{sec:Ext}.

 \subsection{Bosonic and type II strings} \la{sec:SolnGeneBos}
 
In the bosonic as well as the type II case, the winding mode has $\omega_t = \pm 1, n_t = \omega_y = n_y = 0$. Demanding (\ref{condition}), we find that the allowed transformations form a two dimensional Poincare algebra
\begin{equation}
	[\gamma_1 , \gamma_2 ] = 0, \quad [\gamma_3, \gamma_1 ] =  \gamma_2, \quad [\gamma_3 , \gamma_2] =  \gamma_1
\end{equation}
and explicitly we have
 \begin{equation}\label{gammaBos}
\begin{aligned}
    \gamma_1 =   \begin{pmatrix}
    0 & 1 & 0 & 1 \\
    0 & 0 & -1 & 0 \\
    0 & 0 & 0 &  0   \\
    0 & 0 & -1 & 0
    \end{pmatrix} , \quad \gamma_2 =   \begin{pmatrix}
    0 & -1 & 0 & 1 \\
    0 & 0 & -1 & 0 \\
    0 & 0 & 0 & 0 \\
    0 & 0 & 1 & 0
    \end{pmatrix} , \quad \gamma_3 =  \begin{pmatrix}
    0 & 0 & 0 & 0 \\
    0 & 1 & 0 & 0 \\
    0 & 0 & 0 & 0 \\
    0 & 0 & 0 & -1
    \end{pmatrix} .
\end{aligned}
\end{equation}
The generator $\gamma_3$ simply rescales the $y$ circle, so we are not interested in it. We can arrive at a generic charged solution via applying $\gamma_1$ and $\gamma_2$ consecutively. In other words, we choose $\Omega =\exp\left( \frac{v-u}{2}\gamma_2\right) \exp\left(\frac{u+v}{2} \gamma_1\right)$, where $u,v$ are two free parameters. We choose this particular parametrization just so that the formulas are simpler. The transformation (\ref{Mtransf}) relates the new solution to the old  one as
\begin{equation}
\begin{aligned} \la{NewPar}
   ds^2 &=  R^2 d t^2 +   r^2 ( d  y +   a d t)^2 , \quad B_{t y} = b, \\
   R^2 & = \frac{\tilde{r}^2 \tilde{R}^2}{(\tilde{r}^2 + u^2 \tilde{R}^2) (1 + v^2 \tilde{r}^2 \tilde{R}^2)},  \\
   r^2 & = \frac{\tilde{r}^2 +u^2 \tilde{R}^2}{ 1+ v^2 \tilde{r}^2  \tilde{R}^2}, \\
   a & = i {\mu_p  R} =- \frac{u \tilde{R}^2}{\tilde{r}^2 + u^2 \tilde{R}^2 },  \quad ~~~~~~~~~b = i \mu_w R = - \frac{ v  \tilde{r}^2 \tilde{R}^2}{1+ v^2 \tilde{r}^2 \tilde{R}^2}.
\end{aligned}
\end{equation}

The metric in \nref{NewPar} could be viewed as the metric in the time and internal circle direction for any radial position by thinking of $R$ and $r$ as functions of the radial coordinate, say $\rho$. 
If the seed solution is a black hole, then the original circle $\tilde R$ shrinks to zero in a smooth way at the horizon.  From the expression for $R$ in (\ref{NewPar}), we see that when $\tilde R \to 0$ we have that $R \sim 1\cdot \tilde R$ and this ensures that the new metric is smooth at the horizon  if and only if the old one is. 
A result $R\sim c\tilde R$ with a constant $c\not=1$ would map a smooth metric to a metric with a conical singularity.


From now on we will think of the values in the left hand side of \nref{NewPar} as the asymptotic values. 
It is possible to invert these expressions to find $\tilde R$ as a function of the un-tilde variables 
\be \la{Rtil}
 \tilde R = R \sqrt{ \left( 1- { r^2 \mu_p^2 } \right) \left( 1- {\mu_w^2 \over r^2  } \right) }
\ee 
and we can use this to calculate the derivatives 
\be 
{ \partial \tilde R \over \partial \mu_p } = - r  \tilde R   { r\mu_p \over (1 - r^2 \mu_p^2 )} ~,~~~~~~~
{ \partial \tilde R \over \partial \mu_w } = -{ \tilde R \over r }  { \mu_w/r  \over( 1- {\mu_w^2\over r^2} )} ~,~~~~~~~~~
 { \partial \tilde R \over \partial R} =  {\tilde R \over R } .
\ee 


We can then insert these derivatives into the general formulas 
\nref{NewOldF} 
to obtain a concrete expression for the thermodynamic quantities. 
In order to compare to previous literature, it is convenient to introduce the following variables 
 \begin{equation} \la{HyAng}
 \tanh \alpha =  \mu_p r , \quad \tanh\gamma = {\mu_w \over r },
\end{equation}
Then we can write 
\begin{equation} \la{GeneBos}
\begin{aligned}
	R & = \tilde{R} \cosh\alpha \cosh\gamma ,\\
	E &  = e^{ - 2 \phi_D} \left[ \tilde{E}' + \frac{\tilde{S}'}{2\pi \tilde{R}} \left(\sinh^2 \alpha + \sinh^2 \gamma\right)\right] ,\\
	2\pi Q_p & =  r e^{-2 \phi_D}\frac{\tilde{S}'}{ \tilde{R}}   \sinh \alpha \cosh\alpha ,~~~~~~~~
	2\pi Q_w = { 1 \over r} e^{-2 \phi_D}\frac{\tilde{S}'}{  \tilde{R}}  \sinh\gamma \cosh\gamma ,
	\\
	S & = e^{ -2 \phi_D} \tilde{S}' \cosh\alpha \cosh\gamma ,
	\\
	e^{- 2 \phi_D} &= { e^{ - 2 \tilde \phi_D} \over \cosh \alpha \cosh \gamma }  .
\end{aligned}
\end{equation}
We see that the relation between the original and the final dilaton is such that the initial and final entropies are actually equal, which is clear for the black hole case where the entropy is given by the area of the horizon.     $\tilde{S}'$ and $\tilde{E}'$ are defined in \nref{tildep}. 

These formulas are the main result of this subsection. They express the thermodynamic quantities of the charged solution in terms of the ones of the uncharged solution which we used as a seed for the procedure. We emphasize that they apply to both the black hole and the Horowitz-Polchinski solution. The difference between the two merely lies in the relation among $\tilde{E},\tilde{S},\tilde{R}$ of the seed solution. In addition, they should apply to any further interpolating solution. In particular, from the $\alpha'$ corrected thermodynamics of the neutral black hole at large radius, we get the the $\alpha'$ corrected thermodynamics of the charged black hole. We carry this out explicitly in appendix. \ref{app:alpha}.

\subsubsection{Describing the solution generating technique for a general seed CFT} 
\la{CFTarg}

So far, we have only considered the transformations in the asymptotic region. We will now sketch an argument saying that the transformation can be extended to the full conformal field theory. We perform a sequence of tranformations that, in the gravity limit, produces \nref{NewPar}, but which can be applied to an arbitrary seed CFT. 
Our assumption is that we start from a seed CFT with a $t$ shift symmetry. We will call it a $t$ ``rotation'' to emphasize that we do not have the winding symmetry. It acts as a rotation at the horizon of the Euclidean black hole. We only have a single non-holomorphic current associated to this $t$ rotation. 

Our starting theory is the seed CFT times a compact circle $\tilde{y}$ with radius $\tilde{r}$. In other words, the Lagrangian of the $\tilde{y}$ direction has the form $\frac{1}{4\pi}\tilde{r}^2 (\partial \tilde{y})^2$, with $\tilde{y}$ taking value in $[0,2\pi)$. As a starting step, we first decompactify the $\tilde{y}$ direction and let it take value in $(-\infty, \infty)$. The product theory now is the seed CFT times a non-compact line: CFT$_s \times R$. We now perform the following set of operations:
\begin{itemize}
	\item Quotient the theory by a shift $\tilde y \to \tilde y + 2\pi   $ combined with a shift $t\to t+ 2\pi u $.   By the shift in $t$ we mean that we act with the generator of the symmetry of the seed theory. We can think of this symmetry as the exponential of an infinitesimal transformation $U = e^{ i 2\pi Q}$, where $Q$ is a charge given in terms of a current $j_Q$ that is a linear combination of the current performing the shift in $\tilde y$ as well as the current generating the $t$-rotation symmetry of the seed theory. 
	For generic $u$, before we take the quotient, the one-parameter group generated by $Q$ is a copy of $R$.   But after taking the quotient, $Q$ generates
	a $U(1)$ symmetry group.  
	\item Now we perform a T-duality along the direction associated to $Q$ and decompactify the T-dual direction. These two steps can be achieved in one go by gauging the $Q$ symmetry with a compact gauge field $A$ and adding a term to the lagrangian of the form $ i   \hat y d A $, with a non-compact $\hat y$ \cite{Rocek:1991ps}. After this step the theory has a symmetry under $\hat y$ translations and an associated current.  
	\item We now quotient the theory by a $e^{ i 2\pi \hat Q}$ symmetry that acts as $\hat y \to \hat y + 2\pi$ and $t \to t + 2\pi v$. Again we have defined the infinitesimal generator $\hat Q$. 
	\item We now do an ordinary T-duality along the direction given by $\hat Q$. This is achieved by gauging the $\hat Q$ symmetry with a compact gauge field $\hat A$ and adding a term to the Lagrangian of the form $i \check y d \hat A$, with a compact $\check y$.
\end{itemize}

If we apply this sequence of steps to a seed theory that has a gravity description, then we obtain \nref{NewPar},
as we show more explicitly in appendix \ref{App:ExplicitGene}. 



\subsubsection{Generating the charged version of the \HP solution}

We now check that after applying (\ref{GeneBos}) to the Horowitz-Polchinski solution, we get the expected formulas for the entropy of a highly excited string with momentum and winding charges. It is convenient to introduce the left/right moving charges
\begin{equation}\label{QLR}
	Q_{L,R}  = \frac{Q_p}{r} \pm r Q_w  =  e^{-2 \phi_D}\frac{\tilde{S}'}{2\pi \tilde{R}} \left( \sinh \alpha \cosh\alpha \pm \sinh\gamma \cosh\gamma\right) .
\end{equation}
For the neutral Horowitz-Polchinski solution, to the leading order in $\tilde{R} - R_H$, we have $\tilde{S}' = 2\pi R_H \tilde{E}'$, so we can recast the energy and entropy in (\ref{GeneBos}) as  
\begin{equation} \label{recast}
\begin{aligned}
	E & = e^{-2\phi_D} \frac{\tilde{S}'}{2\pi R_H} \left(1 + \sinh^2 \alpha + \sinh^2 \gamma\right), \\
 S & = e^{-2 \phi_D} \tilde{S}' \cosh \alpha \cosh \gamma.
\end{aligned} 
\end{equation}
Combined with (\ref{QLR}) and using properties of the hyperbolic functions, we find that to the leading order in $\tilde{R} - R_H$
\begin{equation}\la{EntcharBos}
  S = 2\pi R_H \half \left(\sqrt{E^2 - Q_L^2} + \sqrt{E^2 - Q_R^2} \right).
\end{equation}
Using $E^2 - Q_{L,R}^2 = 4 N_{L,R}/\alpha'$, it is straightforward to verify that (\ref{EntcharBos}) agrees with the Cardy formula $S = 2\pi \sqrt{cN/6}$, in both the Bosonic and Type II string theory, after using the values of $R_H$ in \nref{MassII}. Of course, for the \HP solution we have also worked out the correction to the $\tilde{S}' - \tilde{M}'$ relation in some cases, and we can combine that with (\ref{GeneBos}) to compute the corrections for the charged solutions.

We should remark that in the extremal limit, where $E \sim { Q_p/r } \pm Q_w r $, and $\alpha, \gamma $ become large, we can write the entropy as 
\be 
S \sim  2\pi { \tilde R \over l_s } \sqrt{|Q_p Q_w|} \la{Exten}
\ee 
where we restored the string length. 
 $Q_p$ and $Q_w$ should be viewed as the values of the quantized charges and are integers.   Then the value of the entropy is completely determined by the value of $\tilde R$ as we approach the extremal limit. If we were to (incorrectly) use just the gravity approximation, then we see that $\tilde R \to 0$ and we get zero entropy. On the other hand, if we note that in the seed solution the value of $\tilde{R}$ should approach $R_H$, then we get the expected answer for a string with momentum and winding.

\subsubsection{Explicit form of the charged version of the \HP solution} 
\la{FullChHP}

In this section we present the form of the charged version of the \HP solution obtained via the solution generating procedure.  
 The initial solution has a non-trivial dilaton and $g_{tt}$ component of the metric specified by a single function $  \varphi$. These are given in (\ref{HPasymp})   
  $g_{tt} \approx  \left(1 + 2   \varphi \right) $ and also $2 \tilde \phi_D = 2 \tilde \phi_{D , \infty} +    \varphi$ from (\ref{HPDil}). 
  We then find that the charged solution has the metric given by (\ref{FinSol}) which we can approximate as 
 \begin{equation} \la{SCM}
\begin{aligned}
& ds^2 =   d\tau^2 { 1 + 2  \varphi  \over (1 - 2   \varphi \sinh^2 \alpha   )(1 - 2   \varphi\sinh^2 \gamma   ) } +   r^2 {  1 - 2  \varphi \sinh^2 \alpha  \over 1- 2  \varphi \sinh^2 \gamma    } \left[ d y -  {i\tanh \alpha \over r } { (1 + 2   \varphi ) d\tau \over    1 -2  \varphi \sinh^2 \alpha}  \right]^2  +  d \vec x^2 , \\
&  B_{y \tau } = - i \tanh \gamma r {   1 + 2   \varphi \over 1 -2    \varphi \sinh^2 \gamma} ,\\
&  e^{ - 2 \phi_D} = e^{ - 2   {\phi}_{\infty, \, D} }  { ( 1 -    \varphi) }  \sqrt{( 1- 2   \varphi \sinh^2 \alpha  )  (1 - 2   \varphi \sinh^2 \gamma  )} ,
\end{aligned}
\end{equation}    
where in this formula we chose the period of the Euclidean time direction as $\tau \sim \tau + 2 \pi R $, which is different from the way we were normalizing it elsewhere. 
Note that $\chi(\vec x)$ remains what it was in the original seed solution. 
The expression (\ref{SCM}) is valid to all orders in $\varphi \sinh^2 \alpha$ and $\varphi \sinh^2 \gamma$, but to  first order in $\varphi$. More precisely there are corrections of the form $ (1+ {\rm (constant)}  \varphi^2 \sinh^2 \alpha )$ which we have not computed. 
Note that the factors of $i$ disappear if we go to the Lorentzian signature. These metrics have no horizon. When we talk about the Lorenzian signature solution we are imagining computing the Euclidean stress tensor from the winding mode $\chi$ and then continuing this stress tensor to Lorentzian signature, where we interpret it as coming from an oscillating string.  Note that we have $\varphi< 0$ in (\ref{SCM}).

\subsection{Heterotic string} \la{sec:SolnGeneHet}

 In the Heterotic case, the winding mode carries winding and momentum $\omega_t = \pm 1, n_t = \pm \frac{1}{2} , \omega_y = n_y = 0$. The transformations obeying (\ref{condition})   can be generated via three generators forming an $SL(2)$ algebra
 \begin{equation}
 	 [\gamma_1 , \gamma_2] = \gamma_3 , \quad [\gamma_3 , \gamma_1 ] = 2\gamma_1 ,  \quad [\gamma_3 , \gamma_2 ] = -2 \gamma_2 ,
 \end{equation}
and explicitly they are
 \begin{equation}\label{gammaHet}
\begin{aligned}
    \gamma_1 =  \begin{pmatrix}
    0 & 0 & 0 & \sqrt{2} \\
    \frac{1}{\sqrt{2}} & 0 & -\sqrt{2} & 0 \\
    0 & 0 & 0 & -\frac{1}{\sqrt{2}} \\
    0 & 0 & 0 & 0
    \end{pmatrix} , \quad \gamma_2 =  \begin{pmatrix}
    0 & \sqrt{2} & 0 & 0 \\
    0 & 0 & 0 & 0 \\
    0 & - \frac{1}{\sqrt{2}} & 0 & 0 \\
    \frac{1}{\sqrt{2}} & 0 & -\sqrt{2} & 0
    \end{pmatrix} , \quad \gamma_3 =  \begin{pmatrix}
    0 & 0 & 0 & 0 \\
    0 & 2 & 0 & 0 \\
    0 & 0 & 0 & 0 \\
    0 & 0 & 0 & -2
    \end{pmatrix} 
\end{aligned}
\end{equation}
The generator $\gamma_3$ simply rescales the $y$ circle, so we are not interested in it. We can arrive at a generic charged solution via applying $\gamma_1$ and $\gamma_2$ consecutively. We choose the transformation $\Omega = \exp \left[ \frac{\sqrt{2} u}{2 - uv} \gamma_2 \right]\exp \left[ \frac{v}{\sqrt{2}} \gamma_1 \right]$, with $u,v$ being two free parameters. Again, the particular parametrization is chosen only to make the formulas simpler. Using this transformation, (\ref{Mtransf}) relates the new solution to the old solution as
\begin{equation}
\begin{aligned} \la{HetVa}
   ds^2 &=  R^2 d t^2 +   r^2 ( d  y +   a d t)^2 , \quad B_{ty} = b, \\
  R^2 & = \left(1 - \frac{uv}{2}\right)^2 \frac{\tilde{r}^2 \tilde{R}^2}{(\tilde{r}^2 + u^2 \tilde{R}^2) (1 + v^2 \tilde{r}^2 \tilde{R}^2)} ,  \\
   r^2 & = \frac{1}{\left(1 - \frac{uv}{2}\right)^2 }\frac{\tilde{r}^2 +u^2 \tilde{R}^2}{ 1+ v^2 \tilde{r}^2  \tilde{R}^2} , \\
   a & = i {   \mu_p R} =  - \left(1 - \frac{uv}{2}\right) \frac{u \tilde{R}^2 + \frac{v}{2} \tilde{r}^2 }{ \tilde{r}^2 + u^2 \tilde{R}^2 },  \quad b = i   \mu_w R=  - \frac{1}{1 - \frac{uv}{2}}\frac{ \frac{u}{2} + v  \tilde{r}^2 \tilde{R}^2}{1+ v^2  \tilde{r}^2 \tilde{R}^2} .
\end{aligned}	
\end{equation}
In (\ref{HetVa}) we can view $R$ and $r$ as well as the corresponding tilded quantities as functions of the radial coordinate $\rho$, though later we will use them mainly as asymptotic values.  For the black hole case $\tilde R$ shrinks smoothly to zero at the horizon. As opposed to what happened in the type II case, we do not
simply have $R\to 1\cdot \tilde R$ for small $\tilde R$, because of the extra factor going like $(1 - uv/2)$ in the expression for $R$ in \nref{HetVa}. This term is very small in most of the solution when the seed solution is weakly curved so that $\tilde R$ is large everywhere except at the horizon. However this extra factor of $(1-uv/2)$ becomes important near the horizon. (Here we imagine that   $u \tilde R/\tilde r$ and $v \tilde R \tilde r$  are at most of order one at infinity.) 
  Therefore the transformation maps a solution with a smooth horizon to one with a conical singularity.
We think that the transformation laws need to be changed near the horizon since \nref{HetVa} are the transformation laws when we have small field gradients, as we have in the asymptotic region. 
 See \cite{Bergshoeff:1995cg,Kaloper:1997ux,Bedoya:2014pma}, and references therein,  for $\alpha'$ corrections that might be relevant. We expect that after including all the relevant $\alpha'$ corrections the metric will turn out to be smooth at the horizon.  Note that in the type II case, we did not need any $\alpha'$ correction to the T-duality in order to preserve the smoothness at the horizon, but it is still possible that there are other type of $\alpha'$ corrections in that case too. 

From now on, we concentrate on the asymptotic region, where \nref{HetVa} is applicable. This is all that is necessary in order to compute the thermodynamic quantities. 
 For computing thermodynamic quantities we need partial derivatives of $\tilde R$ with respect to $R, \mu_p, \mu_w$ keeping $r$ fixed.  
We can easily compute the matrix of the derivatives of the untilde variables with respect to the tilde ones. We invert this matrix to find\footnote{The final expression is simpler than the intermediate steps, so there might be a more direct way to obtain this.}
 \be \la{DerHet}
 { \partial \tilde R \over \partial R } = {\tilde R \over R} { (1+ { uv\over 2 }) \over ( 1- { uv  \over 2   }) } ~,~~~~~~ {\partial \tilde R \over \partial \mu_p } = - r { \tilde R   } { \tilde R u/\tilde r \over 1 + {u^2 \tilde R^2 \over \tilde r^2}} ~,~~~~~~~{\partial \tilde R \over \partial \mu_w } = - { \tilde R  \over  r } {  v\tilde r \tilde R \over 1 +   v^2\tilde r^2 \tilde R^2  }.
 \ee 
 Again, it is convenient to introduce the angles  
  \begin{equation} \la{HyAngse}
 \tanh \alpha =   i { u \tilde R \over \tilde r }  , \quad ~~~~~~~~ \tanh\gamma = i { v \tilde R \tilde r }  ,
\end{equation}
Now, using the general formulas \nref{NewOldF}, the derivatives 
\nref{DerHet}, as well as \nref{HyAngse} we obtain 
\begin{equation}\label{GeneHet}
\begin{aligned}
	R & = \tilde{R} \cosh\alpha \cosh\gamma \left( 1 + { \alpha'\tanh \alpha \tanh \gamma  \over 2 \tilde R^2 } \right),\\
	E &  =
	 e^{ - 2 \phi_D} \left[ \tilde{E}' + \frac{\tilde{S}'}{2\pi \tilde{R}} \left(\sinh^2 \alpha + \sinh^2 \gamma\right)\right] ,
	\\
	2\pi Q_p & = 
	 r e^{-2 \phi_D}\frac{\tilde{S}' }{ \tilde{R}}   \sinh \alpha \cosh\alpha 
	,~~~~~~~~
	2\pi Q_w =
	 { 1 \over r} e^{-2 \phi_D}\frac{\tilde{S}'}{  \tilde{R}}  \sinh\gamma \cosh\gamma ,
	\\
	S & = e^{ -2 \phi_D} \tilde{S}' \cosh\alpha \cosh\gamma\left( 1 - {\alpha' \tanh \alpha \tanh \gamma  \over 2 \tilde R^2 } \right),
	\\
	e^{- 2 \phi_D} &= { e^{ - 2 \tilde \phi_D} \over \cosh \alpha \cosh \gamma \left( 1 + { \alpha' \tanh \alpha \tanh \gamma  \over 2 \tilde R^2 } \right)}. 
\end{aligned}
\end{equation}
 Notice that the answer agrees with \nref{GeneBos} up to the terms that involve $\tanh \alpha \tanh \gamma { \alpha' \over \tilde R^2 }$, where we restored the $\alpha'$ dependence in the formulas to highlight that these terms can be viewed as $\alpha'$ corrections.   Note that we claim that \nref{GeneHet} is exact in $\alpha'$ so that these are all the $\alpha'$ corrections that appear explicitly through the solution generating process. Of course, there are other $\alpha'$ corrections which come in implicitly in (\ref{GeneHet}) through the expressions for $\tilde E $ and $\tilde S$ of the seed solution.

\subsubsection{The charged version of the heterotic \HP solution }\la{ChHet}

In this case, we could also verify that we get the expected entropy for a highly excited string carrying charges by applying (\ref{GeneHet}) to the Horowitz-Polchinski solution. For the seed solution, to the leading order in $\tilde{R} - R_H$ we have $\tilde{S}' = 2\pi R_H \tilde{E}'$, so we can write the energy and entropy in (\ref{GeneHet}) as
\begin{equation}
\begin{aligned}
 E & = e^{-2 \phi_D} \frac{\tilde{S}'}{2\pi R_H} \left[1 + \sinh^2 \alpha + \sinh^2 \gamma \right],\\	
S & = e^{-2\phi_D} \tilde{S}' \cosh \alpha \cosh \gamma \left(1 - \frac{\alpha' \tanh \alpha \tanh \gamma}{2 R_H^2}\right).
\end{aligned}
\end{equation}
Using the definition of $Q_{L,R}$ in (\ref{QLR}) and the specific value $R_H = \left(1+ 1/\sqrt{2} \right)l_s$ for heterotic string theory, one can express the entropy as
\begin{equation}\la{EntHet}
	S = 2\pi l_s \left( \sqrt{E^2 - Q_L^2} + \frac{1}{\sqrt{2}} \sqrt{E^2 - Q_R^2} \right).
\end{equation}
It is straightforward to check that (\ref{EntHet}) agrees with the Cardy formula. The fact that it is asymmetric between the left and right moving parts is a reflection of the difference in the left and right moving central charges. We obtained this asymmetry because the entropy in \nref{GeneHet} contains a term that depends on the relative sign between $\alpha$ and $\gamma$. This term could be viewed as an $\alpha'/\tilde R^2$ correction for large $\tilde R$. But as $\tilde{R}$ approach $R_H$ we get an order one correction as in \nref{EntHet}. In this case, we can also combine the corrections we computed for the neutral Horowitz-Polchinski solution and (\ref{GeneHet}) to get the corrections in the charged cases. 

Note that this asymmetry implies that in the two possible extremal limits $E\to Q_L$ or $E\to Q_R$ we get entropies 
\bea 
S &=& 2\pi 2 \sqrt{|Q_p Q_w|} ~,~~~{\rm for }~~E\sim Q_R, \la{BPSex}
\\ \la{nonBPSex}
S &=& 2\pi \sqrt{2} \sqrt{|Q_p Q_w|}  ~,~~~{\rm for }~~E\sim Q_L .
\eea
Recall that $Q_p$ and $Q_w$ are normalized so they have integer eigenvalues.  
These are, of course just the values we expect for a single string with momentum and winding, though we have obtained them here from the \HP solution. Note that the first limit \nref{BPSex} corresponds to a BPS limit. While the second one is not BPS, it is extremal,  in the sense of taking a minimal value of $E-Q_L$ within the family of solutions we considered. In this second case, there are other configurations with the same charges which are lighter. We could call these ``super extremal'' configurations, and they have been intensively discussed in the context of the weak gravity conjecture. For a recent paper see \cite{Alim:2021vhs}. 
Namely, in the heterotic string we can have states carrying only $Q_L$ with mass $M^2 = Q^2_L -1$. For a single string this is a small correction, but we could have a multiparticle state where each particle has a relatively low value of $Q_L$ so that the ratio of the total quantities, $M^{\rm tot}/Q^{\rm tot }_{L}$,  becomes significantly smaller than one.   In other words,  
the solution-generating procedure generates a particular solution, but it does not need to produce the lowest energy state.

\subsection{The approach to extremality} \la{sec:Ext}

The study of black holes with Kaluza-Klein momentum and fundamental string winding charges was historically interesting \cite{Sen:1995in}. It was a first example of some connection between the entropy of a black hole and that of the extremal strings with momentum and winding. The connection was not perfect because the   entropy of the naive extremal solution is zero. Therefore it was necessary to invoke some string scale corrections to get a finite answer. This also prevented a precise matching of the numerical coefficient \cite{Sen:1995in}. Through the discussion in this section we see that the approach to extremality for the charged black hole is essentially the same as the behavior of uncharged black holes as the size of the black hole becomes small. So the assumptions made in \cite{Sen:1995in} are related to the idea that the solution is generated from an uncharged Schwarzschild black hole of string size and string scale temperature, or that the string size neutral black hole turns into strings \cite{Horowitz:1996nw}. In other words, the assumptions of \cite{Sen:1995in} and \cite{Horowitz:1996nw} are equivalent, once we take into account the relations implied by the solution generation procedure. 

If we knew the precise way in which an uncharged black hole behaves as we decrease its mass, then we would be able to say what happens with a black hole carrying momentum and winding charges as we approach extremality.\footnote{There is a caveat to this statement: if the solution develops a strong coupling region, as it might do at a special point in the type II case, in that case our solution generating technique does not apply.}
 When thinking about the approach to extremality, it is convenient to calculate (both for type II and heterotic)
\be \la{approachext}
  E - { Q_p \over r } - {Q_w r }   =  e^{ - 2 \phi_D} \left[   \tilde{E}' - \frac{\tilde{S}'}{2\pi \tilde{R}} + 
\half { \tilde S' \over 2 \pi \tilde R} ( e^{ - 2 \alpha } + e^{ - 2 \gamma } ) \right] ~,~~~~~~\alpha, \gamma \gg 1 .
  \ee
Notice that in the large $\alpha$ and $\gamma$ limit, the original string coupling $\tilde g$ is very small, see \nref{GeneBos} and \nref{GeneHet}. This ensures some reasonable range of validity for the seed solution in the \HP case. 
Also note that  as $\tilde E' -{\tilde S' \over 2\pi \tilde R } \to  0$, the energy above extremality goes to zero. This is what happens with seed solutions where the energy and the entropy become proportional to each other, as is the case for the  
\HP solution as we approach the free string limit. We will comment on another context where this happens in section \ref{FiveBranes}.

\begin{figure}[t]
    \begin{center}
    \includegraphics[scale=.2]{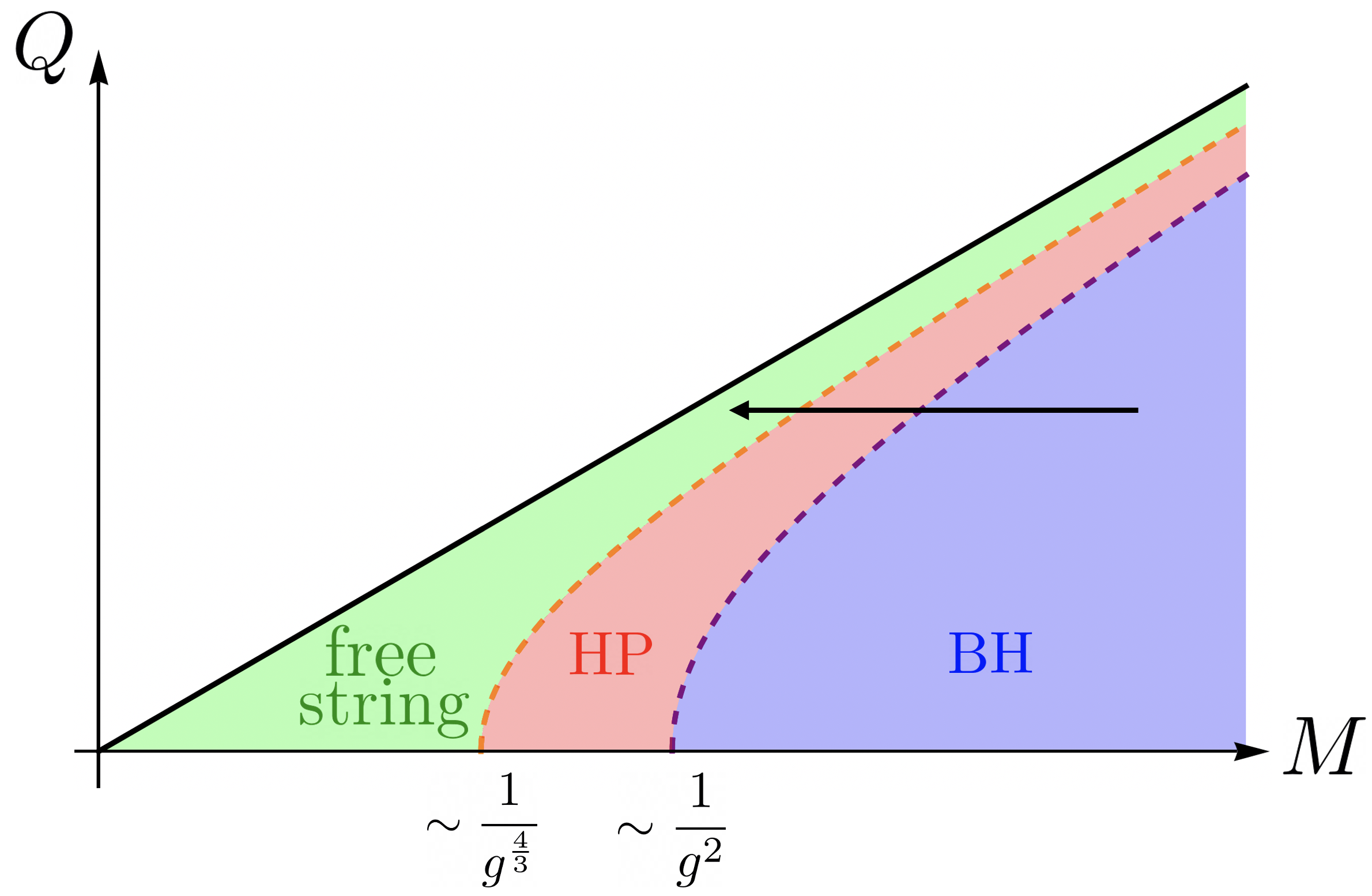}
    \end{center}
    \caption{The approach to extremality for a charged black hole with momentum and winding charges follows the behavior of the seed uncharged black hole as we decrease its mass. Here we display the behavior for $d=3$, or $D=4$.  In other dimensions, such as $d=5$ we expect to have a first order transition to the free string picture as we approach extremality, see figure \ref{fig:phase5d}. 
}
   \label{fig:extremal}
\end{figure}

For example, in $d=3$ dimensions (or $D=4$) we expect a diagram as in figure \ref{fig:extremal} for the behavior of a charged black hole as we approach extremality. As we decrease the mass for a given charge we expect a series of transitions that come from corresponding transitions in the seed solution. In particular, we expect to go from a black hole solution to a charged version of the \HP solution. This solution has no horizon and is describing highly excited strings. Notice that we are near extremality, and the solution can be viewed as a kind of near extremal ``fuzzball'', it has no horizon and the microstates are explicit. But it is not trustworthy in the parameter regime that the black hole is trustworthy.  

 In particular, note that the intrinsic size of the seed solution has a non-monotonic behavior as we decrease the mass, see figure  \ref{fig:phase3d} (c). This translates also to a non-monotonic behavior for a certain notion of ``size'' of the charged solution. It has been speculated that the naive solution is corrected at a string frame distance of order $l_s$ from the naive singularity\cite{Sen:1995in}. Here we are saying that, close to extremality,  the correction happens at larger sizes, since the radius of the seed \HP solution is larger than the string scale. The size increases as we approach extremality along the black arrow while inside the \HP region in figure \ref{fig:extremal}.

  As we further approach extremality   the seed solution  undergoes a transition to a free string. For the uncharged solution, this transition was signaled by the appearance of large fluctations, as mentioned around \nref{ClassVal}. Those fluctuations involve modes whose action is  invariant under the action of the solution generating procedure. So this implies that the charged \HP solution ceases to be valid when $\tilde g$ and $\tilde R - R_H$ gets to the lower bound in \nref{ClassVal}. For $D=4$ this translates into 
\be \la{Trva} 
\tilde E \sim { 1 \over {\tilde g}^{4/3} } ~,~~~~~~~ ({\rm recall~ that}~~~\alpha'=1).
\ee  
  Of course, but using \nref{GeneBos} or \nref{GeneHet} we can express this condition in terms of the quantities, $E,~Q,~g$ of the charged solution. 
  In principle, it might also break down earlier, due to other modes, modes whose action depends on the parameters of the solution generating transformation. We have not analyzed the action of all possible modes\footnote{ Recall that the solution generating procedure leaves the spacetime action invariant, when we consider modes with zero momentum and winding in the internal circles. So the only issue could come from the modes with non-zero momentum in the time or internal circle directions.}, but we will see we get a nice picture assuming that the transition is at \nref{Trva}.  Now, at \nref{Trva} the neutral seed solution is expected to transition into a free string description. At this point that string has a size of order 
  \be \la{Ltra}
  \ell \sim   {\tilde N}^{1/4} \sim \sqrt{\tilde E } \sim {\tilde g}^{-2/3} ~,~~~~~~ \tilde N \equiv { \tilde{E}^2 \over 4 } 
  \ee 
   where we defined and oscillator number and we used the transition value for $\tilde E$ \nref{Trva} as well as \nref{SizeGen}.   
   In terms of the parameters of the charged solution \nref{Ltra} becomes  
   \be \la{SizeBPS}
   \ell \sim ( Q_p Q_w)^{1/4} l_s 
   \ee 
   where we used the expressions for the charges and the dilaton in \nref{GeneBos}, \nref{GeneHet}, as well as $\tilde{S}'/\tilde{R} \sim \tilde{E}' \sim \tilde{g}^2 \tilde{E}$.     This is the size at the point where we transition to a free string. It is the size we would expect for an oscillating BPS string. In such a case, the oscillators are excited at a level of order $N_{\rm eff} \sim Q_pQ_w$ and using the usual arguments we are also led to \nref{SizeBPS} from a free string point of view. In other words, once we get to this transition point we expect that the size continues to stay at \nref{SizeBPS} as we continue to approach the BPS limit.
 We should emphasize that we cannot apply the solution generating transformation to the free string phase since it has large quantum fluctuations. In fact, for the charged solution we expect that the free string description continues to have $N_{\rm eff} \sim Q_pQ_w$ and the size \nref{SizeBPS} as we approach extremality. On the other hand, if we incorrectly apply the solution generation to free string, we would have expected that the size goes to zero as $\tilde E \to 0$ for the uncharged free string. 
 
The conclusion is that we have presented a more accurate picture of the approach to extremality for these charged black holes in $D=4$. Note that even if we cannot say exactly what happens at the transition regions in the figure \ref{fig:extremal}, we can still give a picture well within each region. 

Note that the size we are talking about in \nref{Ltra} is the size of the seed solution. This is not the same as the gravitational radius of the final charged solution, defined as the position where the final metric differs by an order one amount from the Minkowski metric. The latter  contains an additional factor of $\sinh^2 \alpha$ or $\sinh^2 \gamma$ (see e.g. \nref{SCM}). In  the final charged solution,  the size \nref{Ltra} tells us the string frame radial distance when  the solution differs from the naive singular black hole solution, see (\ref{BHcharged}). 

In \cite{Dabholkar:2004yr} it was observed that, by including certain higher curvature corrections, as in \cite{Ooguri:2004zv}, the entropy of the $D=4$ BPS black hole   matches precisely that of the oscillating strings in the heterotic string case, see also \cite{Sen:2004dp}. Given that we expect that the proper description near extremality should be closer to that of the \HP solution, we could view this agreement as some evidence in favor of a continuous transition between the black hole and the \HP solution in the heterotic case. Of course, it is also possible that there is a  phase transition but that the extremal entropy is still preserved for some reason. From our point of view this is a bit magical since the solutions in \cite{Dabholkar:2004yr,Sen:2004dp} do not seem to be the right ones (they are black holes with higher curvature corrections). 

 \subsection{A comment on black holes with non-zero horizon area at extremality} 
 \la{FiveBranes}
 
As a side remark, we note that if we start from the solution for a near extremal five brane, we get a black hole with non-zero horizon area. This is of course a well known way to generate the solution. 
Here we simply want to emphasize the fact that the reason we get a non-zero entropy is connected to the existence of a limiting temperature for the near extremal fivebrane. This property is formally similar to the limiting temperature that we find in the \HP solution, but for different reasons. 

As we said, a near extremal fivebrane   approaches a constant temperature  set by 
$\tilde R \sim l_s \sqrt{N_5}$, for $N_5 \gg 1$,   see e.g. \cite{Maldacena:1997cg}. In this situation, we find that 
\be  \la{NeF}
2 \pi \tilde E=  2\pi \tilde E_e + { \tilde S \over \tilde R } ~,~~~~~~~~~ \tilde R \sim l_s \sqrt{N_5} ,
\ee 
 where $\tilde E_e$ is the energy of the extremal five branes and $\tilde S$ is the entropy. 
 
 We can then see that if we apply the formulas \nref{GeneBos} to this seed solution, we get a solution whose entropy, in the extremal limit is 
 \be 
 S = \tilde R \sqrt{ 2\pi Q_p 2 \pi Q_w } = 2 \pi \sqrt{N_5 Q_p Q_w } .
 \ee 
 This is obtained from a formula as in \nref{Exten} where the
  term involving $\sqrt{N_5}$ comes from  the value of $\tilde R$ of the seed solution in \nref{NeF}. So the formal reason we get a non-zero entropy in the extremal limit is again just that there is a limiting temperature for the seed solution. 
 
 In contrast with the \HP solution, the near extremal fivebrane entropy cannot be given an interpretation in terms of a gas of strings or anything with a direct microstate interpretation.   It has been speculated that a gas of strings in the linear dilaton background could describe such a black hole, see \cite{Jafferis:2021ywg}. Unfortunately in such a picture the strings appear to be strongly coupled. What we are pointing out is that an explicit description of the microstates for the near extremal fivebrane would directly translate into one for a Reissner-Nordstr\"{o}m black hole with non-zero entropy at extremality. 
 

\section{An open string analog of the two solutions} 
\la{OpenString}

In this section we discuss an open string analog of the two solutions, namely the \HP and black hole solutions. 

\begin{figure}[h]
\begin{center}
\includegraphics[scale=.25]{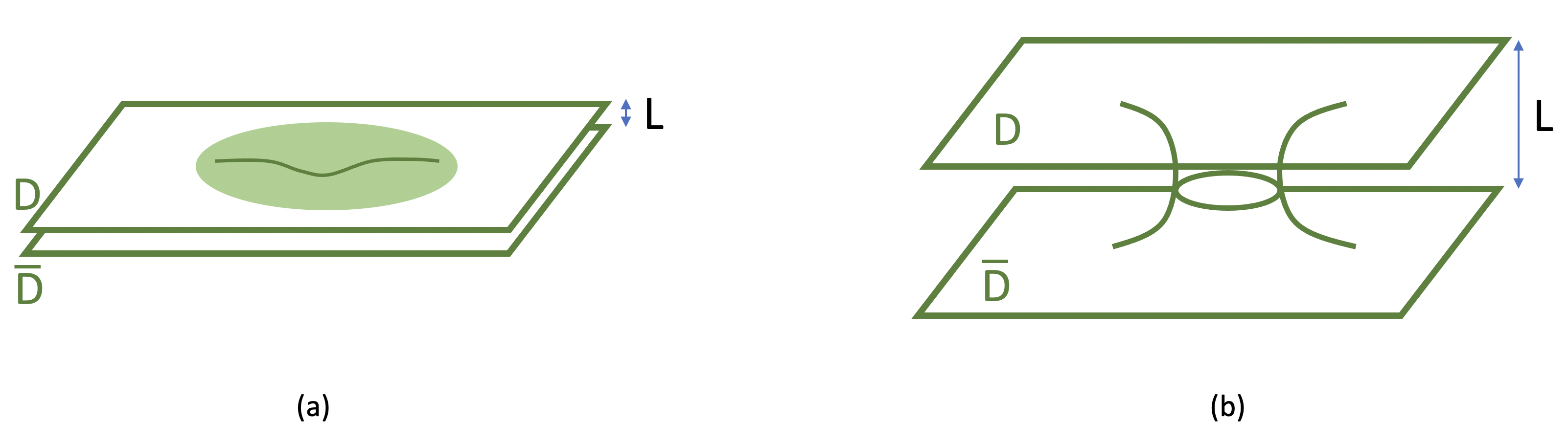}
\caption{We consider a brane and an anti-brane separated by a distance $L$. (a) When $L - L_c \ll l_s$, we have an analogue of the Horowitz-Polchinski solution, where we have a condensate of open string mode and the two branes are deformed and get a bit closer in the middle. (b) When $L\gg l_s$ we have a solution with the two branes are connected.   }
\label{BraneAntiBrane}
\end{center}
\end{figure}

We   start from a D-brane anti-D-Brane pair separated by some distance $L$. This system has an open string mode  with mass 
\be 
m^2 = { L^2 \over ( 2\pi \alpha')^2 } - { 1 \over 2 \alpha'} 
\ee 
which becomes  tachyonic at a small enough $L$, $L < L_c = \pi \sqrt{2} l_s$. This critical value of $L$ is somewhat analogous to the Hagedorn inverse temperature for the closed string case.  When $0<L-L_c \ll l_s$ this mode is light and has a positive mass squared. So we can consider an open string effective action of the form 
\be \la{actopen}
 I = { T L_c^2 \over 2 } \int d^d x\,\left[ (\nabla \varphi)^2 + |\nabla \chi|^2 +  \left( m^2_\infty + { \kappa \over \alpha'}  \varphi \right) |\chi|^2  \right]
\ee 
where $\chi$ is the open string mode and $\varphi$ is related to the distance between the brane and the anti-brane, $L = L_c (1 + \varphi) $. We have truncated the action to include only the light fields that are excited in the solution. 
Since the action has the same form as \nref{Action} \nref{massphi} we will have the same solutions, but with a new interpretation. These solutions exist in the same range of dimensions, $3 \leq d \leq 5$ as the \HP solution, and they  represent a configuration with a localized lump of open string tachyon condensate.\footnote{Brane antibrane annihilation has been discussed from the point of view of the non-abelian Dirac-Born-Infeld action in \cite{Garousi:2005zb}. The degree of freedom corresponding to the transverse motion of the branes was not taken into account in the analysis, leading to different conclusions.}

In addition, when $L \gg l_s$ there is another solution where the brane and anti-brane are connected as in figure \ref{BraneAntiBrane}, for more details see \cite{Callan:1997kz}. The scale size of the solution is of order $L$ and we can trust the solution when $L\gg l_s$. This is somewhat analogous to the black hole solution. 

Both of these solutions can be viewed as classical open string solutions, with a positive euclidean action of order $1/g_s$. As is well known, the brane and anti-brane also attract each other after we consider the one loop open string effects, which are of order one in the $g_s$ expansion \cite{Polchinski:1995mt}. We will be ignoring this effect and we will concentrate on the classical solutions. In particular, we can ask whether the two solutions are continuously connected as we vary $L$. 

The brane/anti-brane system has two separate $U(1)$ gauge symmetries, one $U(1)$ from the brane and the other from the anti-brane. Both of the solutions spontaneously break a linear combination of these two gauge symmetries. 

In principle, we could think of both solutions as providing a bounce solution for the annihilation of the brane and the antibrane. As we mentioned, a more dominant process would be one where the brane and the antibrane just move and get closer to each other. Nevertheless, we could imagine small variations of this setup where we hold the branes at least asymptotically in some fashion, perhaps embedding the whole setup in $AdS$, etc, so that the tunneling process is the dominant decay mode. 

In this case the problem reduces to analyzing the space of conformal invariant boundary conditions. The bulk CFT can be taken to be that of free fields. But the boundary will contain   interactions. The question is whether the two possibilities are continuously connected as we vary $L$. 

\subsection{Boundary linear sigma model}

As in the case of closed strings, we can get some insight by considering a linear sigma model construction. 
  We start with a free bulk theory parametrized by the target space coordinates $(\vec Y, X)$ living in $R^{d+1}$.  Note that now there is a single $X$ coordinate.    We consider Neumann boundary conditions for all coordinates and add a boundary action of the form \cite{Harvey:2000na,Hori:2000ic,Kraus:2000nj,Takayanagi:2000rz} 
\be \la{BdyInt}
 I_{\rm Bdy} =   \int d\tau d\theta \left( i \Lambda D \Lambda + \Lambda W(\XX,\YY) \right)
 \ee  
 where $\Lambda = \eta + \theta h $ is a  fermionic supermultiplet  living at the boundary. It contains a Majorana fermion $\eta$ as well as an auxiliary field $h$.  Here $\theta = \theta_+ =\xi \theta_-$ is the superspace coordinate at the boundary,  and $\xi =\pm 1$ corresponding to the two choices we sum over when implementing the GSO projection. Classically integrating out the auxiliary field, we get a boundary potential of the form $W^2$. Then the fields become localized to the region where $W =0$. For a simple brane we can take $W \sim X$, which would then flow in the IR to a brane at $X =0$. This is related to the idea that we get a D-brane from tachyon condensation on an unstable brane with one more dimension \cite{Sen:1999mg}.
 
 A brane anti-brane pair can be obtained from the function
 \be 
 W = X^2 - b 
 \ee 
 which flows in the IR to    a brane at $X = \sqrt{b}$ and an antibrane at $X = - \sqrt{b} $. 

If we want to describe the boundary CFT in question, we could start from a more general boundary interaction involving  
\be \la{WBdy}
W = ( X^2 -b )(\vec Y^2 + a ) + c 
\ee 
The discussion is now similar to that in section \nref{heterotic}. Depending on the value of $c$ we have a solution with two disconnected surfaces or a single connected surface. There is no new branch opening up at the critical value of $c$. Therefore one would expect that the two should be continuously connected. 

This is also supported by a computation of the supersymmetric index. Here we consider the supersymmetric index for the two dimensional theory on the strip with a boundary interaction \nref{BdyInt} given by \nref{WBdy} and choosing $\xi$ on the left and right boundary such that we are in the Ramond sector. When we are at large volume and the brane has a simple geometric description, the index is just the index of the Dirac operator, since the Ramond ground states are fermions living on the brane, with fermionic indices both in the tangent and normal directions to the brane. For the configurations that we have, which are parity symmetric, we find that the index is zero in both cases.

\section{Conclusions and discussion}\la{concl}

In this paper we have analyzed aspects of black holes in the stringy regime. We have studied the possible connection between black holes and the self gravitating string solution of Horowitz and Polchinski. Each of them defines a worldsheet CFT and we analyzed whether they could be continuously connected. The \HP solution exists in a low number of dimensions and the most favorable case for our discussion is $D=4$.
It is most favorable because we can think of the transition both in the canonical or microcanonical ensembles. We have considered only Euclidean solutions describing configurations in thermal equilibrium, and we have not addressed the real time question such as what happens to a black hole as it evaporates down to string scale.  

For the heterotic string case we found some evidence that they could be continuously connected as CFTs. The first hint comes from the computation of the supersymmetric index which gives the same answer on both branches. More generally, the two solutions are cobordant, which indicates that there will be no known invariants of $(0,1)$ sigma models that can distinguish them. 
The second hint comes from a linear sigma model construction. From the same linear sigma model we find a space of classical vacua which has the topology of the \HP solution or that of the black hole solution depending on the parameters. These are not Ricci flat, but we expect that  once quantum corrections are included we would flow to either solution depending on the value of one parameter, which can be interpreted at the asymptotic value of the temperature. We also expect that we need to fine tune one of the parameters due to the presence of a negative mode on both sides. 

For the type II case, we found an obstruction, the supersymmetric index is different on both sides. This difference is also related to the different spectrum of D-branes on both sides. There are D-branes on the \HP side which can transform into pure fluxes on the black hole side. This can be also expressed in terms of a difference of K-theory invariants. 

For the type II case, it is natural to conjecture that the two regimes are connected via a singular CFT where the dilaton becomes infinite in some region of the target space. It would be nice to make a reasonable guess for this CFT. In particular, one would like to guess the region of the CFT near the singular point, which could be a non-critical string as in some other examples, such as the conifold theory \cite{Ooguri:1995wj}. In the conifold case,  one the infinite liouville-like direction is visible also in the linear sigma model analysis. In our case, all that is happening in the linear sigma model is the appearance (or disappearance) of massive vacua and it is unclear how this is reflected in the conformal limit. Despite the absence of a new branch in the linear sigma model, one might still try to guess that a liouville-like direction emerges. 
For the $D=4$ case, a simple guess would be a single dimension with a linear dilaton and a central charge $\hat c =4$ (or $c=6$). This has the problem that a further Liouville superpotential would be oscillatory. Another possibility is to include also the time direction so that we have a cylinder. This possibility was considered in \cite{Giveon:2006pr} with a related motivation. However, we would like to see that there are two possible resolutions, or deformations, one giving the \HP solution and one the black hole solution.  The only resolution with the right symmetries is to transform the cylinder into the two dimensional black hole, which should be equivalent to a winding condensate as per the FZZ duality \cite{Kazakov:2000pm}, and not a different deformation!  One would also expect that a certain irrelevant deformation of the model would give a pair of massive vacuum on one side and a no extra massive vacua on the other.

 It would be interesting to embed this discussion in $AdS$. One could consider cases where there is a large $N$ gauge theory dual. In such circumstances the Polyakov loop (or thermal Wilson loop) has been studied, and some transitions have been discussed where the eigenvalue distribution develops a cut \cite{Gross:1980he,Wadia:1980cp,Aharony:2003sx}. It has been proposed in \cite{Alvarez-Gaume:2005dvb,Alvarez-Gaume:2006fwd} that the black hole/string transition corresponds to a situation where a non-uniform eigenvalue distribution develops a cut, namely a region where the eigenvalue distribution vanishes. This is consistent with the idea that the Polyakov loop has a non-zero expectation value on both sides of the transition. This scenario could be explored by considering the expectation values of the trace of powers of the Wilson line, $Tr[U^n]$ with $U$ the thermal holonomy.  We can pick up the various Fourier components of the eigenvalue distribution. These Fourier components should decrease exponentially with $n$ on the side where the eigenvalue distribution is smooth and non-vanishing, but could decay as a power of $n$ if there is a cut. However, in string theory the Wilson lines seems suppressed by the area of the worldsheets, which would lead to an exponential suppression also on the black hole side, which was supposed to be the side having the cut.  
 
 It would be also interesting to consider the $AdS_3$ case with an NS flux. In this case, a similar discussion would suggest the existence of a new solution which would start as similar to the \HP solution around a thermal $AdS$ background but which could interpolate to the BTZ black hole.
 
  In our discussion of the solution generating transformation for the heterotic case, we found that the transformation appears to produce a singularity at the horizon of the black hole. We expect that this should be fixed by $\alpha'$ corrections in the transformation law.  It would be interesting to understand how this happens. The transformation we used   makes a concrete prediction for the $\alpha'$ corrections for the thermodynamics of the charged black hole in the heterotic string \ref{app:alphahet}. So a first principle computation of those $\alpha'$ corrections would be an independent check of our formulas.   
  
  A related problem is that the winding mode for the heterotic string has a half unit of momentum. This would naively suggest that the black hole or \HP solution is breaking the Euclidean time translation symmetry. This seems rather peculiar and we expect that it is likely that the proper physical time translation symmetry should be the unbroken generator which also involves the winding charge. But this is an interesting conceptual question which we have not resolved.\footnote{ We suspect that it is related to the non-trivial transformation law for the $B$ field under local lorentz transformations, as implied by the Green-Schwarz anomaly cancellation mechanism \cite{Green:1984sg}.}
  
  We have discussed an open string analog of the transition. Perhaps using  approximate string field theory techniques one can analyze the solution in the transition region. 
  
  Another interesting problem is to understand the behavior of the solution in $D> 6$, where the \HP solution does not exist. We suspect that in the canonical ensemble there might be a black hole like solution all the way to $\beta = \beta_H$. In the microcanonical ensemble it has been argued in \cite{Chen:2021emg} that for large $D$ there is a solution with even higher temperatures. Of course, we also have the   the highly excited free string phase which, for $D>6$, can exist at  any mass \cite{Horowitz:1997jc}. This can compete with the black hole solutions, but the precise location of a possible phase transition requires the knowledge of the mass and entropy of black holes near the Hagedorn temperature.  

\vspace{1cm}
\textbf{Acknowledgments}

 We are grateful to Ahmed Almheiri, Nima Arkani Hamed, Lorenz Eberhardt, Igor Klebanov, Henry Lin  and S. Wadia for discussions.

J.M. is supported in part by U.S. Department of Energy grant DE-SC0009988, the Simons Foundation grant 385600.  

\appendix 
 
\section{Relating thermodynamic quantities through solution generating procedure} \la{app:rel}

In this appendix we derive (\ref{NewOldF}) starting from (\ref{ThQu}). We start with the simple ones, namely the charges $Q_p$ and $Q_w$. We have
\begin{equation}\label{QpDer}
\begin{aligned}
2\pi R Q_p & = e^{-2\phi_D} \partial_{\mu_p} \log Z' \\
& = e^{-2\phi_D} \partial_{\mu_p} \left[ \frac{R}{\tilde{R}} \log \tilde{Z}' (\tilde{R})\right] \\
& = e^{-2\phi_D} R \frac{\partial \tilde{R}}{ \partial \mu_p} \partial_{\tilde{R}} \left[ \frac{1}{\tilde{R}} \log \tilde{Z}' (\tilde{R})\right] \\
& =  - e^{-2\phi_D} \frac{R}{\tilde{R}^2}\frac{\partial \tilde{R}}{ \partial \mu_p}  \tilde{S}'.
\end{aligned}
\end{equation}
From the first line to the second line, we used (\ref{Partfinx}). We then used that $R$ and $\mu_p$ are independent to pull $R$ out of the derivative, and translated $\partial_{\mu_p}$ into $\partial_{\tilde{R}}$ to get to the third line. In the last equality, we used the definition of $\tilde{S}'$
\begin{equation}\la{defSp}
	e^{-2 \tilde{\phi}_D}\tilde{S}' = \tilde{S} = (1 - \tilde{R} \partial_{\tilde{R}}) \log \tilde{Z} = e^{-2 \tilde{\phi}_D}(1 - \tilde{R} \partial_{\tilde{R}}) \log \tilde{Z}'.
\end{equation}
Cancelling out the $R$ factor on both sides of (\ref{QpDer}), we get the first relation in (\ref{NewOldF}). The derivation for $Q_w$ is the same (we only need to change the subscript from $p$ to $w$).

Next we look at the expression for the entropy. From (\ref{ThQu}) we have
\begin{equation}
\begin{aligned}
	S & = e^{-2\phi_D} (1 - R\partial_{R}) \log Z' \\
	& = e^{-2\phi_D} (1- R \partial_{R}) \left[ \frac{R}{\tilde{R}} \log \tilde{Z}' (\tilde{R})\right] \\
 & = - e^{-2\phi_D} R^2 \partial_{R} \left[ \frac{1}{\tilde{R}} \log \tilde{Z}' (\tilde{R}) \right] \\
& =  - e^{-2\phi_D} R^2 \frac{\partial \tilde{R}}{\partial R} \partial_{\tilde{R}}\left[ \frac{1}{\tilde{R}} \log \tilde{Z}' (\tilde{R}) \right] \\
& =   e^{-2\phi_D} \frac{R^2}{\tilde{R}^2} \frac{\partial \tilde{R}}{\partial R}\tilde{S}'.
\end{aligned}
\end{equation}
We used (\ref{Partfinx}) again to get to the second line. From the third line to the fourth line, we traded the derivative with $R$ for the derivative with $\tilde{R}$. We used (\ref{defSp}) again to get to the last line.

Finally, to get the expression for the energy $E$, we can simply use (\ref{Grand}) and the expressions for $S,Q_p,Q_w$ we derived. In other words,
\begin{equation}\label{EnergyE}
\begin{aligned}
	2\pi R E  & = -\log Z + S + 2\pi R \mu_p Q_p + 2\pi R \mu_w Q_w \\
& = e^{-2 \phi_D} \left[ -\log Z' + \frac{R^2}{\tilde{R}^2} \frac{\partial \tilde{R}}{\partial R}\tilde{S}'- \mu_p \frac{R}{\tilde{R}^2}\frac{\partial \tilde{R}}{ \partial \mu_p}  \tilde{S}'  - \mu_w \frac{R}{\tilde{R}^2}\frac{\partial \tilde{R}}{ \partial \mu_w}  \tilde{S}' \right] \\
& = e^{-2 \phi_D} \left[ -\frac{R}{\tilde{R}}\tilde{S}' + 2\pi R \tilde{E}'  + \frac{R^2}{\tilde{R}^2} \frac{\partial \tilde{R}}{\partial R}\tilde{S}' - \mu_p \frac{R}{\tilde{R}^2}\frac{\partial \tilde{R}}{ \partial \mu_p}  \tilde{S}'  - \mu_w \frac{R}{\tilde{R}^2}\frac{\partial \tilde{R}}{ \partial \mu_w}  \tilde{S}' \right].\\
\end{aligned}
\end{equation}
From the second line to the third line, we used (\ref{Partfinx}) as well as
\begin{equation}
 e^{-2\tilde{\phi}_D} \log \tilde{Z}' =  \log \tilde{Z} = \tilde{S} - 2\pi \tilde{R} \tilde{E} = e^{-2\tilde{\phi}_D}\left( \tilde{S}' - 2\pi \tilde{R} \tilde{E}' \right).
\end{equation}
Dividing $R$ on both sides of (\ref{EnergyE}) and reorganizing the terms, we recover the expression in (\ref{NewOldF}).

 \section{$\alpha'$ corrections for charged black holes from the   corrections for the uncharged one}\la{app:alpha}
 
 In this appendix we show how the formulas in sec. \ref{GenCh} can be used to compute the $\alpha'$ corrections of the thermodynamic quantities of the charged black hole given that we know the corrections for the uncharged one. 
 We discuss the bosonic and type II cases as well as the heterotic one. The bosonic case was considered in \cite{Giveon:2009da}, but here we demonstrate that the same results can be reached in a more efficient way using our formulas. With a similar method we could also compute the dilaton ``charge'' or the coefficient of the $1/\rho^{D-3}$ term in the form of the dilaton at large $\rho$. But we will not do it explicitly here.  
 
 \subsection{Bosonic  string corrections} 
 
 In this case we have corrections at first order in $\alpha'$  \cite{Callan:1988hs} 
\bea \la{ESrsq}
 2\pi \tilde E &=& {(D-2) \over (D-3) } \gamma_D \tilde R^{ D-3} \left( 1 -  \epsilon_D { \lambda \over \tilde R^2 } \right) ~,~~~~~~~  \tilde S = \gamma_D \tilde R^{D-2}  \left( 1 -  \sigma_D { \lambda \over \tilde R^2 } \right) ,\\
 \la{gamdef}
\gamma_D &\equiv&  {   \omega_{D-2} \over 4 G_N} \left( { D-3 \over 2 } \right)^{ D-2} ~,~~~~~ \epsilon_D \equiv {  2 ( D-4)(D-2) \over D-3 } ~,~~~~~~~~ \sigma_D \equiv { 2 ( D-5)(D-2)^2 \over (D-3)^2 } .~~~~~~~~~~~~
\eea 
 with $\lambda_{\rm bos} = {\alpha'\over 2 } $. 
%
Using \nref{GeneBos} we get 
 \begin{equation} \la{GeneBosF}
\begin{aligned}
	R & = \tilde{R} \cosh\alpha \cosh\gamma ~,~~~~~~~~~\lambda_{\rm bos} = {\alpha' \over 2 },\\
	2\pi E &  = e^{ - 2 \phi_D} \gamma_D \tilde R^{D-3} \left[ { (D-2 )\over (D-3)} \left( 1 - \epsilon_D { \lambda \over \tilde R^2 } \right) +  \left( 1 - \sigma_D { \lambda \over \tilde R^2 } \right) \left(\sinh^2 \alpha + \sinh^2 \gamma\right)\right] , \\
	2\pi Q_p & =  r e^{-2 \phi_D} \gamma_D \tilde R^{D-3}   \left( 1 -  \sigma_D  { \lambda \over \tilde R^2 } \right) \sinh \alpha \cosh\alpha ,~~~~~~~~
	\\
	2\pi Q_w & = { 1 \over r} e^{-2 \phi_D} \gamma_D \tilde R^{D-3}   \left( 1 - \sigma_D  { \lambda \over \tilde R^2 } \right) \sinh\gamma \cosh\gamma ,
	\\
	S & =  e^{-2 \phi_D} \gamma_D \tilde R^{D-2}   \left( 1 -  \sigma_D  { \lambda \over \tilde R^2 } \right) \cosh\alpha \cosh\gamma .
\end{aligned}
\end{equation}

 \subsection{Type II string corrections} 
 
 For the type II string theory the corrections appear at order $\alpha'^3$ \cite{Chen:2021qrz}
 \bea 
 2 \pi \tilde E &=& {(D-2) \over (D-3)}\gamma_D \tilde R^{D-3} \left( 1 - \hat \epsilon_D {z  \over \tilde R^6}  \right) ,
 \cr  \tilde S &=&   \gamma_D \tilde R^{D-2} \left(1 - \hat \sigma_D { z \over \tilde R^6} \right) ,
 \eea 
 with $\gamma_D$ as in \nref{gamdef}, $ z=  { \zeta(3) \over 16 } {\alpha'}^3$ and 
 \bea
 \hat \epsilon_D &\equiv& { 16 \over 3 }{ (D-8)(330 - 489 D + 242 D^2 - 51 D^3 + 4 D^4) \over (D-3)^5 } ,
 \cr 
\hat \sigma_D &\equiv&  { 16 \over 3 } { (D-9 ) (D-2) (330 - 489 D + 242 D^2 - 51 D^3 + 4 D^4)\over (D-3)^6 }    .
  \eea 
 Then using \nref{GeneBos} we get 
   \begin{equation} \la{GeneIIF}
\begin{aligned}
	R & = \tilde{R} \cosh\alpha \cosh\gamma ,~~~~~~~~~ \\
	2\pi E &  = e^{ - 2 \phi_D} \gamma_D \tilde R^{D-3} \left[ { (D-2 )\over (D-3)} \left( 1 - \hat\epsilon_D { z \over \tilde R^6 } \right) + \left( 1 - \hat \sigma_D { z\over \tilde R^6 } \right) \left(\sinh^2 \alpha + \sinh^2 \gamma\right)\right], \\
	2\pi Q_p & =  r e^{-2 \phi_D} \gamma_D \tilde R^{D-3}   \left(  1 - \hat \sigma_D { z\over \tilde R^6 } \right) \sinh \alpha \cosh\alpha ,~~~~~~~~
	\\
	2\pi Q_w & = { 1 \over r} e^{-2 \phi_D} \gamma_D \tilde R^{D-3}   \left(  1 - \hat \sigma_D { z\over \tilde R^6 } \right) \sinh\gamma \cosh\gamma ,
	\\
	S & =  e^{-2 \phi_D} \gamma_D \tilde R^{D-2}   \left(  1 - \hat \sigma_D { z\over \tilde R^6 }  \right) \cosh\alpha \cosh\gamma .
\end{aligned}
\end{equation}

 \subsection{Heterotic string corrections} \la{app:alphahet}
 
 For the heterotic string we again have corrections at first order in $\alpha'$ \cite{Callan:1988hs}. They are the same as in \nref{ESrsq}, but with $\lambda_{\rm het} = \alpha'/4$ (and the same values of $ \epsilon_D$ and $ \sigma_D$ as in \nref{gamdef}).
  Then using \nref{GeneHet} we get 
 \begin{equation}\label{GeneHetF}
\begin{aligned}
	R & = \tilde{R} \cosh\alpha \cosh\gamma \left( 1 + { \tanh \alpha \tanh \gamma  \over 2 \tilde R^2 } \right) ~,~~~~~\lambda_{\rm het} = {\alpha' \over 4 } ,\\
	2\pi E &  =  e^{ - 2 \phi_D} \gamma_D \tilde R^{D-3} \left[ { (D-2 )\over (D-3)} \left( 1 -  \epsilon_D { \lambda \over \tilde R^2 } \right) + \right. 
	\cr  & ~~~~~~~~~~~~~\left.+ \left( 1 - \sigma_D { \lambda \over \tilde R^2 } \right) \left(\sinh^2 \alpha + \sinh^2 \gamma\right)\right], \\
		2\pi Q_p & =  r e^{-2 \phi_D} \gamma_D \tilde R^{D-3}   \left( 1 - \sigma_D  { \lambda \over \tilde R^2 } \right) \sinh \alpha \cosh\alpha ,~~~~~~~~
	\\
	2\pi Q_w & = { 1 \over r} e^{-2 \phi_D} \gamma_D \tilde R^{D-3}   \left( 1 -  \sigma_D  { \lambda \over \tilde R^2 } \right) \sinh\gamma \cosh\gamma ,
	\\
	S & = e^{ -2 \phi_D}  \gamma_D \tilde R^{D-2}   \left( 1 - \sigma_D  { \lambda \over \tilde R^2 } \right)  \cosh\alpha \cosh\gamma\left( 1 - { \tanh \alpha \tanh \gamma  \over 2 \tilde R^2 } \right) ,
\end{aligned}
\end{equation}
where in the expression for the entropy we can expand to first non-trivial order in $1/\tilde R^2$. 
These formulas do not agree with the ones in  \cite{Giveon:2009da}, since those do not include the terms involving the ${\tanh \alpha \tanh \gamma \over 2 \tilde R^2} $ corrections. 

\section{Explicit steps for generating the charged solution in the  sigma model regime} \la{App:ExplicitGene}
 
In this appendix we check that the steps outlined in (\ref{CFTarg}) indeed lead to (\ref{NewPar}), when applied to a CFT with a gravity limit. The procedure here applies to the bosonic and the Type II case. Our starting point is the $(D+1)$-dimensional metric
\begin{equation}
	ds^2 = \tilde{R}^2 dt^2 + \tilde{r}^2 d\tilde{y}^2 + ds_{D-1}^2,\quad  ~~~~t \sim t+ 2\pi, \, ~~~~~~\tilde{y} \in (-\infty, \infty).
\end{equation}
The functions $\tilde{R}^2,\tilde{r}^2$ as well as the dilaton $\tilde{\phi}_D$ only depend on the coordinates of the $(D-1)$ dimensions. There is no non-zero $B$ field to start with. The transformations below do not change the $(D-1)$-dimensional part of the metric, so we leave it implicit and will omit it from the formulas below. Now we follow the steps in (\ref{CFTarg}):

\begin{itemize}
\item In step one, we quotient the $t,\tilde{y}$ directions by a combined translation $t \sim t + 2\pi u,~ \tilde{y} \sim \tilde{y} + 2\pi$. This is just one transformation changing both coordinates and not two independent transformations. We can also write it as 
$(t,\tilde y) \sim (t,\tilde y) + 2\pi ( u,1)$. As part of the original identification we also have $(t, \tilde y) \sim (t , \tilde y) + 2 \pi ( 1,0)$. This identification goes along for the ride and we will not discuss it further. 
\item 
We now do a T-duality along the same direction that we identified. In practice a T-duality can be implemented by gauging the symmetry in question  and adding to the lagrangian a term of the form $ i \hat { y}  d A$ where $\hat {  y} $ is the dual variable \cite{Rocek:1991ps}. In our particular case this leads to a Lagrangian of the form\footnote{We use the short hand notation 
$(\partial y - A)^2 = \eta^{\alpha \beta} ( \partial_\alpha y - A_\alpha) (\partial_\beta y - A_\beta)$ and $dA = \epsilon^{\alpha \beta} \partial_\alpha A_\beta  $ where $\alpha, \beta$ are two dimensional indices. We omitted an overall $1/(4\pi)$ factor multiplying the Lagrangian. } 
\be 
 \tilde R^2 ( \partial t - u A)^2 + \tilde r^2 (\partial \tilde y - A)^2 + 2 i \hat y d A 
 \ee 
 where under a gauge transformation of $A$, $A\to A + d \epsilon$,  we have  $(t,\tilde y) \to (t,\tilde y) + \epsilon( u, 1)$. Here we take $A$ to be a compact gauge field.  We will choose $\hat y$ to be non-compact at this step. So this is a T-duality followed by a decompactification of the dual circle. This is not equivalent to the original theory, but this is not a problem. The point is  that we can generate a new theory from the original one.  We now choose the gauge $\tilde y=0$ and we integrate out $A$.  We find 
 \begin{equation}
\begin{aligned} \la{equtd}
ds^2 
& = \frac{\tilde{R}^2 \tilde{r}^2}{ \tilde{r}^2 + u^2 \tilde{R}^2 } dt^2 + { 1 \over \tilde{r}^2 + u^2 \tilde{R}^2  } d \hat y^2 
 \\
 B_{  t \hat y } &=  \frac{u \tilde{R}^2}{ \tilde{r}^2 + u^2 \tilde{R}^2  }  \\
 e^{ - 2 \phi } &= { e^{-2 \tilde \phi}   (\tilde{r}^2 + u^2 \tilde{R}^2 ) }  .
\end{aligned}
\end{equation} 
On this step we have started to generate a non-trivial dilaton, with $\tilde \phi$ being the original dilaton.
 We also have generated a $B$ field.
 %
\item 
We now quotient by a symmetry of the form $(\hat y ,   t) \to (\hat y,   t) + 2 \pi (1,v)$. 
 \item 
   We now do a T-duality again, using the same generator that was used for the quotient in the last step.  In practice, this involves adding a gauge field again, call it $\hat A$, so that from \nref{equtd} we are led to a lagrangian of the form 
   \be 
   \frac{\tilde{R}^2 \tilde{r}^2}{ \tilde{r}^2 + u^2 \tilde{R}^2 } ( \partial t -  v \hat A)^2 + { 1 \over (\tilde{r}^2 + u^2 \tilde{R}^2 ) } (\partial  \hat y - \hat A)^2 +  \frac{2i u \tilde{R}^2}{ \tilde{r}^2 + u^2 \tilde{R}^2  } (d t -v \hat A) \wedge (d \hat y - \hat A) -2 i \check y d\hat A 
\ee
In this case we choose $\check y$ to be compact. We choose the gauge $\hat y=0$ and integrate out $\hat A$ to find the final form of the metric and the $B$ field 
%
   \bea 
   ds^2 &=&  { \tilde R^2  \tilde r^2 \over (\tilde{r}^2 + u^2  \tilde R^2) (1 + v^2 \tilde R^2 \tilde r^2 )} d   t^2+   { (\tilde r^2 + u^2 \tilde R^2) \over (1 + \tilde R^2 \tilde r^2 v^2) } \left[ d \check y - { u  \tilde R^2 \over (\tilde r^2 + u^2 \tilde R^2) } d  t \right]^2    
   \cr 
   & ~& B_{  t  \check y } = - { v \tilde r^2 \tilde R^2 \over (1 + \tilde R^2 \tilde r^2 v^2) } ~,~~~~~~~
    e^{ -2\phi} = { e^{ -2 \tilde \phi}   (1 + \tilde R^2 \tilde r^2 v^2 ) } \la{FinSol}
\eea

Of course, we could have implemented all these steps in one go by applying the O(2,2) transformation. We chose to do them step by step to argue that we can generalize the procedure to a general seed theory.  
\end{itemize}
The dilaton in \nref{FinSol} is the $D+1$ dimensional dilaton. In previous formulas we have considered the $D$ dimensional one. To obtain that one, we need to multiply by the appropriate factor of the radius of the coefficient in brackets in the first line of \nref{FinSol}.  


More explicitly, choosing the seed solution as
\begin{equation}
 ds^2 = fdt^2 + \frac{d\rho^2}{f} + \rho^2 d\Omega_{d-1}^2 + \tilde r^2 d\tilde{y}^2, \quad f = 1-\frac{\mu}{\rho^{D-3}},
\end{equation}
where now $\tilde r$ is just a constant, we arrive at the final form of the metric \cite{Sen:1994eb}:
\begin{equation}
\begin{aligned} \la{BHcharged}
 & ds^2 = f_w^{-1} f_p^{-1} f d\tau^2 + \frac{d\rho^2}{f} + \rho^2 d\Omega_{d-1}^2 +  r^2 \frac{f_p}{f_w} \left[ d\check y - \frac{i\tanh \alpha f d\tau}{f_p} \right]^2 \\
 & B_{\tau \check y } = -i \tanh \gamma r \frac{f}{f_w}, \quad e^{-2\phi_{D+1}} = e^{-2\phi_{D+1, \, \infty}} f_w,
\end{aligned}
\end{equation}
where
\begin{equation}
	f_p = 1+ \frac{\mu \sinh^2 \alpha}{\rho^{D-3}}, \quad f_w 	= 1+ \frac{\mu \sinh^2 \gamma}{\rho^{D-3}}, \quad r = \tilde r \frac{\cosh\gamma}{\cosh\alpha}, \quad \tau \sim \tau + \frac{4\pi \mu^{\frac{1}{D-3}}}{(D-3) \cosh\alpha\cosh\gamma}  
\end{equation}
where the angles $\alpha, ~\gamma$ are defined via relations analogous to \nref{HyAng}, namely 
 \be 
 \tanh \alpha = i{u \tilde R \over \tilde r }   ~,~~~~~~~~~~
 \tanh \gamma = i{ v \tilde r \tilde R } . 
 \ee 

\small
\bibliographystyle{ourbst}

 \bibliography{DraftHP}
\end{document}